\documentclass[12pt]{article}
\usepackage{amsmath,epsfig,graphicx,verbatim,subfigure}
\usepackage{epstopdf}
\usepackage{color}

\usepackage{multirow}
\usepackage{array}

\include{epsf} 
\def\ltap{\raisebox{-.6ex}{\rlap{$\,\sim\,$}} \raisebox{.4ex}{$\,<\,$}} 
\def\gtap{\raisebox{-.6ex}{\rlap{$\,\sim\,$}} \raisebox{.4ex}{$\,>\,$}}

\newcommand\as{\alpha_{\mathrm{S}}} 
 
\def\beq{\begin{equation}} 
\def\eeq{\end{equation}} 
\def\beeq{\begin{eqnarray}} 
\def\eeeq{\end{eqnarray}} 
 
\def\to{\rightarrow}

\def\mgg{M_{\gamma \gamma}}

\def\Dpgg{\Delta \Phi_{\gamma\gamma} }
\def\dy{\Delta y_{\gamma \gamma}}
\def\lpt{d\sigma^{(l)}/dp_{T \gamma \gamma}}
\def\spt{d\sigma^{(s)}/dp_{T \gamma \gamma}}
\def\ETmax{E_{T\,{\rm max}}}

\newcommand{\percent}{{\%}}

\begin{document} 
\begin{titlepage}
\renewcommand{\thefootnote}{\fnsymbol{footnote}}
\begin{flushright}
ZH-TH 06/18
     \end{flushright}
\par \vspace{2mm}
\vspace*{-1cm}
\begin{center}
{\Large \bf 
Diphoton production at the LHC:\\[0.2cm]
a QCD study up to NNLO }
\end{center}

\par \vspace{2mm}
\begin{center}
{\bf Stefano Catani${}^{(a)},$ Leandro Cieri${}^{(b,c)},$
Daniel de Florian${}^{(d)}$,\\[0.2cm]
Giancarlo Ferrera${}^{(e)}$}
and
{\bf Massimiliano Grazzini${}^{(c)}$
}\\

\vspace{5mm}

$^{(a)}$ INFN, Sezione di Firenze and Dipartimento di Fisica e Astronomia,\\ 
Universit\`a di Firenze,
I-50019 Sesto Fiorentino, Florence, Italy

$^{(b)}$ INFN, Sezione di Milano-Bicocca,\\
Piazza della Scienza 3, I-20126 Milano, Italy

$^{(c)}$ Physik-Institut, Universit\"at Z\"urich, 
CH-8057 Zurich, Switzerland

${}^{(d)}$ International Center for Advanced Studies (ICAS), 
ECyT-UNSAM, \\ Campus 
Miguelete, 25 de Mayo y Francia, (1650) Buenos Aires, Argentina 

${}^{(e)}$ Dipartimento di Fisica, Universit\`a di Milano and\\ INFN, 
Sezione di Milano,
I-20133 Milan, Italy

\vspace{2mm}

\end{center}

\par \vspace{2mm}
\begin{center} {\large \bf Abstract} \end{center}
\begin{quote}
\pretolerance 10000

We consider the production of prompt-photon pairs
at the LHC 
and we report on a study of QCD radiative corrections up to the 
next-to-next-to-leading order (NNLO).
We present a detailed comparison of next-to-leading order (NLO) results obtained 
within the standard and smooth cone isolation criteria, by studying
the dependence on the isolation parameters.
We highlight the role of different partonic subprocesses within the two 
isolation criteria, and we show that they produce large radiative corrections 
for both criteria.
Smooth cone isolation is a consistent procedure 
to compute QCD radiative corrections at NLO and beyond.
If photon isolation is sufficiently tight, 
we show that 
the NLO results for the 
two isolation procedures are consistent with each other
within their perturbative uncertainties.
We then extend our study to NNLO by using smooth cone isolation.
We discuss the impact of the NNLO corrections and the corresponding perturbative
uncertainties for both fiducial cross sections and distributions, and we comment 
on the comparison with some LHC data. Throughout our study we remark on the main 
features that are produced by the kinematical selection cuts 
that are applied to the photons.
In particular, we examine soft-gluon singularities that appear in the perturbative 
computations of the invariant mass distribution of the photon pair,
the transverse-momentum spectra of the photons, and the fiducial cross section
with asymmetric and symmetric photon transverse-momentum cuts,
and we present their behaviour in analytic form.

\end{quote}

\vspace*{\fill}
\begin{flushleft}
February 2018

\end{flushleft}
\end{titlepage}

\setcounter{footnote}{2}
\renewcommand{\thefootnote}{\fnsymbol{footnote}}
\section{Introduction}

The production of photon pairs (diphotons) with high invariant mass at high-energy
hadron colliders is a very relevant process in the context of both 
Standard Model (SM) studies 
and searches for new-physics signals.

Experimentally a pair of photons is a very clean final state, and photon
energies and momenta can be measured with high precision 
in modern electromagnetic calorimeters.
Since photons do not interact strongly with other final-state particles, 
\emph{prompt} photons represent ideal probes to test
the properties of the SM and corresponding theoretical predictions
(see Refs.~\cite{Aaltonen:2011vk}--\cite{Aaboud:2017vol}
for recent experimental analyses of Tevatron and LHC diphoton data).
Measurements involving a pair of isolated photons have played a crucial role 
in the 
discovery at the LHC \cite{Aad:2012tfa,Chatrchyan:2012xdj}
of a Higgs boson, whose properties are compatible with those of the SM one.
Studies of the Higgs boson properties in the diphoton decay mode have been
performed \cite{Khachatryan:2014ira,Aad:2014eha}.
Diphoton measurements
(see, e.g.,
Refs.~\cite{:2012afa}--\cite{Khachatryan:2015qba})
are also important in many new-physics scenarios, 
including
searches for extra dimensions  
or supersymmetry. 
The relevance of LHC measurements of the diphoton invariant-mass spectrum 
is highlighted by the recent observation 
\cite{ATLAS750}--\cite{Khachatryan:2016hje}
of an excess of events with invariant mass of about 750~GeV that might have
indicated the presence of resonances over the diphoton SM background.
That observation raised a great deal of attention from December 2015 till the
time of the 2016 Summer Conferences \cite{Khachatryan:2016yec,Aaboud:2017yyg}.

Owing to its physics relevance, the study of diphoton production requires accurate
theoretical calculations which, in particular, include QCD radiative corrections at
high perturbative orders.
In high-energy collisions, final-state prompt photons with high transverse momentum
can be originated through {\em direct} production from hard-scattering subprocesses
and through {\em fragmentation} subprocesses of QCD partons.
The theoretical computation of fragmentation subprocesses requires 
non-perturbative information, in the form of parton fragmentation functions 
of the photon, 
which typically has large associated uncertainties.
However, the effect of fragmentation
contributions is significantly reduced by the photon {\em isolation criteria} 
that are necessarily
applied in hadron collider experiments to suppress the very large reducible 
background of `non-prompt' photons (e.g., photons that are faked by jets
or produced by hadron decays). 
Two such criteria are the so-called `standard' cone isolation and the `smooth'
cone isolation proposed by Frixione \cite{Frixione:1998jh}.
The standard cone isolation is the criterion that is typically
used by experimental analyses. This criterion can be experimentally implemented
in a relatively straightforward manner, but it only suppresses part 
of the fragmentation contribution.
By contrast, the smooth cone isolation 
(formally) eliminates the entire fragmentation contribution, although, due to 
the finite granularity of the detectors, it cannot be directly
applied at the experimental level in its original form.
Owing to the absence of fragmentation contributions, theoretical calculations
are much simplified by using the smooth cone isolation, and it is relatively
simpler to compute radiative corrections at high perturbative orders.
Considering calculations at the next-to-leading order (NLO) 
in the QCD coupling $\alpha_S$, 
in Ref.~\cite{Butterworth:2014efa} it was shown that, if the isolation 
is `tight enough', the two isolation criteria lead to theoretical NLO results 
that are quantitatively very similar for various observables in
diphoton production processes at high-energy hadron colliders.

In the present paper we deal with the diphoton production process 
$pp\rightarrow \gamma \gamma X$, where $p$ is a colliding proton and
$X$ denotes the inclusive final state that accompanies the $\gamma \gamma$ pair. 
QCD radiative corrections to this process were first computed up to the 
NLO
in Ref.~\cite{Binoth:1999qq}. This is a complete NLO calculation of both the 
direct and fragmentation (specifically, single- and
double-fragmentation) components of the cross section. The calculation is
implemented at the fully-differential level
in the numerical Monte Carlo code \texttt{DIPHOX}, which can be used
to perform computations with any infrared and collinear safe isolation criteria
(including the standard and smooth cone criteria). The \texttt{DIPHOX}
calculation also includes the so-called {\em box contribution} 
\cite{Dicus:1987fk} to the partonic channel $gg \to \gamma \gamma$
(which is formally a contribution of next-to-next-to-leading order (NNLO)
type), since this contribution can be quantitatively enhanced by the large
gluon--gluon parton luminosity at high-energy hadron--hadron colliders.
The next-order gluonic corrections to the box contribution 
(which are part of the N$^3$LO QCD corrections to diphoton production)
were computed in Ref.~\cite{Bern:2002jx}
(and implemented in the numerical program \texttt{GAMMA2MC}) 
and found to have a moderate quantitative effect.
An independent NLO calculation \cite{Campbell:2011bn}
is implemented in the code \texttt{MCFM}, which, however, includes the
fragmentation component only at the leading order (LO), while the box
contribution is treated according to the \texttt{GAMMA2MC} code.
A complete calculation at the NNLO of {\em both} the direct and fragmentation
components still nowadays remains computationally very challenging.

Fragmentation contributions are absent by considering 
smooth cone isolation. In the context of smooth cone isolation, diphoton
production at the LO (i.e. at $\mathcal{O}(\alpha_S^0)$) 
emerges \textit{via} 
the quark--antiquark annihilation subprocess $q\bar{q}\rightarrow \gamma\gamma$.
The NLO QCD corrections 
are due to
quark--antiquark annihilation and to new partonic channels (\textit{via} 
the subprocess $qg \rightarrow \gamma \gamma q$
and ${\bar q}g \rightarrow \gamma \gamma {\bar q}$)
with an initial-state colliding gluon.
At the NNLO
the $gg$ channel 
starts to contribute
and it (the entire channel and not only its box contribution)
can be fully consistently included in the perturbative QCD calculation.
The large value of the gluon parton distribution function
(PDF) of the colliding proton
makes the gluon
initiated channels important, especially at small and intermediate
values of the diphoton invariant mass.
The scattering amplitudes that are needed to evaluate the NNLO QCD corrections
were computed in 
Ref.~\cite{Dicus:1987fk,Barger:1989yd,Bern:1994fz,Anastasiou:2002zn}, 
and they were used in Ref.~\cite{Catani:2011qz} to perform the first
NNLO calculation of diphoton production in hadronic collisions. The NNLO
calculation at the fully-differential level is based on the 
$q_T$ subtraction method~\cite{Catani:2007vq} and it
was implemented \cite{Catani:2011qz} in the numerical code 
\texttt{2$\gamma$NNLO}.
More recently, an independent calculation of diphoton production at the NNLO
has been performed in Ref.~\cite{Campbell:2016yrh} by using the method of
$N$-jettiness subtraction \cite{Boughezal:2015dva, Gaunt:2015pea}.
The NNLO calculation of Ref.~\cite{Campbell:2016yrh} is implemented 
\cite{Boughezal:2016wmq} in the \texttt{MCFM} program.
An independent NNLO computation of diphoton production at the NNLO, 
based on the $q_T$ subtraction method, has been implemented in the general purpose
NNLO generator \texttt{MATRIX} \cite{matrix}.

The transverse-momentum spectrum of the photon pair is sensitive to
logarithmically-enhanced contributions at high perturbative orders.
Transverse-momentum resummation at next-to-next-to-leading logarithmic (NNLL) 
accuracy for inclusive diphoton production was implemented \cite{Balazs:2007hr}
in the \texttt{ResBos} code, and more recently the complete NNLL calculation has been
combined with the NNLO contributions and implemented in the numerical program 
\texttt{2$\gamma$Res} \cite{Cieri:2015rqa}.
 
Using smooth cone isolation,  
associated production of diphotons and jets has been computed up to NLO
for various jet multiplicities, namely, 
$\gamma \gamma + 1$~jet \cite{DelDuca:2003uz, Gehrmann:2013aga}
(the fragmentation component at LO is included in the calculations of 
Refs.~\cite{Gehrmann:2013aga,Campbell:2014yka}),
$\gamma \gamma + 2$~jets \cite{Gehrmann:2013bga, Bern:2014vza, Badger:2013ava},
$\gamma \gamma + 2~\!b$-jets \cite{Faeh:2017fpp}
and $\gamma \gamma + 3$~jets \cite{Badger:2013ava}.
Owing to the computational simplifications of smooth cone isolation
(with respect to standard cone isolation), some hadron collider
processes with one
final-state photon, such as associated $Z\gamma$ 
\cite{Grazzini:2013bna, Grazzini:2015nwa, Campbell:2017aul}
and $W\gamma$ \cite{Grazzini:2015nwa} production
and inclusive single-photon production \cite{Campbell:2016lzl}, 
have also been computed up to
the NNLO in QCD perturbation theory.

Lowest-order electroweak radiative corrections to diphoton production at LHC
energies have been computed in Refs.~\cite{Bierweiler:2013dja,Chiesa:2017gqx}.
They produce small effects on NNLO QCD results for inclusive diphoton
production. The effects can increase by selecting photons with 
transverse momenta in the~TeV region.

In this paper we present studies of NLO and NNLO QCD radiative corrections to
inclusive diphoton production at LHC energies. The results at the NNLO
are based on the theoretical calculation of Ref.~\cite{Catani:2011qz},
as implemented in the fully-differential Monte Carlo program 
\texttt{2$\gamma$NNLO}, and on the independent NNLO calculation implemented in \texttt{MATRIX} \cite{matrix}.
The first version of \texttt{2$\gamma$NNLO},
which was originally used in Ref.~\cite{Catani:2011qz}, 
had a numerical
implementation bug that was subsequently corrected.
A main result of Ref.~\cite{Catani:2011qz} is that NNLO radiative corrections are substantial 
for diphoton kinematical configurations of interest at high-energy hadron
colliders. The selected quantitative results that were presented in 
Ref.~\cite{Catani:2011qz} are mostly related to diphoton kinematical
configurations that are typically examined in Higgs boson searches and studies
at the LHC. In this paper we consider more general kinematical configurations.
In particular, we present NLO and NNLO studies
related to photon isolation, and we discuss various detailed features of the
theoretical results up to NNLO.

The results of the code \texttt{2$\gamma$NNLO} have been
used by experimental 
collaborations in some of their data/theory comparisons. The analyses performed
at the
Tevatron ($\sqrt{s}=1.96$~TeV) by the CDF~\cite{Aaltonen:2012jd} 
and D0~\cite{Abazov:2013pua} collaborations 
and at the LHC ($\sqrt{s}=7$~TeV and 8~TeV)
by the ATLAS~\cite{Aad:2012tba, Aaboud:2017vol} 
and CMS \cite{Chatrchyan:2014fsa}
collaborations 
show that the inclusion of the NNLO corrections greatly improves the
description of diphoton production data.
The NNLO predictions have also been used in data analyses related to
new-physics searches in high-mass diphoton events at 
$\sqrt{s}=13$~GeV~\cite{CMS750}.
In summary, these measurements prove that
the NNLO results 
are important to understand phenomenological aspects of diphoton production.


The paper is organized as follows.
Section~\ref{sec:isol} is devoted to a comprehensive study of photon isolation.
In particular, we perform a detailed comparison of the standard and smooth
isolation criteria in the context of perturbative QCD results at LO and NLO.
We discuss the role of different partonic subprocesses within the two different
isolation criteria. We also remark on the effects that are produced by the
isolation parameters and by the kinematical selection cuts that are
applied to the photons. Specifically, in Sect.~\ref{sec:criteria}
we introduce the photon isolation criteria and comment on some of
their features. The QCD calculations that are used in our study are briefly
described in Sect.~\ref{subsec:theory}. The quantitative results for total and
differential cross sections are presented in Sect.~\ref{subsec:res}.
Section~\ref{sec:sigmatot} is devoted to the LO and NLO results for total
cross sections. Results for differential cross sections at LO and NLO are
presented in Sects.~\ref{sec:diffLO} and \ref{sec:diffNLO}, respectively.
In Sect.~\ref{sec:results} we present detailed NNLO results for diphoton 
production within the smooth isolation prescription.
We study the perturbative stability of the results and we discuss related
theoretical uncertainties by considering several observables that are relevant 
in diphoton production at hadronic colliders.
We also discuss the comparison of 
the NNLO predictions to recent LHC data \cite{Aad:2012tba}. 
Specifically, results for total cross sections and differential cross sections
are presented in Sects.~\ref{sec:totalNNLO} and \ref{sec:diffNNLO},
respectively. In Sect.~\ref{sec:ispar} we present some results on the dependence
on the isolation parameters. The comparison with LHC data is discussed 
in Sect.~\ref{sec:datavsth}. In Sect.~\ref{sec:ptcut} we discuss the effects of
(asymmetric and symmetric) photon transverse-momentum cuts on total and
differential cross sections and, in particular, we comment on related 
perturbative (soft-gluon) instabilities.
Finally, 
in Sect.~\ref{sec:summa} we summarize our results.

\section{Photon isolation}
\label{sec:isol}

\subsection{Isolation criteria}
\label{sec:criteria}

Hadron collider experiments at the Tevatron and the LHC do not perform 
{\it inclusive} photon measurements. The background of secondary photons 
coming from the decays of $\pi^{0}$, $\eta$, etc. overwhelms the signal 
by several orders of magnitude and the experimental selection of prompt 
diphotons requires {\em isolation} cuts (or criteria) to reject this background. 
The \textit{standard cone} isolation and the 
\textit{smooth cone} isolation
are two 
of these criteria. Both criteria consider the amount of hadronic (partonic)
transverse energy\footnote{For each four-momentum $p_i^\mu$, the corresponding
transverse momentum ($p_{Ti}$),  transverse energy ($E_{Ti}$), 
rapidity ($y_i$) and azimuthal angle ($\Phi_i$) are defined in the
centre--of--mass frame of the colliding hadrons. Angular distances 
$R_{i \gamma}$ are defined in rapidity--azimuthal angle space ($R_{i \gamma}^2=
(y_i-y_\gamma)^2 + (\Phi_i-\Phi_\gamma)^2$).}
$E_{T}^{had}(r)= \sum_i E_{T\,i}^{had} \;\Theta(r-R_{i \gamma})$
inside a cone of radius $r$ around the direction of the photon momentum 
${\bf p}_\gamma$. Then the isolated photons are selected by limiting the value
of $E_{T}^{had}(r)$.

The standard cone isolation criterion fixes the size $R$ of the radius of the
isolation cone and it requires
\begin{equation}
\label{Eq:Isol_standard}     
E_{T}^{had}(R) \leq \,\ETmax \;\;,    
\end{equation}
where the isolation parameter $\ETmax$ can be either a fixed value of
transverse energy or a function of the photon transverse momentum 
$p_{T \gamma}$ (i.e., 
$\ETmax = \epsilon\, p_{T \gamma}$ with a fixed parameter $\epsilon$).
A combination of these two options is also possible: for instance,
$\ETmax= 0.05\, p_{T \gamma} + 6$~GeV is used
in the study of
Refs.~\cite{ATLAS750,Aaboud:2016tru}.

Provided $\ETmax$ is finite (not vanishing) standard cone isolation leads
to infrared-safe cross sections \cite{Catani:2002ny} 
in QCD perturbation theory.
Parton radiation exactly collinear with the direction of the photon
momentum is allowed by the constraint in Eq.~(\ref{Eq:Isol_standard}) and,
as a consequence, the treatment of standard cone isolation within perturbative
QCD requires the introduction of parton to photon fragmentation functions.
Decreasing the value of $\ETmax$ reduces and suppresses the effect
of the fragmentation function (and of the corresponding partonic subprocesses).

The smooth cone isolation criterion \cite{Frixione:1998jh} (see also 
Refs.~\cite{Frixione:1999gr,Catani:2000jh})
also fixes the size $R$ of the isolation
cone and it requires
\begin{eqnarray}
\label{Eq:Isol_frixcriterion} 
 E_{T}^{had}(r) \leq \,\ETmax ~\chi(r;R) \;, \quad
 \mbox{ in {\it all} cones with} \;r \leq R 
\;\;,    
\end{eqnarray}
with a suitable choice of the $r$ dependence of the isolation
function $\chi(r;R)$. 
The two key properties \cite{Frixione:1998jh} of the isolation
function are: $\chi(r;R)$ has to smoothly vanish
as the cone radius $r$ vanishes 
($\chi(r) \rightarrow 0 \;,\; \mbox{if} \;\; r \rightarrow 0\,$), and it
has to fulfil the condition 
\mbox{$\; 0<\chi(r;R) \leq 1$} (in particular, $\chi$ must not vanish)
for any finite (non-vanishing) value of $r$.
Since $E_{T}^{had}(r)$ does not increase by decreasing $r$,
in practice the requirement in Eq.~(\ref{Eq:Isol_frixcriterion}) is 
effective
only if $\chi(r;R)$
monotonically decreases as $r$ decreases.

The smooth cone isolation criterion
implies that, closer to the photon, less hadronic activity 
is allowed. The amount of energy deposited by parton radiation at angular
distance $r=0$ from the photon is required to be exactly 
equal to zero, and the fragmentation component (which has a purely collinear 
origin in perturbative QCD) of the cross section vanishes completely.
The cancellation of perturbative QCD soft divergences still takes place 
as in ordinary infrared-safe cross sections, since parton radiation is not
forbidden in any finite region of the phase space \cite{Frixione:1998jh}. 
It is also preferable to
choose isolation functions $\chi(r;R)$ with a sufficiently smooth dependence on
$r$ over the entire range $0< r < R$. In particular, large discontinuities of 
$\chi(r;R)$ at finite values of $r$ are potential sources of instabilities
\cite{Catani:1997xc} in fixed-order perturbative calculations. Small
discontinuities of the function $\chi(r;R)$ (such as those in the discretized
version \cite{Binoth:2010nha}
of smooth cone isolation) are instead acceptable.

A customary choice of the isolation function $\chi(r;R)$ is 
\begin{equation}
\label{Eq:Isol_chinormal}
\chi(r;R) = \left( \frac{1-\cos (r)}{1-\cos (R)} \right)^{n}\;,
\end{equation}
where the value of the power $n$ is typically set to $n=1$.
We also consider the following isolation function:
\begin{equation}
\label{Eq:Isol_chirpow}
\chi(r;R) = \left( \frac{r}{R} \right)^{2n}\;,
\end{equation}
whose value depends on the ratio $r/R$ (rather than $r$ and $R$,
independently).
The two functions in Eqs.~(\ref{Eq:Isol_chinormal}) and 
(\ref{Eq:Isol_chirpow}) are equal at the isolation cone boundary $r \to R$
($\chi(r;R) \to 1$) and they behave similarly as $r\to 0$ 
($\chi(r;R) \propto r^{2n}$).

Comparing the isolation requirements in Eqs.~(\ref{Eq:Isol_standard})
and (\ref{Eq:Isol_frixcriterion}) by using the same values of $R$ and $\ETmax$ in both
equations, we see that smooth cone isolation is more restrictive than standard
cone isolation. Therefore, the following physical constraint applies:
\begin{equation}
\label{eq:a}
d\sigma_{\rm smooth}(R;\ETmax) < d\sigma_{\rm standard}(R;\ETmax)
\;\;,
\end{equation}
where $d\sigma$ generically denotes total cross sections and differential cross
sections with respect to photon kinematical variables, and the subscripts
`smooth' and `standard' refer to smooth and standard isolation, respectively.
Note that 
the isolation parameters $R$ and $\ETmax$ are 
set at the same values
in the two isolated cross sections, 
$d\sigma_{\rm smooth}$ and $d\sigma_{\rm standard}$, that are compared
in the inequality (\ref{eq:a})
(e.g., the inequality is not necessarily valid if smooth isolation at a given
value of $\ETmax$ is compared with standard isolation at a different and
smaller value of $\ETmax$).
An analogous reasoning applies to the cross section dependence on the isolation
parameters $\ETmax$ and $R$, since the isolation
requirement can become more or less restrictive by varying these parameters.
Therefore, we have the following physical behaviour:
\begin{equation}
\label{eq:b}
d\sigma_{\rm is}(R;\ETmax) \;\;
\mbox{monotonically decreases as}~\ETmax~\mbox{decreases}
\;\;(R~\mbox{fixed}) \;\;,
\end{equation}
\begin{equation}
\label{eq:c}
d\sigma_{\rm is}(R;\ETmax) \;\;
\mbox{monotonically increases as}~R~\mbox{decreases}
\;\;(\ETmax~\mbox{fixed}) \;\;,
\end{equation}
\begin{equation}
\label{eq:d}
d\sigma_{\rm smooth}(R;\ETmax;n) \;\;
\mbox{monotonically decreases as}~n~\mbox{increases}
\;\;(R~\mbox{and}~\ETmax~\mbox{fixed}) \;\;,
\end{equation}
and the subscript `is' equally applies to both isolation criteria
(e.g., `is'=`smooth' or `is'=`standard'). 
The relation (\ref{eq:d}) refers to the dependence on the power $n$ in the case
of the isolation function in Eqs.~(\ref{Eq:Isol_chinormal}) or 
(\ref{Eq:Isol_chirpow}) (a similar relation applies to the cross section
dependence by considering two isolation functions $\chi_1(r)$ and
$\chi_2(r)$ such that $\chi_1(r) > \chi_2(r)$).

The standard cone isolation criterion is simpler and, 
as stated in the Introduction, it is the criterion that is
used in experimental analyses at hadron colliders
(the actual experimental selection of isolated photons, including isolation
requirements, is definitely much more involved than the relatively
straightforward implementation of the criterion).
The experimental implementation of smooth cone isolation
(in its strict original form)
is complicated\footnote{There is activity related to the experimental 
implementation~\cite{Binoth:2010nha,AlcarazMaestre:2012vp,Blair:1379880,
Wielers:2001suas} 
of the discretized version of smooth cone isolation. 
An experimental implementation of the smooth isolation criterion was 
done by the OPAL collaboration~\cite{Abbiendi:2003kf} in a study of
isolated-photon production in
photon--photon collisions at LEP.
A discretized version of smooth isolation is used in the QCD calculation 
of Ref.~\cite{Frye:2015rba}.} especially 
by the finite granularity of the Tevatron and LHC detectors.

Independently of the intrinsic differences between the standard and smooth 
cone isolation criteria, Eq.~(\ref{eq:a}) implies that a {\em reliable}
(theoretical or experimental) cross section result with smooth cone isolation
represents a lower bound on the corresponding (i.e., with the same values 
of $\ETmax$ and $R$) cross section result with standard cone isolation
(this statement is valid within reliable estimates of related theoretical 
or experimental uncertainties). In particular, this also implies
that a comparison between theoretical smooth isolation results and experimental
standard isolation results leads to meaningful and valuable information.

The comparison between smooth and standard isolation can be sharpened
by considering tight isolation requirements.
As expected on general grounds, the differences between the two isolation
criteria should be reduced in the case of tight isolation cuts. This expectation
is confirmed by the diphoton studies in 
Refs.~\cite{Butterworth:2014efa, Badger:2016bpw}, 
which show theoretical NLO results that
are similar for the two isolation criteria if the isolation parameters
are tight enough (e.g., $R=0.4$ and $\ETmax < 5~$GeV or 
$\ETmax= \epsilon\, p_{T \gamma}$ with $\epsilon < 0.1$ as in Sect.~11
of Ref.~\cite{Butterworth:2014efa} or Sect.~4.3.1 
of Ref.~\cite{Badger:2016bpw}).

In Sect.~\ref{subsec:res} we present detailed results for smooth and 
standard isolation.
We study the dependence on the isolation parameter $\ETmax$ and its effects
on both total cross sections and differential cross sections in various
kinematical regions.

\subsection{Perturbative QCD calculations}
\label{subsec:theory}

In this paper we present quantitative results on QCD
radiative corrections to diphoton production by using both smooth cone and
standard cone isolations.
We consider total cross sections (more precisely, fiducial cross sections)
and differential cross sections as functions of various kinematical variables
of the prompt photons.
The kinematical variables that we use are the diphoton invariant mass 
$M_{\gamma \gamma}$, the photon polar angle $\theta^*$ in the Collins--Soper
rest frame \cite{Collins:1977iv} of the diphoton system, the azimuthal angle
separation $\Delta \Phi_{\gamma \gamma}$ between the two photons, the diphoton
transverse momentum $p_{T \gamma \gamma}$ and the transverse momenta
$p_{T \gamma}^{hard}$ and $p_{T \gamma}^{soft}$ 
($p_{T \gamma}^{hard} > p_{T \gamma}^{soft}$) of the harder and softer photon
of the diphoton pair. As previously specified,
azimuthal angles and transverse momenta are defined in the centre--of--mass
frame of the colliding hadrons.

\setcounter{footnote}{2}

The QCD results on smooth cone isolation are obtained by using the numerical
programs \texttt{2$\gamma$NNLO} \cite{Catani:2011qz} and \texttt{MATRIX} \cite{matrix},
which include perturbative QCD contributions up to
NNLO. The NNLO calculations are based on the $q_T$
subtraction method \cite{Catani:2007vq}, which is applicable to hadroproduction
processes of generic high-mass systems of colourless particles and it has been
already applied to explicit NNLO calculations of several specific processes
(Higgs boson \cite{Catani:2007vq,Grazzini:2008tf} and vector boson 
\cite{Catani:2009sm,Catani:2010en} production, associated production of a Higgs
boson and a vector boson \cite{Ferrera:2011bk,Ferrera:2014lca}, diboson
production processes such as $Z\gamma$ 
\cite{Grazzini:2013bna,Grazzini:2015nwa}, $W^\pm \gamma$ \cite{Grazzini:2015nwa},
$ZZ$ \cite{Cascioli:2014yka,Grazzini:2015hta}, 
$W^+W^-$ \cite{Gehrmann:2014fva,Grazzini:2016ctr} and 
$W^{\pm}Z$ \cite{Grazzini:2016swo} production, 
Higgs boson pair production \cite{deFlorian:2016uhr},
associated production of a Higgs boson pair and a $W$ or a 
$Z$ boson \cite{Li:2016nrr}).
The $q_T$ subtraction method has recently been extended to heavy-quark
production and applied to the NNLO calculation of top quark pair production
\cite{Bonciani:2015sha}. 

In the case of diphoton production, the method combines the NLO cross section
$d\sigma_{\gamma \gamma + {\rm jets}}$ for the production of the photon pair plus one
or two jets (partons) with an appropriate subtraction counterterm (its explicit
expression is given in Ref.~\cite{Bozzi:2005wk})
and a hard-virtual contribution (see Ref.~\cite{Catani:2013tia})
for diphoton production with no additional final-state jets (partons).
In the code \texttt{2$\gamma$NNLO} the cross section 
$d\sigma_{\gamma \gamma + {\rm jets}}$ is computed up to NLO by using the
dipole-subtraction method \cite{Catani:1996vz}
as implemented in the diphoton calculation of Ref.~\cite{DelDuca:2003uz}.
In \texttt{MATRIX} the cross section $d\sigma_{\gamma \gamma + {\rm jets}}$ is computed at NLO by using
the fully automated implementation of the dipole-subtraction method within the Monte Carlo program \texttt{MUNICH}\footnote{\texttt{MUNICH} is the 
abbreviation of ``MUlti-chaNnel Integrator at Swiss~(CH) precision'' --- an automated parton-level NLO 
generator by S.~Kallweit.},
and all (spin- and colour-correlated)
tree-level and one-loop amplitudes are obtained from \texttt{OpenLoops} \cite{Cascioli:2011va}.
The combination of $d\sigma_{\gamma \gamma + {\rm jets}}$ and its counterterm
is numerically finite, although the two contributions 
are separately divergent in
the limit of vanishing transverse momentum $p_{T \gamma \gamma}$ of the photon
pair. In \texttt{2$\gamma$NNLO} the cancellation of the separate divergences is numerically treated by
introducing a lower limit $q_{T \rm cut}$ on $p_{T \gamma \gamma}$
($p_{T \gamma \gamma} > q_{T \rm cut}$) and considering small values of 
$q_{T \rm cut}$ (formally performing the limit 
$q_{T \rm cut} \to 0$). 
Decreasing the value of $q_{T \rm cut}$
reduces the size of systematical errors (due to the finite value of 
$q_{T \rm cut}$) at the expense of increasing computational time and resources
to handle numerical instabilities. As a practical compromise, 
in our study we use
a finite value of $q_{T \rm cut}$ in the range $q_{T \rm cut}=0.1$~GeV--0.2~GeV.
The NNLO generator \texttt{MATRIX} implements instead a lower limit $r_{\rm cut}$ on the ratio $p_{T\gamma\gamma}/M_{\gamma\gamma}$ ($p_{T \gamma \gamma} > r_{\rm cut} M_{\gamma\gamma}$) \cite{matrix}, and we use values in the range $r_{\rm cut}=0.08\%$--$0.15\%$.

Owing to the finite values of $q_{T \rm cut}$ or $r_{\rm cut}$,
a systematical uncertainty of about $\pm {\cal O}(1\%)$ affects
all the NNLO results\footnote{At NLO the generator \texttt{MATRIX} can also use
the dipole-subtraction method \cite{Catani:1996vz}, which does not require the
regularization parameter $r_{\rm cut}$. Our diphoton NLO results have been 
quantitatively cross-checked by numerical comparisons of the calculations that use
$q_T$ subtraction (\texttt{2$\gamma$NNLO}, \texttt{MATRIX}) and dipole subtraction
(\texttt{MATRIX}).} 
presented in this
paper.
The quoted systematical uncertainty is sufficient for 
the purpose of the studies that we present throughout the paper.
It is substantially smaller than 
the size of additional (NLO) NNLO theoretical uncertainties that we find, for
instance, by performing customary variations of the factorization and
renormalization scales at (NLO) NNLO.
We also note that, at fixed value of $q_{T \rm cut}$ ($r_{\rm cut}$), the NNLO systematical
error on differential cross sections tends to be larger than the corresponding
error on total cross sections.
In particular,
some specific observables (and, more precisely, their value
in restricted kinematical regions) that are very sensitive to the {\em detailed}
shape of the $p_{T \gamma \gamma}$ distribution in the limit 
$p_{T \gamma \gamma} \to 0$ can be affected by larger NNLO systematical errors
due to the finite value of $q_{T \rm cut}$ ($r_{\rm cut}$): these observables may require
refined numerical studies with smaller values of $q_{T \rm cut}$ ($r_{\rm cut}$). Nonetheless,
these same specific observables are (expected to be) affected by increased
theoretical uncertainties due to large higher-order perturbative corrections
(instabilities).
Examples of these observables are the differential cross section
$d\sigma/d\Delta \Phi_{\gamma \gamma}$ at $\Delta \Phi_{\gamma \gamma} \simeq
\pi$, and the differential cross sections $d\sigma/dM_{\gamma \gamma}$,
$d\sigma/dp_{T \gamma}^{hard}$ and $d\sigma/dp_{T \gamma}^{soft}$ 
(or related integrated quantities) in soft-gluon
sensitive kinematical regions (see Sect.~\ref{sec:ptcut}).

We use \texttt{2$\gamma$NNLO} and \texttt{MATRIX} to obtain results at LO, NLO and NNLO in QCD
perturbation theory. At each order our calculation includes all the terms
(and only those terms) that contribute to the total cross section at the
corresponding perturbative order according to the formal expansion in powers of
$\alpha_S$. Therefore (in the context of smooth cone isolation), only the
initial-state $q{\bar q}$ partonic channel contributes at LO, the 
initial-state $qg$ and ${\bar q}g$ channels start to contribute at NLO, and the 
initial-state $gg$ channel starts to contribute at NNLO. In particular the box
contribution $gg \to \gamma\gamma$ \cite{Dicus:1987fk}
only enters at NNLO, together with all the partonic subprocesses
(e.g., including the gluon initiated subprocess $gg \to \gamma\gamma q{\bar q}$) 
that contribute at the same order. We do not include the NLO correction
\cite{Bern:2002jx} to the box contribution, since it is part of the complete
(and still unknown) N$^3$LO corrections to inclusive diphoton production.
All the partonic subprocesses are treated in the massless-quark framework with
$N_f=5$ quark flavours. In particular, we do not include 
NNLO contributions
with virtual top quarks since they are not yet fully known
(e.g., the loop top quark contribution to the two-loop 
scattering amplitude $q{\bar q} \to \gamma\gamma$ has not
yet been computed in complete form \cite{Becchetti:2017abb}). 
The effect of including the NLO correction 
\cite{Bern:2002jx} and the top quark correction to the (massless-quark)
box contribution
$gg \to \gamma\gamma$ has been considered in the diphoton calculation of
Ref.~\cite{Campbell:2016yrh}. The top quark correction to the box contribution
is also studied in Refs.~\cite{Kawabata:2016aya} and \cite{Jain:2016kai}.
 
The results on standard cone isolation are obtained by using the  
\texttt{DIPHOX} calculation \cite{Binoth:1999qq}, which includes QCD radiative
corrections up to the NLO.
The NLO calculation that is implemented in \texttt{DIPHOX} is based on a 
combined use
\cite{Binoth:1999qq,Chiappetta:1996wp} of the subtraction and phase-space 
slicing methods. The slicing approximation is applied \cite{Binoth:1999qq}
to the phase-space region where the minimum transverse-momentum of the three
final-state partons at NLO is smaller than $p_{Tm}$
(formally considering the limit $p_{Tm} \to 0$).
Our numerical results are obtained by using $p_{Tm}=0.1$~GeV,
which is the default value of the slicing parameter (regulator) $p_{Tm}$ 
that is
suggested in the \texttt{DIPHOX} program.
Since we are interested in an order-by-order comparison with smooth cone 
isolation
results, the box contribution $gg \to \gamma\gamma$ is not included in the 
\texttt{DIPHOX} NLO results.
Note, however, that at LO and NLO all the initial-state partonic channels
($q{\bar q}, qg, {\bar q}g, gg$) contribute in \texttt{DIPHOX} because of the
presence of a non-vanishing fragmentation component (though the 
fragmentation component is quantitatively suppressed by the 
isolation procedure).
Owing to the double-fragmentation component, 
even the initial-state $gg$ channel
is not vanishing at LO (at NLO the initial-state $gg$ channel contributes also
through the single-fragmentation component). 

We note that (due to the limited number of final-state partons in fixed-order
computations) LO
calculations do not cover the entire kinematical region for inclusive diphoton
production. LO calculations give non-vanishing cross sections only in limited 
LO {\em kinematical subregions}. Outside these subregions, 
only the NLO results start
to give non-vanishing cross section contributions. Therefore, outside the
LO kinematical subregions the NLO (NNLO) results effectively represent 
an LO (NLO) perturbative QCD prediction. In spite of the effective meaning 
of the
results, we always use the default labels LO, NLO and NNLO according to the 
perturbative order
in which the results contribute to the inclusive (total) cross section.
For example, in the case of the $\Delta \Phi_{\gamma \gamma}$ distribution, the
LO kinematical subregion has $\Delta \Phi_{\gamma \gamma}=\pi$: the region of
small values of $\Delta \Phi_{\gamma \gamma}$ receives contributions only from
the NLO and NNLO results that represent effective LO and NLO predictions,
respectively. We also note that the LO kinematical subregions can be different
in the context of smooth cone and standard cone isolation. For instance,
at LO $p_{T \gamma \gamma}=0$ and $p_{T \gamma}^{hard} = p_{T \gamma}^{soft}$
in the case of smooth cone isolation, whereas
$p_{T \gamma \gamma}$ can be different from zero 
and $p_{T \gamma}^{hard} \geq p_{T \gamma}^{soft} 
\geq p_{T \gamma}^{hard} - \ETmax$ in the case of
standard cone isolation
(non-vanishing values of $p_{T \gamma \gamma}$ are produced through the LO
fragmentation component of the cross section).
We also comment about the dependence on the isolation parameters $R$ and
$\ETmax$ (the comment has similarities with our previous comment on
LO kinematical subregions). The LO results are independent of the size $R$
of the isolation cone. The LO cross section depends on $\ETmax$
in the case of standard cone isolation (the $\ETmax$ dependence is due to
the fragmentation component), while it is independent of the value of 
$\ETmax$ in the case of smooth cone isolation.

In Eqs.~(\ref{eq:a})--(\ref{eq:d}) we have listed some constraints on isolated
photon cross sections. These are physical constraints (properties) in the sense
that they apply to physical (`positive definite') events: if the isolation
requirements are more (less) restrictive, the selected number of isolated 
photons
decreases (increases). Such properties do not `a priori' apply to theoretical
calculations based on perturbative  QCD (of course, if the constraints are not
fulfilled, the perturbative QCD calculation is not a reliable approximation of
the physical result) since, beyond the effective LO approximation, they involve
negative-weight contributions (which are due to virtual radiative corrections
and to subtraction contributions related to the factorization procedure of
parton distribution functions and fragmentation functions).

As is well known, fixed-order perturbative results can show unphysical behaviour
in specific kinematical regions. In the case of diphoton production, for
instance, it is known \cite{Binoth:1999qq}
that the differential cross sections at small values of
$p_{T \gamma \gamma}$ ($p_{T \gamma \gamma} \to 0$) or large values of 
$\Delta \Phi_{\gamma \gamma}$ ($\Delta \Phi_{\gamma \gamma} \to \pi$)
cannot be reliably computed in fixed-order perturbation theory: the disease is
due to large logarithmic corrections that have to be resummed to all 
perturbative
orders \cite{Balazs:2007hr,Cieri:2015rqa}
to recover the physical behaviour and predictivity.

In the presence of photon isolation cuts, perturbative computations of total
cross sections can also show some misbehaviour. Isolated (with both smooth and
standard isolation) photon cross sections fulfil the physical constraint 
$\sigma_{\rm is}(R;\ETmax) < \sigma_{\rm inc}$, where 
$\sigma_{\rm inc}$ is the inclusive (non-isolated) cross section. Nonetheless,
in the case of standard cone isolation at NLO this constraint is violated
\cite{Catani:2002ny}
at very small values ($R \ltap 0.1$) of the radius $R$ of the isolation cone.
The violation is due to large logarithmic ($\ln R$) corrections, and the
physical behaviour is recovered through a corresponding all-order resummation
procedure \cite{Catani:2013oma}. We expect a (qualitatively) similar $\ln R$
dependence in the case of smooth cone isolation, though in the present paper we
do not consider very small values of $R$. 

Violation of expected properties is not necessarily related to
logarithmically-enhanced
QCD corrections: it can be simply a consequence of a slowly convergent
(toward a reliable estimate of the physical result) perturbative expansion.
The LO, NLO and NNLO results obtained in Ref.~\cite{Catani:2011qz}
with smooth cone isolation certainly indicate that we are not dealing with a fastly
convergent perturbative expansion in the case of diphoton production at
high-energy hadron colliders.
Additional complications can occur in direct comparisons 
(as in Eq.~(\ref{eq:a})) between calculations with smooth and standard
isolation. The complications follow from the fact that such a comparison
does not involve ingredients that are in one-to-one correspondence. 
Owing to the presence of the fragmentation component, 
at each perturbative order the standard isolation result depends on the photon
fragmentation function, on the corresponding factorization scale $\mu_{frag}$
and on related partonic subprocesses. The poorly known fragmentation function
certainly affects the standard isolation results and, especially, the comparison
with smooth isolation results (which have no analogue of the 
fragmentation function). 

Throughout the paper
we comment on these and additional points related to physical behaviour and
perturbative QCD calculations 

\setcounter{footnote}{2}

\subsection{Quantitative results}
\label{subsec:res}

In our theoretical study of standard and smooth isolation
we consider isolated diphoton production in $pp$
collisions at the centre--of--mass energy $\sqrt{ s}=7$~TeV.
We apply the following kinematical cuts on photon transverse momenta and
rapidities: $p_{T \gamma}^{\rm hard} \geq 25$~GeV, 
$p_{T \gamma}^{\rm soft}\geq 22$~GeV and the rapidity of both photons 
is limited in the range $|y_\gamma|<2.37$.
The minimum angular distance
between the two photons is $R_{\gamma \gamma}^{\rm min}=0.4$.

These are basically the kinematical cuts used in the ATLAS Collaboration study of
Ref.~\cite{Aad:2012tba}.
The analysis of Ref.~\cite{Aad:2012tba}
is restricted to a smaller rapidity region since it excludes the rapidity
interval $1.37 < |y_\gamma|< 1.52$, which is outside the acceptance of the
electromagnetic calorimeter. For the sake of simplicity, in this section we do
not consider such additional rapidity restriction; 
the rapidity restriction is instead
applied in the results of the following Sect.~\ref{sec:results}.

In the perturbative calculation,
the QED coupling constant $\alpha$ is fixed at the value $\alpha=1/137$.
We use the 
MMHT 2014 sets \cite{Harland-Lang:2014zoa} of parton distribution functions 
(PDFs),
with parton
densities and $\as$ evaluated at each corresponding perturbative order
(i.e., we use the $(k+1)$-loop running $\as$ at N$^k$LO, with $k=0,1,2$).
We use the NLO photon fragmentation functions of Ref.~\cite{Bourhis:1997yu},
and specifically the BFG set~II (we have checked that BFG set~I leads to very small
quantitative differences). The central value $\mu_0$ of the 
renormalization scale ($\mu_R$), PDF factorization scale ($\mu_F$) and 
fragmentation function scale ($\mu_{frag}$) is set to be equal to the
invariant mass of the diphoton system, $\mu_0=\mgg$.
We compute scale variation uncertainties by considering the two asymmetric scale
configurations with 
$\{ \mu_R=\mu_0/2$, $\mu_F=\mu_{frag}=2\mu_0 \}$ and 
$\{ \mu_R=2\mu_0$ , $\mu_F=\mu_{frag}=\mu_0/2 \}$.

More precisely,  we have considered independent scale variations by a factor of two
up and down with respect to the central value $\mu_0$. We find a common overall
behaviour of the cross sections: they increase by decreasing $\mu_R$  and
decrease by decreasing $\mu_F$ or $\mu_{frag}$. Therefore the two 
asymmetric scale configurations are those that maximize the scale dependence
within the considered scale variation range.
The sole exception regards the invariant mass cross section at large values of
$\mgg$ ($\mgg \gtap 200$~GeV): 
this kinematical region is
sensitive to larger values of the parton momentum fraction $x$ in the PDFs and,
as a consequence, it turns out that
the cross section decreases by increasing~$\mu_F$.

\begin{figure}[htb]
\begin{center}
\begin{tabular}{cc}
\includegraphics[width=0.49\textwidth]{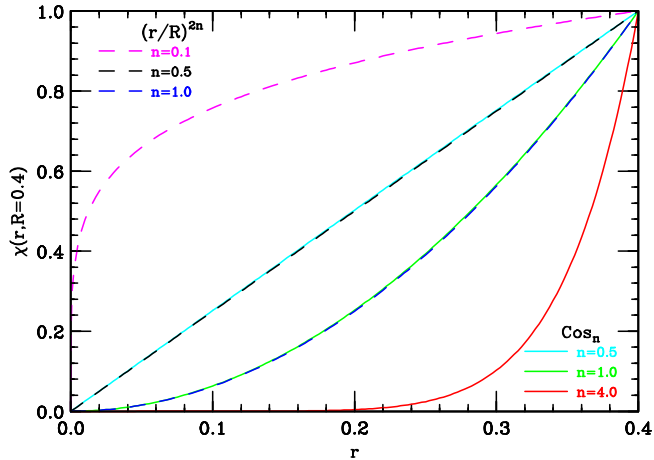}
\end{tabular}
\end{center}
\caption{\label{fig:Chir}
{\em The different shapes of the isolation functions $\chi(r;R)$ 
for selected values of the power $n$ and with $R=0.4$. 
The functions $\chi(r;R)$ in Eqs.~(\ref{Eq:Isol_chinormal})
and (\ref{Eq:Isol_chirpow}) are labelled as $(r/R)^{2n}$ (dashed lines)
and Cos$_n$ (solid lines), respectively.
}}
\end{figure}

The radius $R$ of the isolation cone is fixed at $R=0.4$. We study the isolation
parameter dependence by considering values of $\ETmax$  in the range
between  2~GeV and 10~GeV 
(in this section we mainly show
results at the extreme
values in this range). In the case of smooth cone isolation, we use the isolation
function $\chi(r;R)$ in Eq.~(\ref{Eq:Isol_chirpow})
and we examine the cross section dependence on the power $n$ 
that controls the shape of $\chi(r;R)$.
In Fig.~\ref{fig:Chir} we show the $r$ dependence of the isolation function 
for some selected values of the power $n$ and the fixed value $R=0.4$
of the radius of the isolation cone.
We note that the value $R=0.4$ is sufficiently small so that
the two isolation functions in Eqs.~(\ref{Eq:Isol_chinormal})
and (\ref{Eq:Isol_chirpow}) quantitatively coincide at 
the percent level in the case with $n=1$.

We present perturbative QCD
results for both standard and smooth isolation and, in particular,
we perform comparisons by using the same value of the isolation parameter 
$\ETmax$ for both criteria. The comparison between the two criteria
can also be performed differently. For instance, having fixed the values of
$R$ and $\ETmax$ for standard isolation, one can use different values of
isolation parameters ($R, \ETmax$ and $n$) for smooth isolation to the
purpose of trying to obtain similar quantitative results for the two criteria, 
as a pragmatic approach to mimic the standard cone isolation that is used in
experimental conditions.
We think that our comparison with the same value of $\ETmax$ (and $R$) is
more informative to investigate and understand differences and similarities
between perturbative QCD results for the two criteria.

The QCD results on standard cone isolation depend on the parton-to-photon
fragmentation function
$D_{a/\gamma}(z;\mu_{frag})$ $(a=q,{\bar q},g)$, $z$ being the photon momentum
fraction with respect to the momentum of the fragmenting parton $a$.
Owing to the isolation procedure, the value of $z$ is bounded by a minimum value
$z_{\rm min}$ ($1 \geq z \geq z_{\rm min}$), and this leads to a
quantitative suppression of the fragmentation component of the diphoton
cross section. The typical value of $z_{\rm min}$ is 
$z_{\rm min} \sim p_{T \gamma}/(p_{T \gamma}+\ETmax)$, $p_{T \gamma}$ being
the transverse momentum of the photon that is involved in the fragmentation
process. In our quantitative study we use relatively-large values of 
$p_{T \gamma}$ (i.e., typically, $p_{T \gamma} > 22$~GeV) and relatively-small
values of $\ETmax$. Therefore, $z_{\rm min}$ is always large 
($z_{\rm min} \gtap 0.9$ at $\ETmax=2$~GeV, and still 
$z_{\rm min} \gtap 0.7$ at $\ETmax=10$~GeV), and the 
suppression factor~\footnote{At the formal level 
$\frac{\as}{\alpha} D_{a/\gamma}$ is the order of magnitude of the ratio between
the fragmentation component and the direct component.}
due to $\frac{\as}{\alpha} D_{a/\gamma}$ is sizeable
(roughly one order of magnitude or more,
depending on $\ETmax$) \cite{Bourhis:1997yu}.
We note that at such high values of $z$ the quark (or antiquark)
fragmentation function $D_{q/\gamma}$ (or $D_{{\bar q}/\gamma}$) is much larger
(roughly by more than a factor of ten) than the gluon fragmentation function
$D_{g/\gamma}$ \cite{Bourhis:1997yu}. In our calculation we consistently 
(according to the formal perturbative expansion) include all the fragmentation
functions. However, due to the dominance of $D_{q/\gamma}$ and
$D_{{\bar q}/\gamma}$, in all our qualitative (or semi-quantitative) comments we
neglect the effect of $D_{g/\gamma}$ (i.e., we can assume that only
$D_{q/\gamma}$  and $D_{{\bar q}/\gamma}$ contribute). We also note that,
because of QCD scaling violation, 
at high values of $z$, $D_{a/\gamma}(z;\mu_{frag})$ increases (although weakly)
by increasing $\mu_{frag}$.

\subsubsection{Total cross sections at LO and NLO}
\label{sec:sigmatot}

We begin the presentation of our quantitative results by considering the total
cross section (namely, the fiducial cross section for the applied kinematical
cuts on the photons). Values of total cross sections at LO and NLO
for both smooth and standard isolation are reported in 
Table~\ref{Table:Isol1}.

\begin{table}

\begin{center}

\renewcommand{\arraystretch}{1.5}

\begin{tabular}{ |c||c|c|c|c| }

\hline

\multirow{2}{*}{ } & \multicolumn{2}{ |c| }{$\ETmax = 2$~GeV} &
\multicolumn{2}{ |c| }{$\ETmax = 10$~GeV}\\

\cline{2-5}

& $\sigma^{{\rm LO}}$~~(pb) & $\sigma^{{\rm NLO}}$~~(pb) & $\sigma^{{\rm
LO}}$~~(pb)  & $\sigma^{{\rm NLO}}$~~(pb) \\\hline

\multirow{1}{*}{Standard} & $12.15~^{+14.5\, \percent }_{-14.3\,
\percent }$

 & $31.1~^{+12.8\, \percent }_{-12.3\, \percent }$&$19.51~^{+25.0\,
\percent }_{-20.8\, \percent }$  & $33.3~^{+12.3\, \percent }_{-11.3\,
\percent }$ \\

\multirow{1}{*}{$\left[ {\rm direct} \right]$} & $10.56~^{+10.7\,
\percent }_{-12.0\, \percent }$  & $27.30~^{+7.8\, \percent }_{-9.2\,
\percent }$ & $10.56~^{+10.7\, \percent }_{-12.0\, \percent }$  &
$18.45~^{-10.3\, \percent }_{+3.8\, \percent }$ \\

\hline

\multirow{1}{*}{Smooth} & $10.56~^{+10.7\, \percent }_{-12.0\, \percent
}$  & $31.92~^{+12.6\, \percent }_{-12.1\, \percent }$ &
$10.56~^{+10.7\, \percent }_{-12.0\, \percent }$ & $33.91~^{+13.0\,
\percent }_{-12.6\, \percent }$ \\

\hline

\end{tabular}


\caption{\label{Table:Isol1}
{\em Results for LO and NLO total cross sections with two values of 
$\ETmax$ and the photon kinematical cuts described in the text
(beginning of Sect.~\ref{subsec:res}). The results are obtained by using smooth
and standard isolation (the direct component of the standard isolation cross
section is also reported). 
The NLO smooth isolation results use the isolation function
$\chi(r;R)=\left( r/R \right)^{2n}$ with $n=1$.
The central values of $\sigma$ are obtained with the
scale choice $\mu_R=\mu_F=\mu_{frag}=\mgg$ and the scale dependence corresponds
to $\{ \mu_R=\mgg/2$, $\mu_F=\mu_{frag}=2\mgg \}$ (upper values) and
$\{ \mu_R=2\mgg$ , $\mu_F=\mu_{frag}=\mgg/2 \}$ (lower values). The last digit
of each cross section value has an error of one unit
from the statistical uncertainty of the numerical calculation.
}}
\renewcommand{\arraystretch}{1}

\end{center}

\end{table}

Using smooth cone isolation, the total cross section at LO is
$\sigma^{LO}_{\rm smooth}= 
10.56~{\rm pb}~^{+10.7\, \percent }_{-12.0\, \percent }$~(scale),
where the percentage variation 
refers to the scale dependence of the result\footnote{Throughout 
the paper, any quantitative
statements about scale dependence refer to scale variation effects that are
computed as described in the text at the beginning of Sects.~\ref{subsec:res} 
and \ref{sec:results}, respectively.}.
We note that $\sigma^{LO}_{\rm smooth}$ is independent of $\ETmax$ and 
of the isolation function $\chi(r;R)$. 
The cross section is produced
only by the initial-state $q{\bar q}$ channel through the partonic subprocess
$q {\bar q} \to \gamma \gamma$,
and the scale dependence of 
$\sigma^{LO}_{\rm smooth}$ (which is entirely due to variations of $\mu_F$
in the quark and antiquark PDFs) is quite small. All these features are a very
crude approximation of the diphoton production dynamics.

Using standard cone isolation, the LO cross section $\sigma^{LO}_{\rm standard}$
depends on $\ETmax$ and it also depends on all the three scales $\mu_F, \mu_R$
and $\mu_{frag}$ ($\mu_R$ and $\mu_{frag}$ enter through the fragmentation
component).
The scale dependence of $\sigma^{LO}_{\rm standard}$ is relatively similar 
to that of $\sigma^{LO}_{\rm smooth}$: we find 
$^{+14.5\, \percent }_{-14.3\, \percent }$~(scale)
and 
$^{+25.0\, \percent }_{-20.8\, \percent }$~(scale)
with $\ETmax=2$~GeV and $\ETmax=10$~GeV, respectively.
The $\ETmax$ dependence of $\sigma^{LO}_{\rm standard}$ is instead more
`surprising'. Considering $\ETmax=2$~GeV (very tight isolation),
$\sigma^{LO}_{\rm standard}$ is slightly larger, by about 15\%, than 
$\sigma^{LO}_{\rm smooth}$ (actually the two cross sections are very similar
within the corresponding scale dependence). However, 
$\sigma^{LO}_{\rm standard}$ increases by a factor of about 1.6 in going from 
$\ETmax=2$~GeV to $\ETmax=10$~GeV and, at $\ETmax=10$~GeV,  
$\sigma^{LO}_{\rm standard}$ is roughly a factor of 2 larger than 
$\sigma^{LO}_{\rm smooth}$.

Since $\sigma^{LO}_{\rm smooth}$ does not depend on the isolation parameters
($R,n,\ETmax$), there is obviously no way to approximate (or, mimic) the
quantitative value of $\sigma^{LO}_{\rm standard}$ at $\ETmax=10$~GeV
by using smooth cone isolation with different isolation parameters (e.g., 
a value
of $\ETmax$ larger than 10~GeV).

The value $\ETmax=10$~GeV is not particularly large and it cannot be regarded
as a very loose isolation parameter. Therefore, on physical grounds, we do not
expect two actual features of the LO result: the large difference between 
$\sigma^{LO}_{\rm standard}$ and $\sigma^{LO}_{\rm smooth}$ at 
$\ETmax=10$~GeV, and the large $\ETmax$ dependence of 
$\sigma^{LO}_{\rm standard}$ in going from 
$\ETmax=2$~GeV to $\ETmax=10$~GeV. We mean that physical events are not
expected to have these features, since such features cannot be regarded as a
physical consequence of the hadronic activity inside the photon
isolation cones and of
its detailed structure (which leads to an ensuing sensitivity to the isolation
criteria). These observed LO features require some explanation, 
which we are going to discuss.

At the LO, in the context of standard isolation, the distinction between direct
and fragmentation components is unambiguous, and the direct component {\em
exactly} coincides with the entire contribution to the smooth isolation result.
The double-fragmentation component always gives a small contribution
(few percent at $\ETmax=10$~GeV and few permill at $\ETmax=2$~GeV)
to $\sigma^{LO}_{\rm standard}$. Therefore, the $\ETmax$ dependence of
$\sigma^{LO}_{\rm standard}$ is due to the single-fragmentation component, whose
contribution to $\sigma^{LO}_{\rm standard}$ is small (of the order of 10\%)
at $\ETmax=2$~GeV (because of the suppression due to the small value of 
$\ETmax$) and sizeable at $\ETmax=10$~GeV. At the larger value of 
$\ETmax$ the direct and single-fragmentation components have the same size,
but this is not due to the fact that the fragmentation function is particularly
large. At the LO the direct component only involves the partonic process
\beq
\label{qqbarlo}
q {\bar q} \to \gamma \gamma \;\;,
\eeq
whereas the (single) fragmentation component
also involves the partonic process  
\beq
\label{qglofrag}
q g \to \gamma q  \quad\quad (q \to \gamma +X) \;\;,
\eeq
where the notation $q \to \gamma +X$ denotes the fragmentation of the
final-state quark $q$
into a photon through $D_{q/\gamma}$ (a similar notation is used in subsequent
equations for partonic subprocesses).
The PDF luminosity (the detailed definition of PDF luminosities is presented in
Eq.~(\ref{lum}))
${\cal L}_{q {\bar q}}$ of the initial-state direct
subprocess is sizeably smaller than the luminosity ${\cal L}_{q g}$
of the initial-state fragmentation subprocess (this follows from the smaller
size of the antiquark PDF with respect to the gluon PDF at small values,
such as $x \sim 10^{-2}$,
of parton momentum fraction $x$ in the PDFs), and the suppression
due to the isolated fragmentation function
$D_{q/\gamma}$ is compensated by the increased size of ${\cal L}_{q g}$
with respect to ${\cal L}_{q {\bar q}}$
(the suppression from $D_{q/\gamma}$ is much stronger by decreasing $\ETmax$,
because $z > z_{\rm min}$ and $z_{\rm min}$ increases 
by decreasing $\ETmax$).

According to our discussion, the presence of 
the fragmentation process of Eq.~(\ref{qglofrag})
in the LO standard cone isolation explains the quantitative dependence of 
$\sigma^{LO}_{\rm standard}$ on $\ETmax$ and the quantitative differences 
between $\sigma^{LO}_{\rm standard}$ and $\sigma^{LO}_{\rm smooth}$. At the same
time,
our discussion is useful to anticipate expected features of the NLO results.
The NLO calculation within smooth cone isolation includes 
the partonic process
\beq
\label{qgnlo}
q g \to \gamma \gamma q \;\;.
\eeq
In the kinematical configuration where the
final-state quark is inside the isolation cone of one photon, 
the process in Eq.~(\ref{qgnlo})
roughly corresponds to the LO fragmentation process 
of Eq.~(\ref{qglofrag}):
therefore we expect that the NLO smooth isolation
result receives a large NLO correction from this kinematical configuration of
this process, in such a way to reduce the observed LO `deficit' with respect
to standard cone isolation
(in other words, $q g \to \gamma \gamma q$ is suppressed with respect to 
$q {\bar q} \to \gamma \gamma$ by an extra power of $\as$, but this suppression
is compensated by the increased luminosity of the partonic initial state).
Moreover, the kinematical configuration where the final-state quark is outside
the isolation cones of both photons contributes through the process
of Eq.~(\ref{qgnlo})
to the NLO calculation of both smooth and standard
isolations. This is not a very limited kinematical region (the size $R$ of the
isolation cones is not large) and, due to the large value of the 
${\cal L}_{q g}$ luminosity, it gives a sizeable and $\ETmax$ independent
NLO contribution to both isolation prescriptions. It follows that,
for {\em both} isolation criteria, we expect large NLO corrections and a much
reduced $\ETmax$ dependence with respect to the LO result.
As we are going to show shortly, the expectations of our discussion are
confirmed by the actual NLO results.

We note that the LO results that we have presented are obtained by using LO
PDFs and the NLO BFG fragmentation functions, since LO fragmentation functions
are not readily available in the default setup of the \texttt{DIPHOX} code.
This mismatch (at the strictly formal level) of perturbative order in the
fragmentation functions should not strongly affect the main features of the LO
standard isolation results. More generally, standard cone isolation results are
certainly affected by an additional uncertainty, which is due to the poorly
known fragmentation functions, that is difficult to be estimated at the
quantitative level.
The recent Ref.~\cite{Kaufmann:2017lsd} presents a very brief overview on 
prospects for
improving the determination of the photon fragmentation functions.

\setcounter{footnote}{2}

\begin{figure}[htb]
\begin{center}
\begin{tabular}{cc}
\hspace*{-4.5mm}
\includegraphics[width=0.48\textwidth]{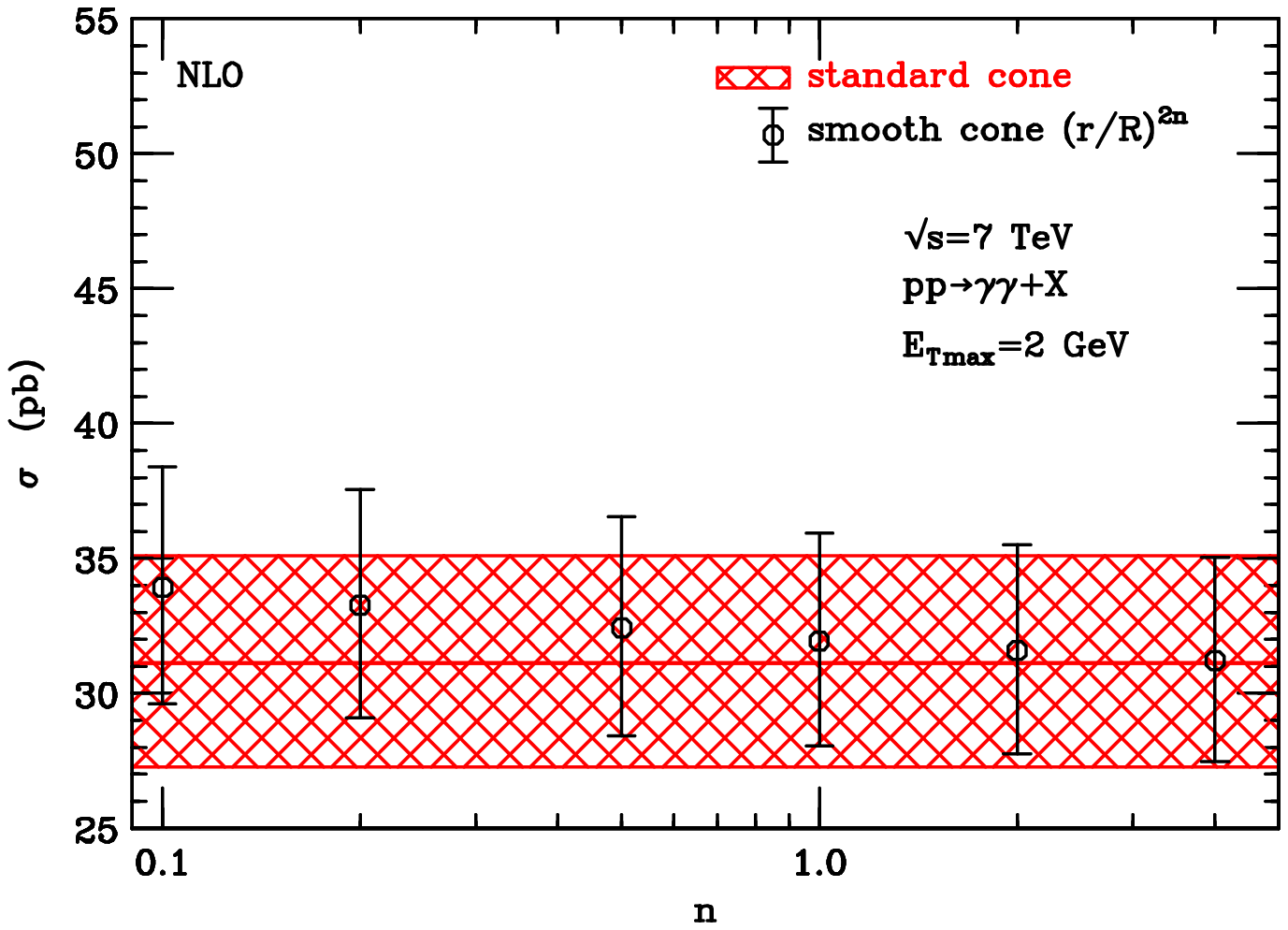}
& \includegraphics[width=0.48\textwidth]{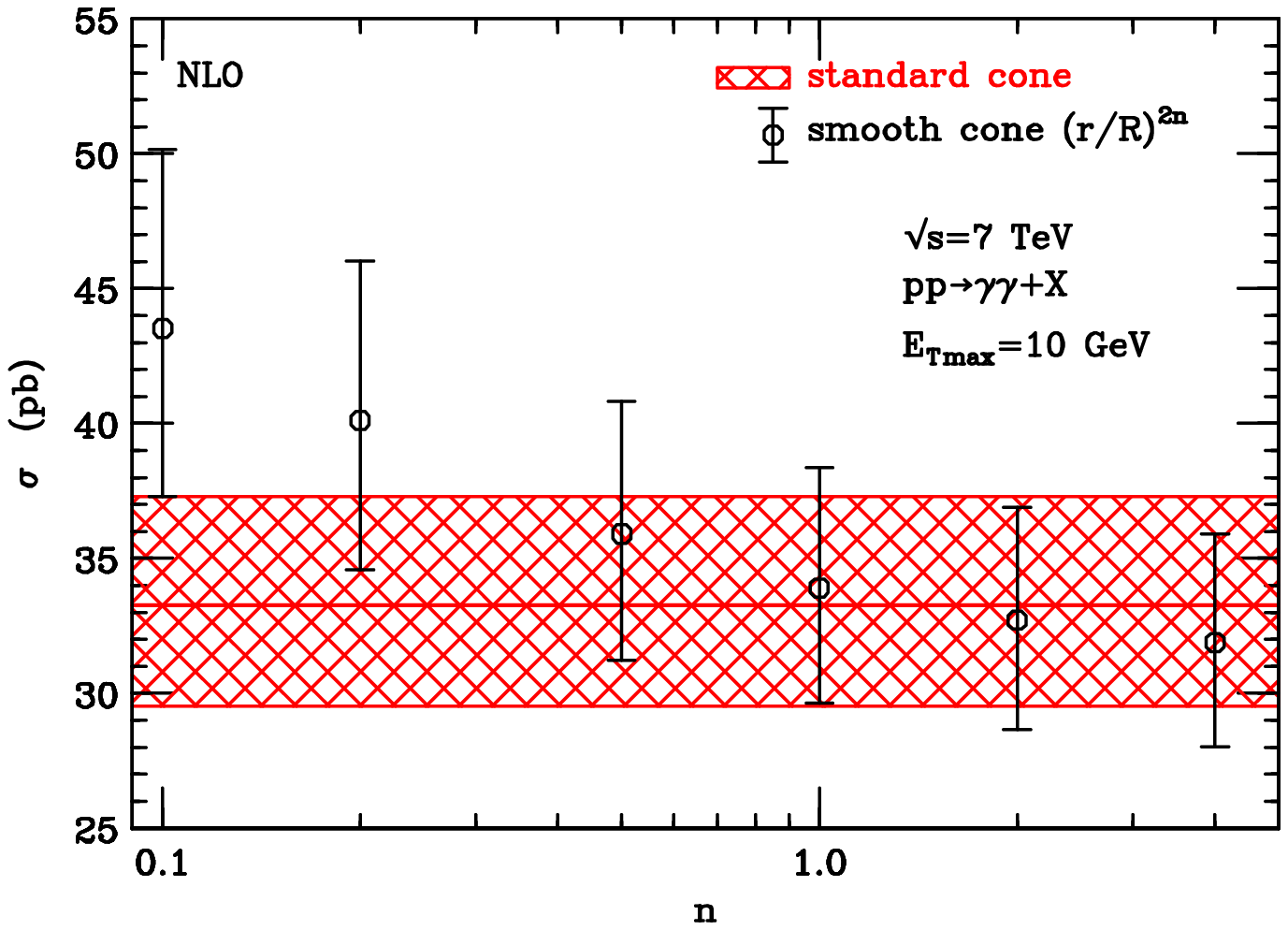}
\\
\end{tabular}
\end{center}
\caption{\label{fig:Xtot}
{\em Value of the NLO total cross section, 
including scale variation dependence, for the standard (red line and band) 
and smooth (black error bars) isolation criteria.
The photon kinematical
cuts are described in the text 
(beginning of Sect.~\ref{subsec:res}).
The results are obtained for two different values of $\ETmax = 2$~GeV
(left panel)  and 10~GeV (right panel).
In the case of smooth cone isolation, some different values of the 
power $n$ ($n=0.1, 0.2, 0.5, 1, 2, 4$)
in the isolation function
$\chi(r;R)=\left( r/R \right)^{2n}$ are considered. 
}}
\end{figure}

The value of the NLO total cross section $\sigma^{NLO}$, including its
corresponding scale variation dependence, is reported in Table~\ref{Table:Isol1}
and Fig.~\ref{fig:Xtot}
for two different values of $\ETmax$: 
$\ETmax= 2$~GeV (Fig.~\ref{fig:Xtot}-left) and
$\ETmax= 10$~GeV (Fig.~\ref{fig:Xtot}-right).

In the case of smooth cone isolation, the NLO result depends on the power $n$
of the isolation function $\chi(r;R)=\left( r/R \right)^{2n}$.
We postpone the discussion of the $n$ dependence and we consider the case with
$n=1$.
The values of the cross sections are
$\sigma^{NLO}_{\rm smooth}= 
31.92~{\rm pb}~^{+12.6\, \percent }_{-12.1\, \percent }$~(scale)
with $\ETmax = 2$~GeV and
$\sigma^{NLO}_{\rm smooth}= 
33.91~{\rm pb}~^{+13.0\, \percent }_{-12.6\, \percent }$~(scale) 
with $\ETmax = 10$~GeV. 
The standard isolation cross section $\sigma^{NLO}_{\rm standard}$
is very similar to $\sigma^{NLO}_{\rm smooth}$ at both values of 
$\ETmax$ (see Table~\ref{Table:Isol1} and Fig.~\ref{fig:Xtot}): 
the differences are at most 
at the level of 2--3\%
and they are sizeably smaller than the scale dependence of 
$\sigma^{NLO}$.
The $\ETmax$ dependence is small: the NLO cross section for smooth 
(standard) isolation
increases by a factor
of 1.06 (1.07) in going from $\ETmax= 2$~GeV to $\ETmax= 10$~GeV.
These features are in qualitative agreement with physical expectations.

The NLO radiative corrections are large (as expected from our previous
discussion). Considering the ratio
$K^{NLO}=\sigma^{NLO}/\sigma^{LO}$, the value of $K^{NLO}_{\rm smooth}$ is
roughly 3
and the values of $K^{NLO}_{\rm standard}$ are approximately
2.6 ($\ETmax = 2$~GeV) and 1.7 ($\ETmax = 10$~GeV). In the case of the
smooth isolation criterion, a sizeable part (roughly 50\%) of the NLO
total cross section is due to the $qg$ initial-state partonic channel
(which is absent at the LO), and also the LO $q{\bar q}$ channel receives 
sizeable (roughly 50\%) NLO corrections that increase the total cross section.
In the case of standard cone isolation, at the NLO the distinction between
direct and fragmentation components is no longer unambiguous (`physical'):
it depends on the factorization scheme and on the fragmentation scale 
$\mu_{frag}$. Within the $\overline {\rm MS}$ factorization scheme and the scale
variation range that we use, the direct component contributes about 45--65\%
(85--90\%) of $\sigma^{NLO}_{\rm standard}$ with $\ETmax = 10$~GeV
($\ETmax = 2$~GeV). Since $\sigma^{NLO}_{\rm standard}$ and 
$\sigma^{NLO}_{\rm smooth}$ are very similar, the similarity is the consequence of a
non-trivial interplay between the direct and fragmentation contributions to
$\sigma^{NLO}$ (especially in the case with $\ETmax = 10$~GeV) and, in
particular,
the LO equivalence between the smooth cone result and the direct component of
the standard cone result is definitely lost at the NLO.

We note that the scale dependence of the total cross section has a similar size at LO and at
NLO, and it is much smaller than the size of the NLO corrections. This implies
that the scale dependence of $\sigma^{NLO}$ cannot be consistently regarded
as a reliable estimate of uncalculated higher-order radiative corrections
to $\sigma$: the `true' theoretical uncertainty of $\sigma^{NLO}$ is {\em
certainly
larger} than the NLO scale dependence that we have computed.
A similar comment applies to the scale dependence of the results for 
the differential
cross sections that we present in the following.

We now discuss the dependence of the smooth isolation cross section  
$\sigma^{NLO}_{\rm smooth}$ on the power $n$ of the isolation function
$\chi(r;R) = (r/R)^{2n}$ (see Eq.~(\ref{Eq:Isol_chirpow})).
The NLO results of the total cross section for some selected values of $n$ in
the range $0.1 \leq n \leq 4$ are reported in Fig.~\ref{fig:Xtot}.
We note that the $n$ dependence of $\sigma^{NLO}_{\rm smooth}$ is small
(in particular, it is smaller than the scale dependence) within this range of
values of $n$. Specifically, considering the interval
$1/2 < n < 2$, the central value of $\sigma^{NLO}_{\rm smooth}$ varies by about
3\% (10\%) if $\ETmax = 2$~GeV ($\ETmax = 10$~GeV).
Most of the qualitative features of the results in Fig.\ref{fig:Xtot} are
consistent with physical expectations. The photons are more isolated by
increasing $n$ at fixed $\ETmax$
and, consequently, $\sigma^{NLO}_{\rm smooth}$ monotonically
decreases (in agreement with the physical behaviour in Eq.~(\ref{eq:d})).
Moreover, by decreasing $\ETmax$ the photons are more isolated and 
consequently the total cross section is less sensitive to variations of the
power $n$. Nonetheless, we note that, by decreasing the value of $n$,
$\sigma^{NLO}_{\rm smooth}$ tends to become larger than 
$\sigma^{NLO}_{\rm standard}$, thus violating the physical constraint in
Eq.~(\ref{eq:a}). This feature deserves some comments, which are presented
below.

The perturbative dependence of $\sigma_{\rm smooth}$ at very small ($n \ll 1$)
or very large ($n \gg 1$) values of $n$ can be understood in relatively-simple
terms. If $n$ is very small, the isolation function  $\chi(r;R) = (r/R)^{2n}$
is approximately equal to unity with the exception of the angular region of very
small values of $r$: therefore, the $n$ dependence of $\sigma_{\rm smooth}$
is very sensitive to radiation of partons that are {\em collinear} to the photon
direction. If $n$ is very large, the isolation function very strongly 
suppresses parton radiation inside the photon isolation cone: therefore, the
$n$ dependence of $\sigma_{\rm smooth}$ is very sensitive to {\em soft} parton
radiation. The dominant effects of soft and collinear radiation can be easily
computed at NLO (see Refs.~\cite{Frixione:1998jh,Catani:2002ny, Gordon:1994ut}).
We consider the NLO correction 
$\delta^{NLO} = (\sigma^{NLO} - \sigma^{LO})/\sigma^{LO}$, and we limit 
ourselves to
sketch the dominant $n$ dependence of the soft 
($\delta^{NLO, {\rm soft}}_{\rm smooth}$)
and collinear ($\delta^{NLO, {\rm coll}}_{\rm smooth}$) contribution to
$\delta^{NLO}$ within smooth isolation. We have
\beeq
\label{soft}
\delta^{NLO, {\rm soft}}_{\rm smooth} &\propto& 
- \; \as \,R^2 \left( \ln\left(\frac{Q}{\ETmax}\right) + n \right) \;,
\quad\, (n \gg 1) \;\;, \\
\label{coll}
\delta^{NLO, {\rm coll}}_{\rm smooth} &\propto&
+ \; \frac{\as}{n} \;\frac{\ETmax}{Q} \;\;,
\quad \quad \quad \quad \quad \quad \quad \quad
(n \ll 1) \;\;,
\eeeq
where $Q$ is the typical hard scale of the cross section
(the scale is of the order of the minimum value of $p_{T \gamma}^{hard}$)
and we have considered small values of $\ETmax$ (by neglecting relative
corrections of ${\cal O}(\ETmax/Q)$).

The proportionality factor that is not explicitly denoted on the right-hand side
of Eq.~(\ref{soft}) depends on the LO cross section for the partonic process
$q{\bar q} \to \gamma \gamma$.
The soft contribution in Eq.~(\ref{soft}) is negative. It is due to a strong
kinematical
mismatch between (negative) soft virtual radiation (one-loop corrections in the
subprocess $q{\bar q} \to \gamma \gamma$), which is not affected by isolation,
and (positive) soft real radiation (the subprocess 
$q{\bar q} \to g \gamma \gamma$ at the tree level), which is strongly 
suppressed by isolation. We note that $\delta^{NLO, {\rm soft}}_{\rm smooth}$
is proportional to $n$, so that eventually $\sigma^{NLO}_{\rm smooth}$
diverges to `$-~\infty$' in the limit $n \to +\infty$.
This NLO divergence is the perturbative signal of the infrared unsafety of the
isolated cross section in the limit of completely isolated photons (no
accompanying transverse energy inside the isolation cone). We observe that 
$\delta^{NLO, {\rm soft}}_{\rm smooth}$ is proportional to $R^2$,
so that the soft contribution is strongly
suppressed if the photon isolation cone has a small radius $R$. 
We also note that $\delta^{NLO, {\rm soft}}_{\rm smooth}$ is due to subprocesses
with a $q{\bar q}$ initial state. The subprocess $qg \to q \gamma \gamma$ is
formally subdominant in the soft limit ($n \gg 1$), but it represents 
a sizeable quantitative correction to 
$\delta^{NLO, {\rm soft}}_{\rm smooth}$
because of the increased PDF luminosity of the $qg$ initial state.
The $R^2$-suppressed dependence of  
$\delta^{NLO, {\rm soft}}_{\rm smooth}$ and the large size of the (positive)
correction to it from the  $qg$ initiated subprocess explain why the results for
$\sigma^{NLO}_{\rm smooth}$ in Fig.~\ref{fig:Xtot}
have a small dependence on $n$ at relatively-large values of $n$ (e.g., $n
\simeq 4$).

The collinear contribution in Eq.~(\ref{coll}) is relevant to discuss the 
$n$ dependence of the results in Fig.~\ref{fig:Xtot} at small values of $n$.
We note that the NLO contributions from the initial-state $q{\bar q}$ channel
(e.g., $q{\bar q} \to g \gamma \gamma$) are subdominant in the limit 
$n \ll 1$. The contribution in Eq.~(\ref{coll}) is
due to real radiation of a collinear quark or antiquark inside the photon
isolation cone through the partonic processes $qg \to q\gamma \gamma$ and 
${\bar q} g \to {\bar q}\gamma \gamma$
(the proportionality factor that is not explicitly denoted in the right-hand
side of Eq.~(\ref{coll}) depends on the LO cross section for the partonic
processes $qg \to q\gamma$ and ${\bar q} g \to {\bar q}\gamma$). Therefore, 
$\delta^{NLO, {\rm coll}}_{\rm smooth}$ is positive and independent of $R$. Moreover, $\delta^{NLO, {\rm coll}}_{\rm smooth}$
is proportional to $\ETmax$, so that its induced $n$ dependence
is reduced by decreasing $\ETmax$ (in agreement with the results in 
Fig.~\ref{fig:Xtot} at small $n$).
Owing to its dependence on $n$, $\delta^{NLO, {\rm coll}}_{\rm smooth}$
sizeably increases by decreasing $n$ at fixed $\ETmax$ and, eventually, 
$\delta^{NLO, {\rm coll}}_{\rm smooth}$ (and, consequently,
$\sigma^{NLO}_{\rm smooth}$) diverges to `$+ \infty$' in the limit $n \to 0$.
Since $\sigma^{NLO}_{\rm smooth}$ becomes arbitrarily large by decreasing $n$,
it is obvious that at sufficiently small values of $n$ the physical requirement
$\sigma_{\rm smooth} < \sigma_{\rm standard}$ (see Eq.~(\ref{eq:a}))
is unavoidably violated. This misbehaviour of
$\sigma^{NLO}_{\rm smooth}$ at small values of $n$ implies that beyond-NLO
contributions are relevant. Indeed, each higher-order contribution is equally
misbehaved at $n \ll 1$: the N$^k$LO correction is proportional to
$(\as/n)^k$ (because of multiple collinear radiation inside the photon isolation
cone) an it cannot be regarded as a small correction if $n \ltap \as$ (i.e.,
$n \ltap 0.1$). In principle, the perturbative treatment at small values of $n$
can be improved by a proper all-order resummation of these collinear
contributions of ${\cal O}((\as/n)^k)$. However, such resummation treatment
cannot be pursued for arbitrarily small values of $n$ since it unavoidably fails
in the limit $n \to 0$, because of non-perturbative photon fragmentation effects
(smooth isolation with $n=0$ requires photon fragmentation functions since it is
equivalent to standard isolation).

We can draw some conclusions from our discussion on small values of $n$.
Owing to the physical requirements in Eqs.~(\ref{eq:a}) and 
(\ref{eq:d}), in principle cross sections for standard and smooth isolation 
tend to agree at very small values of $n$. However, fixed-order QCD
computations for smooth isolation are not reliable if $n \ll 1$ (they are
affected by large higher-order corrections) and, in particular,
they can violate the physical constraint in Eq.~(\ref{eq:a}). 
In practice, to the purpose of approximating the standard isolation criterion
within fixed-order QCD calculations it is more appropriate to consider smooth
isolation with values of $n$ that are not too small. From the results in 
Fig.~\ref{fig:Xtot}, we can conclude that the total cross sections
$\sigma^{NLO}_{\rm smooth}$ and  
$\sigma^{NLO}_{\rm standard}$ quantitatively agree if $n \simeq 1$.

In the following we consider differential cross sections with respect to various
kinematical variables and we limit ourselves to present smooth isolation
results with $n=1$.
We have checked that the shape of the various NLO kinematical distributions 
is very
little affected by variations of $n$ within the range $0.1 \ltap n \ltap 4$. 
At the NLO, variations 
of $n$ basically produce
overall normalization effects, whose size corresponds to the $n$ dependence of  
$\sigma^{NLO}_{\rm smooth}$ that is observed in Fig.~\ref{fig:Xtot}.

\setcounter{footnote}{2}

\subsubsection{Differential cross sections at the LO}
\label{sec:diffLO}

\begin{figure}[htb]
\begin{center}
\begin{tabular}{cc}
\hspace*{-4.5mm}
\includegraphics[width=0.485\textwidth]{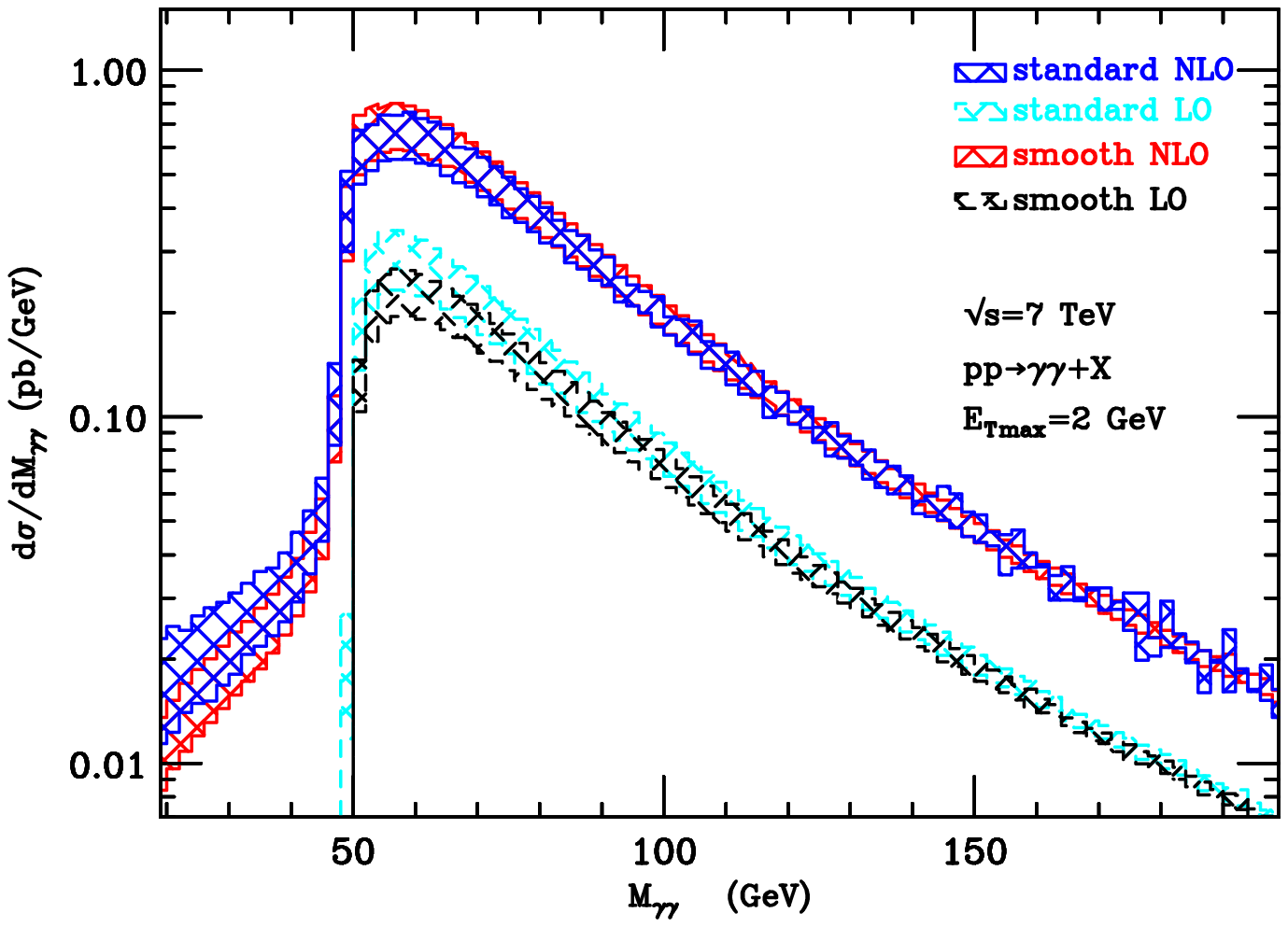}
& \includegraphics[width=0.485\textwidth]{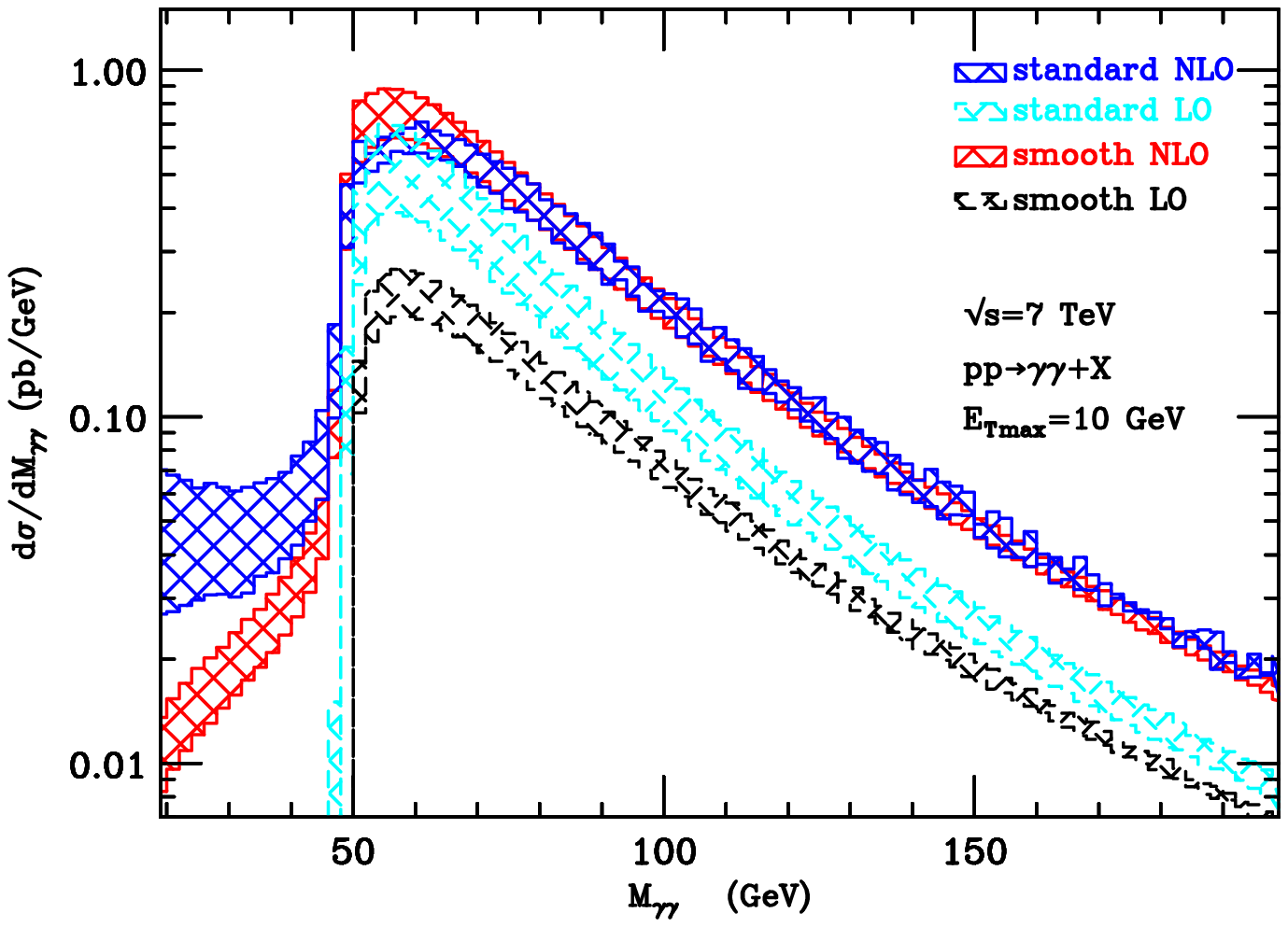}
\\
\end{tabular}
\end{center}
\caption{\label{fig:dm}
{\em The $M_{\gamma \gamma}$  differential cross section 
for $\ETmax=2$~GeV (left panel) and $\ETmax=10$~GeV (right panel)
and with the same photon kinematical cuts as in Fig.~\ref{fig:Xtot}.
The scale variation bands of the LO and NLO results for smooth and standard
isolation are as follows: 
LO smooth isolation (black dashed),
LO standard isolation (light-blue dashed),
NLO smooth isolation (red solid) and
NLO standard isolation (blue solid).}}
\end{figure}

\begin{figure}[htb]
\begin{center}
\begin{tabular}{cc}
\hspace*{-4.5mm}
\includegraphics[width=0.48\textwidth]{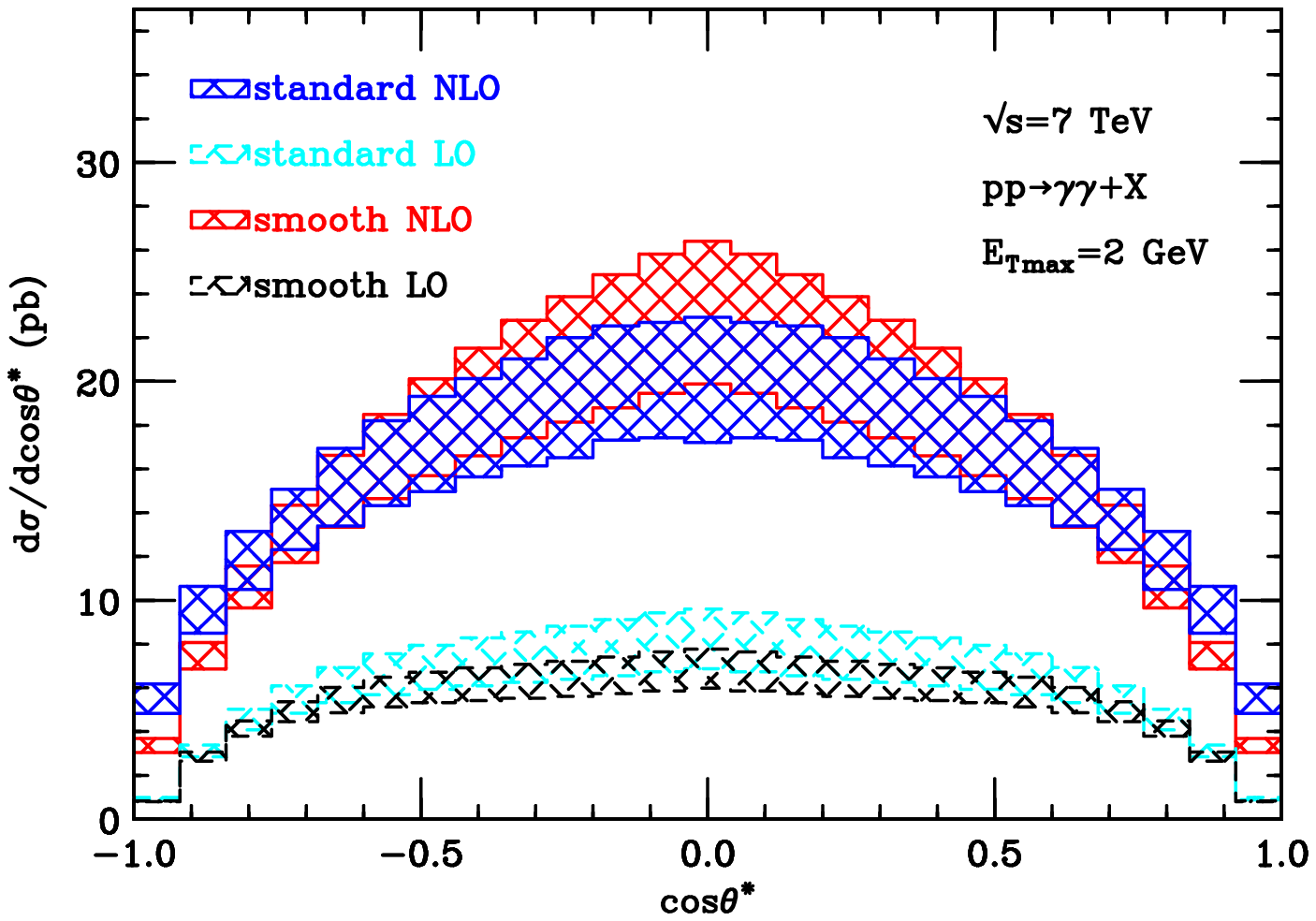}
& \includegraphics[width=0.48\textwidth]{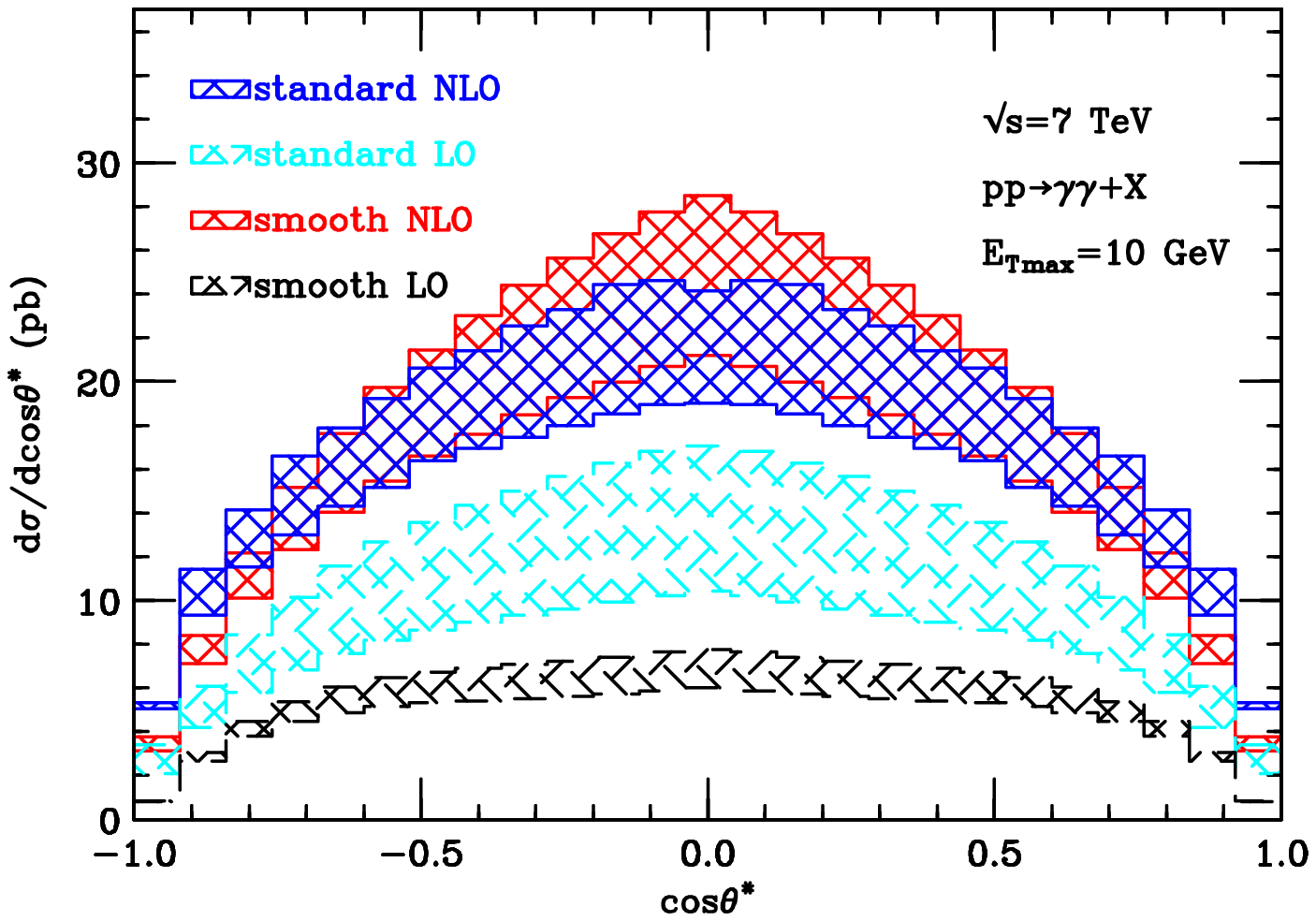}
\\
\end{tabular}
\end{center}
\caption{\label{fig:dcos}
{\em 
The $\cos \theta^{*}$ differential cross section 
for $\ETmax=2$~GeV (left panel) and $\ETmax=10$~GeV (right panel)
and with the same photon kinematical cuts as in Fig.~\ref{fig:Xtot}.
The scale variation bands of the LO and NLO results for smooth and standard
isolation are as follows: 
LO smooth isolation (black dashed),
LO standard isolation (light-blue dashed),
NLO smooth isolation (red solid) and
NLO standard isolation (blue solid).
}}
\end{figure}

In Fig.~\ref{fig:dm} we present the LO and NLO results (including their scale
variation dependence) for the differential cross section with respect to the
diphoton invariant mass $M_{\gamma \gamma}$. We consider two different values,
2~GeV (Fig.~\ref{fig:dm}-left) and 10~GeV (Fig.~\ref{fig:dm}-right), of
$\ETmax$ and we use both standard and smooth isolation.
In Fig.~\ref{fig:dcos} we present the analogous results for the differential 
cross section with respect to the angular variable 
$\cos \theta^{*}$ in the Collins--Soper rest frame.
The results in Figs.~\ref{fig:dm} and \ref{fig:dcos} are obtained by numerical
integration over small bins in $M_{\gamma \gamma}$ and $\cos \theta^{*}$,
respectively: we use bins of constant size equal to 2~GeV for 
$M_{\gamma \gamma}$ and 0.08 for $\cos \theta^{*}$.
In the following we discuss LO and NLO differential cross sections in turn.
At the LO, we preliminarily note that standard and smooth isolation results for
the differential cross sections have qualitatively similar shapes, with
differences of the overall normalization that are quite similar to the
quantitative differences (which we have previously discussed) of the
corresponding LO total cross sections.

We first consider smooth cone isolation.
Owing to transverse-momentum conservation in the corresponding LO partonic
process $q{\bar q} \to \gamma \gamma$, the diphoton azimuthal separation is
$\Dpgg= \pi$ and the transverse momentum of the diphoton
system is $p_{T \gamma \gamma}=0$. The corresponding differential cross
sections $d\sigma^{LO}_{\rm smooth}/dx$, with $x=\Dpgg$ or 
$x=p_{T \gamma \gamma}$, are simply proportional to a $\delta$-function
($\delta(\Dpgg- \pi)$ or $\delta(p_{T \gamma \gamma})$)
and the proportionality factor is the LO total cross section 
$\sigma^{LO}_{\rm smooth}$.
The double differential cross section with respect to $M_{\gamma \gamma}$
and the scattering angle $\theta_S$ is
\begin{equation}
\label{lodir}
\frac{d\sigma^{LO}_{\rm smooth}}{d M_{\gamma \gamma}^2 \;d|\!\cos \theta_S|}
\propto \frac{\alpha^2}{s^2} \; g_{q{\bar q}}(\theta_S) \;\sum_q e_q^4 
\;{\cal L}_{q{\bar q}}(\tau;\mu_F) \;\;,
\end{equation}
where $\sqrt s$ is the centre--of--mass energy of the hadronic collision
(e.g., $\sqrt s=7$~TeV as in Fig.~\ref{fig:dm}), $\tau=M_{\gamma \gamma}^2/s$
and the photon scattering angle $\theta_S$ is defined in the 
centre--of--mass frame of the LO partonic collision, and it is related to the
diphoton rapidity separation $\dy$:
\begin{equation}
\label{dy}
|\!\cos \theta_S|= 
\tanh
\left( \frac{|\dy|}{2}\right) \;\;.
\end{equation}
We remark that, in the case of the LO smooth isolation cross section in
Eq.~(\ref{lodir}), $\theta_S$ and the Collins--Soper polar angle $\theta^{*}$
actually coincides ($|\!\cos \theta_S|=|\!\cos \theta^*|$), since 
$p_{T \gamma \gamma}=0$. In the right-hand side of Eq.~(\ref{lodir}) we have
not denoted an overall proportionality factor of order unity, which is
independent of the kinematical variables $s, M_{\gamma \gamma}$ and 
$\theta_S$. The angular dependent function $g_{q{\bar q}}(\theta_S)$ is
\begin{equation}
\label{cosdir}
g_{q{\bar q}}(\theta_S) = 
\frac{2\,(1+\cos^2 \theta_S)}{1-\cos^2 \theta_S} \;\;,
\end{equation}
and it is specific for the angular distribution that is dynamically produced by
the partonic process $q{\bar q} \to \gamma \gamma$.
The function ${\cal L}_{q{\bar q}}(\tau;\mu_F)$ is the $q{\bar q}$ PDF
luminosity and $e_q$ is the quark electric charge in units of the positron
charge ($e_q=2/3$ for up-type quarks). The PDF luminosity for the collision
of two partons $a$ and $b$ is defined as
\begin{equation}
\label{lum}
{\cal L}_{ab,h_1h_2}(\tau;\mu_F) \equiv 
\int_0^1 \frac{dx_1}{x_1} \int_0^1 \frac{dx_2}{x_2} 
\left[ f_{a/h_1}(x_1,\mu_F) f_{b/h_2}(x_2,\mu_F) + 
\bigl( a \leftrightarrow b \bigr) \right] \;
\delta(\tau - x_1x_2) \;\;,
\end{equation}
where $f_{a/h}(x,\mu_F)$ is the PDF of the parton $a$ in the hadron $h$.

A distinctive feature of the right-hand side of Eq.~(\ref{lodir}) and, hence,
of the double differential cross section is its exactly {\em factorized}
dependence on $M_{\gamma \gamma}$ and $\cos \theta_S$. The $M_{\gamma \gamma}$
dependence is controlled by the luminosity ${\cal L}_{q{\bar q}}$:
by decreasing $M_{\gamma \gamma}$, ${\cal L}_{q{\bar q}}$ rapidly increases.
The $\cos \theta_S$ (or $\cos \theta^*$) dependence is controlled by 
$g_{q{\bar q}}(\theta_S)$: by increasing $|\!\cos \theta_S|$,
$g_{q{\bar q}}$ sharply increases and it becomes singular at 
$|\!\cos \theta_S|=1$ (because of the `unphysical' behaviour of $2\to2$
`massless' parton scattering).
The computation of the single differential cross sections
$d\sigma/dM_{\gamma \gamma}$ and $d\sigma/d\cos \theta^{*}$
requires the application of kinematical cuts to select the hard-scattering
regime (i.e., values of $M_{\gamma \gamma}$ in the perturbative region and
values of $\cos \theta^{*}$ sufficiently far 
from the forward/backward scattering singularity). 
The simplest type of kinematical cuts is a minimum value of 
$M_{\gamma \gamma}$ and a maximum value of $|\!\cos \theta_S|$. These
kinematical cuts, which preserve the factorized structure of Eq.~(\ref{lodir})
with respect to the  $M_{\gamma \gamma}$ and $\cos \theta_S$ dependence,
lead to differential cross sections $d\sigma/dM_{\gamma \gamma}$ and 
$d\sigma/d\cos \theta^{*}$ that are simply proportional to 
${\cal L}_{q{\bar q}}(\tau)$ and $g_{q{\bar q}}(\theta^*)$, respectively.
The shapes of these differential cross sections are different (especially in the
case of the $\cos \theta^{*}$ distribution) from those observed in 
Figs.~\ref{fig:dm} and \ref{fig:dcos}.
The differences originate from the kinematical cuts on the photon transverse
momenta and rapidities that are described at the beginning of 
Sect.~\ref{subsec:res},
and that are actually used in the computation of the differential cross sections
of Figs.~\ref{fig:dm} and \ref{fig:dcos}.

We first discuss the effect of the transverse-momentum ($p_T$) cuts
$p_{T \gamma}^{hard} \geq p_H$ (specifically $p_H=25$~GeV) and
$p_{T \gamma}^{soft} \geq p_S$ (specifically $p_S=22$~GeV).
Since $p_{T \gamma \gamma}=0$ (i.e., $p_{T \gamma}^{hard}=p_{T \gamma}^{soft}$),
only the value of $p_H$ is effective, and we have the LO constraint
\begin{equation}
\label{ptcut}
M_{\gamma \gamma} \;\sin \theta_S \geq M^{LO}_{\rm dir} \;\;,
\end{equation}
where $M^{LO}_{\rm dir}=2p_H$ (specifically $M^{LO}_{\rm dir}=50$~GeV)
and we have simply used 
$2p_{T \gamma}^{hard}= M_{\gamma \gamma} \;\sin \theta_S$.
Since $\sin \theta_S <1$ and $M_{\gamma \gamma}^2 < s$, 
the constraint in
Eq.~(\ref{ptcut}) implies a lower boundary 
$M_{\gamma \gamma} \geq M^{LO}_{\rm dir}$ on $M_{\gamma \gamma}$ 
(the LO smooth isolation cross section in Fig.~\ref{fig:dm}
vanishes for $M_{\gamma \gamma} < 50$~GeV) and an upper
boundary $|\!\cos \theta_S| < \sqrt{1-(M^{LO}_{\rm dir})^2/s}$ on
$|\!\cos \theta_S|$ 
(this corresponds to $1-|\!\cos \theta_S| \ltap 3 \cdot 10^{-5}$ 
in Fig.~\ref{fig:dcos}). More importantly, the constraint in Eq.~(\ref{ptcut})
{\em correlates} the $M_{\gamma \gamma}$ and $\cos \theta_S$ dependencies, thus
destroying the factorized structure in the right-hand side of Eq.~(\ref{lodir})
and leading to relevant effects on the shape of the single differential cross
sections. 

A relevant effect regards $d\sigma/dM_{\gamma \gamma}$ in the region
close to the LO threshold $M^{LO}_{\rm dir}$. At fixed values of 
$M_{\gamma \gamma}$, the constraint in Eq.~(\ref{ptcut}) leads to an upper
limit, $|\!\cos \theta_S| < \sqrt{1-(M^{LO}_{\rm dir}/M_{\gamma \gamma})^2}$, 
that strongly
suppresses the integration region over $\cos \theta_S$ if 
$M_{\gamma \gamma} \to M^{LO}_{\rm dir}$. The increase of 
$d\sigma^{LO}/dM_{\gamma \gamma}$ (due to ${\cal L}_{q{\bar q}}$) for decreasing
values of $M_{\gamma \gamma}$ is thus damped by this phase space suppression:
$d\sigma^{LO}_{\rm smooth}/dM_{\gamma \gamma}$ reaches a maximum value 
(in the region close to $M_{\gamma \gamma} \sim M^{LO}_{\rm dir}$) and then it
sharply decreases and it vanishes (proportionally to 
$\sqrt{M_{\gamma \gamma}^2- (M^{LO}_{\rm dir})^2}\;$) at 
$M_{\gamma \gamma} = M^{LO}_{\rm dir}$. This LO behaviour of 
$d\sigma/dM_{\gamma \gamma}$ is 
visible in the smooth isolation 
results of Fig.~\ref{fig:dm}.
Actually, the vanishing behaviour of $d\sigma^{LO}/dM_{\gamma \gamma}$
in the limit $M_{\gamma \gamma} \to M^{LO}_{\rm dir}$ is so steep that it is not
clearly recognizable in the invariant-mass bin 50~GeV$ < M_{\gamma \gamma} <
52$~GeV closest to $M^{LO}_{\rm dir}=$50~GeV. This vanishing behaviour is more
evident by using $M_{\gamma \gamma}$-bins with a much smaller bin size (see
Fig.~\ref{fig:dmth}
at the end of Sect.~\ref{sec:diffNLO}).

The effect of the $p_T$ cut constraint in
Eq.~(\ref{ptcut}) on $d\sigma/d\cos \theta^{*}$ is even much more relevant
than the effect on $d\sigma/dM_{\gamma \gamma}$.
In the computation of $d\sigma^{LO}_{\rm smooth}/d\cos \theta_S$ from
Eq.~(\ref{lodir}), the PDF luminosity 
${\cal L}_{q{\bar q}}(\tau=M_{\gamma \gamma}^2/s)$ is integrated over 
$M_{\gamma \gamma}$ up to a lower limit, 
$M_{\gamma \gamma} > M^{LO}_{\rm dir}/\sin \theta_S$, that depends on
$\theta_S$: therefore large values of $|\!\cos \theta_S|$ are suppressed, while
small values of $|\!\cos \theta_S|$ are relatively enhanced by the increasing
size of the PDF luminosity for decreasing values of $M_{\gamma \gamma}$.
This PDF modulation of the $\cos \theta_S$ dependence has exactly the opposite
qualitative effect with respect to the $\cos \theta_S$ dependence of the
partonic angular distribution $g_{q{\bar q}}(\theta_S)$. It turns out that the
PDF modulation effect quantitatively dominates, and 
$d\sigma/d\cos \theta^{*}$ has a bell (concave) shape 
(see Fig.~\ref{fig:dcos}) rather than the inverse-bell (convex) shape of the
angular distribution $g_{q{\bar q}}(\theta_S)$ of the underlying partonic
process. We note that, varying $\cos \theta_S$ over a wide range around 
the central region (say, $|\!\cos \theta_S| \ltap 0.5$), the lower limit 
on the PDF integration varies in a restricted range 
($M^{LO}_{\rm dir} < M_{\gamma \gamma} \ltap M^{LO}_{\rm dir}/0.8$): this
implies that the results of Fig.~\ref{fig:dcos} for 
$d\sigma/d\cos \theta^{*}$ in the region $|\!\cos \theta_S| \ltap 0.5$
are quite sensitive to the behaviour of the corresponding double differential
cross section in a very limited range 
(50~GeV$ < M_{\gamma \gamma} \ltap 60$~GeV) of $M_{\gamma \gamma}$.

\begin{figure}[htb]
\begin{center}
\begin{tabular}{cc}
\includegraphics[width=0.49\textwidth]{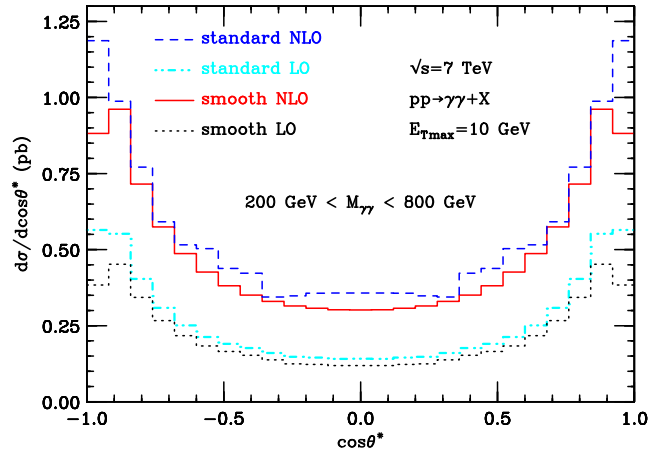}
\\
\end{tabular}
\end{center}
\caption{\label{fig:dcosnew}
{\em 
The $\cos \theta^{*}$ differential cross section 
for $\ETmax=10$~GeV and with the constraint
$200~GeV < M_{\gamma \gamma} < 800$~GeV in addition to the same 
photon kinematical cuts as in Fig.~\ref{fig:dcos}.
The results, which are obtained with the scales $\mu_R=M_{\gamma \gamma}/2$
and $\mu_F=\mu_{frag}= 2 M_{\gamma \gamma}$, are  
as follows: 
LO smooth isolation (black dotted),
LO standard isolation (light-blue dash-dotted),
NLO smooth isolation (red solid) and
NLO standard isolation (blue dashed).
}}
\end{figure}

From our discussion it follows that the PDF modulation effect on the shape of
$d\sigma/d\cos \theta^{*}$ can be reduced by applying an additional kinematical
cut on $M_{\gamma \gamma}$, namely, 
$M_{\gamma \gamma} > M_{\gamma \gamma}^{\rm min}$ with a fixed ($\theta_S$
independent) value $M_{\gamma \gamma}^{\rm min}$, and by selecting sufficiently
large values of $M_{\gamma \gamma}^{\rm min}$.
In particular, increasing $M_{\gamma \gamma}^{\rm min}$ one can eventually
recover the qualitative $\theta_S$ dependence due to $g_{q{\bar q}}(\theta_S)$.
For illustrative purpose, in Fig.~\ref{fig:dcosnew} we present the results for
$d\sigma/d\cos \theta^{*}$ with the same kinematical cuts as in 
Fig.~\ref{fig:dcos}-right ($\ETmax=10$~GeV) and the additional constraint
200~GeV$< M_{\gamma \gamma} < 800$~GeV (i.e., 
$M_{\gamma \gamma}^{\rm min}=200$~GeV). We see that the shape of 
$d\sigma/d\cos \theta^{*}$ in Fig.~\ref{fig:dcosnew} is much different from that
in Fig.~\ref{fig:dcos}-right and it is qualitatively more similar to the shape
of $g_{q{\bar q}}(\theta_S)$. The constraint $M_{\gamma \gamma} < 800$~GeV
has a negligible quantitative effect on the shape of 
$d\sigma/d\cos \theta^{*}$. At the LO, the additional constraint 
$M_{\gamma \gamma} > 200$~GeV implies that the $p_T$ cuts have no effect on the
shape of $d\sigma^{LO}_{\rm smooth}/d\cos \theta^{*}$ in the region where
$|\!\cos \theta^*| \ltap 0.97$: within this $\cos \theta^*$ region, 
$d\sigma^{LO}_{\rm smooth}/d\cos \theta^{*}$ follows the shape of 
$g_{q{\bar q}}(\theta_S)$, modulo a PDF effect that is due to the kinematical
cuts on the photon rapidities (the effect of the rapidity cuts is discussed
below). 
We can also comment on the 
diphoton production study that is presented in 
Ref.~\cite{Butterworth:2014efa} (see Sect.~III.11 therein). 
That study uses the kinematical cuts
$p_{T \gamma}^{hard} \geq 40$~GeV,  
$p_{T \gamma}^{soft} \geq 30$~GeV and 100~GeV$< M_{\gamma \gamma} < 160$~GeV,
which correspond to $M^{LO}_{\rm dir}= 80$~GeV and 
$M_{\gamma \gamma}^{\rm min}=100$~GeV (note that $M_{\gamma \gamma}^{\rm min}$
is much closer to $M^{LO}_{\rm dir}$ with respect to the cuts considered in 
Fig.~\ref{fig:dcosnew}). The corresponding differential cross section 
$d\sigma/d\cos \theta^{*}$ (see Fig.~III.50
in Ref.~\cite{Butterworth:2014efa})
has a maximum value
at $|\!\cos \theta^*| \sim 0.6$, and a shape that is somehow intermediate
between those in Figs.~\ref{fig:dcos} and \ref{fig:dcosnew}: this behaviour is
in agreement with the expectation from our discussion.

Our discussion and the results in Figs.~\ref{fig:dcos} and \ref{fig:dcosnew}
evidently show that the shape of $d\sigma/d\cos \theta^{*}$ can be strongly
affected by the applied kinematical cuts as the consequence of a non-trivial
interplay between underlying hard-scattering dynamics and PDF behaviour.

The results in Figs.~\ref{fig:dm} and \ref{fig:dcos}
are obtained by also including the photon rapidity cut $|y_\gamma| < y_M$
(specifically $y_M=2.37$) in addition to the $p_T$ cuts (see Eq.~(\ref{ptcut}))
that we have just discussed. The photon rapidity cut reduces the size of the
cross sections but, since the value of $y_M$ is sufficiently large, the overall
qualitative shape of the differential cross sections is basically unchanged.
More precisely, the rapidity cut leads to the LO upper boundaries
$|\!\cos \theta_S| < {\rm tanh}(y_M)$
($|\!\cos \theta_S| \ltap 0.98$ in Fig.~\ref{fig:dcos}) and 
$M_{\gamma \gamma} < e^{-y_M} {\sqrt s}$ ($M_{\gamma \gamma} \ltap 650$~GeV
with $y_M=2.37$ and ${\sqrt s}=7$~TeV) on $\cos \theta_S$ and 
$M_{\gamma \gamma}$, respectively, and it modifies the form of the PDF
luminosity contribution in Eq.~(\ref{lodir}). The modification amounts to the
replacement ${\cal L}_{ab}(\tau) \to {\cal L}_{ab}(\tau;y_{max})$, where the
customary PDF luminosity ${\cal L}_{ab}(\tau)$ is replaced by a PDF luminosity
with `rapidity restriction'. The rapidity restricted luminosity
${\cal L}_{ab}(\tau;y_{max})$ is simply obtained by inserting the constraint 
$|\ln(x_1/x_2)| < 2 y_{max}$ in the $\{x_1, x_2 \}$ integration region of
Eq.~(\ref{lum}) (at the LO, the rapidity $y_{\gamma \gamma}$ of the diphoton
system is $2 y_{\gamma \gamma}= \ln(x_1/x_2)$). The value of $y_{max}$ is
related to the photon rapidity cut, $y_{max}=y_M-|\dy|/2$, and it
depends on the diphoton rapidity separation $\dy$ and, hence, on 
$\cos \theta_S$ (through Eq.~(\ref{dy})).
Therefore, the rapidity restriction produces a suppression of the PDF luminosity
contribution, and the suppression is larger at larger values of 
$|\!\cos \theta_S|$.

We now consider LO kinematical distributions within standard cone isolation. 
The LO differential cross sections $d\sigma^{LO}_{\rm standard}$ are obtained by
combining the direct and fragmentation components,
$d\sigma^{LO}_{\rm standard}= d\sigma^{LO}_{\rm dir} + d\sigma^{LO}_{\rm frag}$,
and the direct component contribution $d\sigma^{LO}_{\rm dir}$ is exactly equal
to $d\sigma^{LO}_{\rm smooth}$. 

Many features of the fragmentation component
contribution $d\sigma^{LO}_{\rm frag}$ are similar to those of 
$d\sigma^{LO}_{\rm smooth}$, and we only note the main differences.
In the fragmentation component the photon is accompanied by collinear hadronic
fragments and, therefore, we have $\Dpgg=\pi$ (as in the case of 
smooth cone isolation at LO), while $p_{T \gamma \gamma} \neq 0$ (at variance
with respect to smooth cone isolation). Moreover, due to the isolation procedure,
we have $p_{T \gamma \gamma} < \ETmax$. Since $p_{T \gamma \gamma} \neq 0$,
$\cos \theta_S$ (see Eq.~(\ref{dy})) is not exactly equal to the Collins--Soper
variable $\cos \theta^{*}$: the relation between the two angular variables is 
$\cos \theta^{*}= 
\sqrt{M_{\gamma \gamma}^2/(M_{\gamma \gamma}^2 + p_{T \gamma \gamma}^2)} \,
\cos \theta_S$ (which is valid for $\Dpgg=\pi$). Since we are
considering relatively-small values of $\ETmax$, we still have
$p_{T \gamma \gamma} \ll M_{\gamma \gamma}$ and $\cos \theta^{*} \simeq 
\cos \theta_S$. The LO double differential cross section for the fragmentation
component is
\begin{equation}
\label{lofrag}
\frac{d\sigma^{LO}_{\rm frag}}{d M_{\gamma \gamma}^2 \;d|\!\cos \theta_S|}
\propto \frac{\alpha \,\as(\mu_R)}{s^2} \; g_{qg}(\theta_S) \;\sum_q e_q^2 
\;{\cal D}_{qg}(\tau;z_{\rm min};\mu_F, \mu_{frag}) + \dots \;\;,
\end{equation}
where
\begin{equation}
\label{cosfrag}
g_{qg}(\theta_S) = 
\frac{5 -\cos^2 \theta_S}{2(1-\cos^2 \theta_S)} \;\;,
\end{equation}
\begin{equation}
\label{fraglum}
{\cal D}_{qg}(\tau;z_{\rm min};\mu_F, \mu_{frag}) =
\int_{z_{\rm min}}^1 \frac{dz}{z} \;D_{q/\gamma}(z;\mu_{frag})
\;{\cal L}_{qg}(\tau/z;\mu_F) \;\;.
\end{equation}
As in the case of Eq.~(\ref{lodir}), the right-hand side of Eq.~(\ref{lofrag})
does not include an overall proportionality factor of order unity and we have
explicitly written only the single-fragmentation contribution
due to the initial-state $qg$ partonic channel (the dots in the  
right-hand side of Eq.~(\ref{lofrag}) stand for all the other 
single-fragmentation and double-fragmentation contributions). As previously
remarked in the context of our discussion of the total cross section
$\sigma^{LO}_{\rm standard}$, when the fragmentation component is large
(i.e., with a similar size as the direct component) the 
single-fragmentation contribution from the $qg$ initial state gives the bulk of
the entire fragmentation component. The angular distribution 
$g_{qg}(\theta_S)$ in Eq.~(\ref{cosfrag}) is due to $qg \to q \gamma$ scattering
and its $\cos \theta_S$ dependence is similar to that of 
$g_{q{\bar q}}(\theta_S)$ in Eq.~(\ref{cosdir}).

The function ${\cal D}_{qg}$ in Eq.~(\ref{fraglum}) is an 
`effective' (with isolation) partonic luminosity, which is obtained by
convoluting the $qg$ PDF luminosity ${\cal L}_{qg}$ with the quark-to-photon
fragmentation function $D_{q/\gamma}$. The boundary value $z_{\rm min}$ in the
convolution is due to photon isolation (in the case of the single fragmentation
component, $z_{\rm min}$ is due to the isolation requirement
$z > p_{T \gamma}^{soft}/(p_{T \gamma}^{soft}+\ETmax)\,$) and it increases as
$\ETmax$ decreases, thus leading to an increasing suppression effect of the
fragmentation component as $\ETmax$ decreases (see Figs.~\ref{fig:dm} 
and~\ref{fig:dcos}).

Despite the isolation suppression, we have already remarked that the effective
partonic luminosity 
$\frac{\as}{\alpha} {\cal D}_{qg}(\tau;z_{\rm min})$
can still be quantitatively similar to ${\cal L}_{q{\bar q}}$
(and, hence, $d\sigma^{LO}_{\rm frag}$ and $d\sigma^{LO}_{\rm dir} =
d\sigma^{LO}_{\rm smooth}$ can have similar size) because ${\cal L}_{qg}$ is
larger than ${\cal L}_{q{\bar q}}$. Increasing the value of 
$M_{\gamma \gamma}$, the photon transverse momenta and, consequently, 
$z_{\rm min}$ tend to increase (unless $|\dy|$ and, correspondingly,
$|\!\cos \theta_S|$ have large values), therefore reducing the size of the
fragmentation component. The effect is visible in the LO results of 
Fig.~\ref{fig:dm}, which show that the relative difference between  
$d\sigma^{LO}_{\rm smooth}$ and $d\sigma^{LO}_{\rm standard}$ is reduced at high
values of $M_{\gamma \gamma}$. The effect is also visible in the comparison
between the LO results of Figs.~\ref{fig:dcos}-right and \ref{fig:dcosnew}: the
invariant-mass cut $M_{\gamma \gamma} > 200$~GeV strongly reduces the relative
contribution of the fragmentation component, unless $|\!\cos \theta^{*}|$
is large. We note that, increasing the value of $M_{\gamma \gamma}$, the relative
effect of the fragmentation component also decreases because ${\cal L}_{qg}$
and ${\cal L}_{q{\bar q}}$ become quantitatively more similar for increasing
values, $x \sim {\sqrt \tau}=M_{\gamma \gamma}/{\sqrt s}$, of the parton momentum
fraction $x$. 

\begin{figure}[htb]
\begin{center}
\begin{tabular}{cc}
\includegraphics[width=0.49\textwidth]{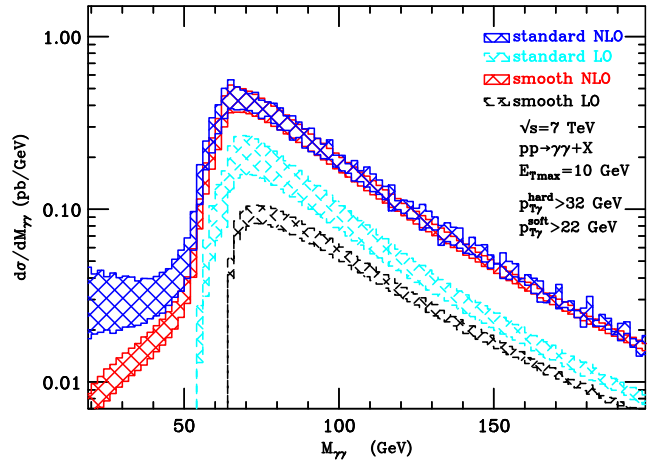}
\\
\end{tabular}
\end{center}
\caption{\label{fig:dmnew}
{\em The $M_{\gamma \gamma}$  differential cross section 
for $\ETmax=10$~GeV 
with the same photon kinematical cuts as in Fig.~\ref{fig:dm}-right and
an increased cut on the minimum value of $p_{T \gamma}^{hard}$
($p_{T \gamma}^{hard} > 32$~GeV).
The scale variation bands of the LO and NLO results for smooth and standard
isolation are as follows: 
LO smooth isolation (black dashed),
LO standard isolation (light-blue dashed),
NLO smooth isolation (red solid) and
NLO standard isolation (blue solid).}}
\end{figure}

Since 
$p_{T \gamma \gamma} = p_{T \gamma}^{hard} - p_{T \gamma}^{soft} \neq 0$
(in particular, $p_{T \gamma}^{hard} - p_{T \gamma}^{soft} < \ETmax$),
both values, $p_H$ and $p_S$, of the $p_T$ cuts are effective in the case of the
fragmentation component. In particular, they still lead to an LO kinematical
boundary,
$M_{\gamma \gamma}~>~M^{LO}_{\rm frag}$, on 
$M_{\gamma \gamma}$ but we have $M^{LO}_{\rm frag} < M^{LO}_{\rm dir}$. The
boundary value is $M^{LO}_{\rm frag} = 2 \sqrt{p_H (p_H - \ETmax)}\;$ if
$p_H - \ETmax > p_S$, and 
$M^{LO}_{\rm frag} = 2 \sqrt{p_H p_S}\;$ if $p_S > p_H - \ETmax$.
The vanishing of the LO standard isolation cross section at 
$M_{\gamma \gamma}~<~M^{LO}_{\rm frag}$ is visible in Fig.~\ref{fig:dm}-left
($M^{LO}_{\rm frag} \simeq 48$~GeV) and Fig.~\ref{fig:dm}-right
($M^{LO}_{\rm frag} \simeq 47$~GeV). The presence of two different LO thresholds,
$M^{LO}_{\rm frag}$ and $M^{LO}_{\rm dir}$, for standard and smooth isolation 
is more evident in Fig.~\ref{fig:dmnew}
($M^{LO}_{\rm frag}\simeq 53$~GeV and $M^{LO}_{\rm dir}=64$~GeV), which presents
the results for $d\sigma/dM_{\gamma \gamma}$ with the same kinematical
cuts as in Fig.~\ref{fig:dm} but with an increased value of $p_H$
($p_{T \gamma}^{hard} > p_H= 32$~GeV).

The shape of $d\sigma^{LO}_{\rm frag}/dM_{\gamma \gamma}$ near the LO threshold
is qualitatively 
similar to that of $d\sigma^{LO}_{\rm smooth}/dM_{\gamma \gamma}$: the maximum
value and the sharp decrease of $d\sigma^{LO}_{\rm frag}/dM_{\gamma \gamma}$
for $M_{\gamma \gamma} \sim M^{LO}_{\rm frag}$ are produced by the $p_T$ cuts
through the same kinematical mechanism that we have described in the case of the
smooth isolation result. In the case of the single-fragmentation component,
using $2p_{T \gamma}^{soft}= 2 z p_{T \gamma}^{hard} = 
{\sqrt z} M_{\gamma \gamma} \sin \theta_S$, 
we can express $z_{\rm min}$ as a
function of $\ETmax/(M_{\gamma \gamma} \sin \theta_S)$ and the $p_T$ cuts
lead to the constraints 
$M_{\gamma \gamma} \sin \theta_S > 2 p_H {\sqrt z}$
and ${\sqrt z} M_{\gamma \gamma} \sin \theta_S > 2 p_S$.
Note that the integration region over the photon momentum fraction $z$ is limited
also by the effect of the $p_T$ cuts. In the vicinity of the LO invariant-mass
threshold, $M_{\gamma \gamma} \sim M^{LO}_{\rm frag}$, 
the phase space integration region over $z$
is strongly suppressed by these cuts, and the $M_{\gamma \gamma}$ distribution
vanishes proportionally to $(M_{\gamma \gamma} - M^{LO}_{\rm frag})^{3/2}$.
In particular, this vanishing behaviour is stronger and smoother (by a factor of 
$M_{\gamma \gamma} - M^{LO}_{\rm frag}$) than the LO vanishing behaviour of 
$d\sigma/dM_{\gamma \gamma}$ for smooth cone isolation.

The effect of the rapidity cut $|y_\gamma| < y_M$ is analogous to the case of
smooth isolation: it leads to the replacement 
${\cal L}_{qg}(\tau) \to {\cal L}_{qg}(\tau;y_{max})$
($y_{max}=y_M-|\dy|/2$) in the right-hand side of Eq.~(\ref{fraglum}).

\subsubsection{Differential cross sections at the NLO}
\label{sec:diffNLO}

\begin{figure}[htb]
\begin{center}
\begin{tabular}{cc}
\hspace*{-4.5mm}
\includegraphics[width=0.48\textwidth]{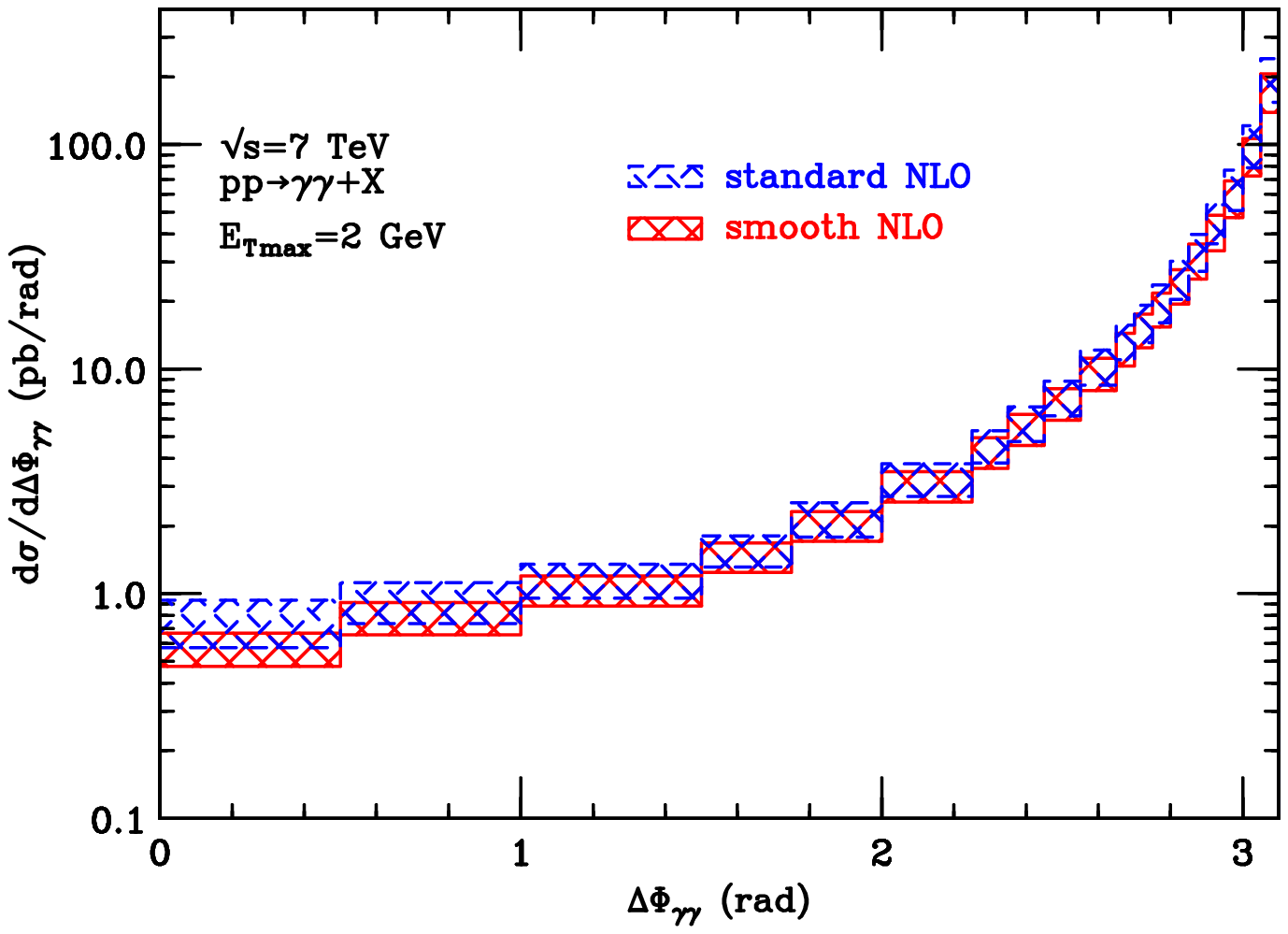}
& \includegraphics[width=0.48\textwidth]{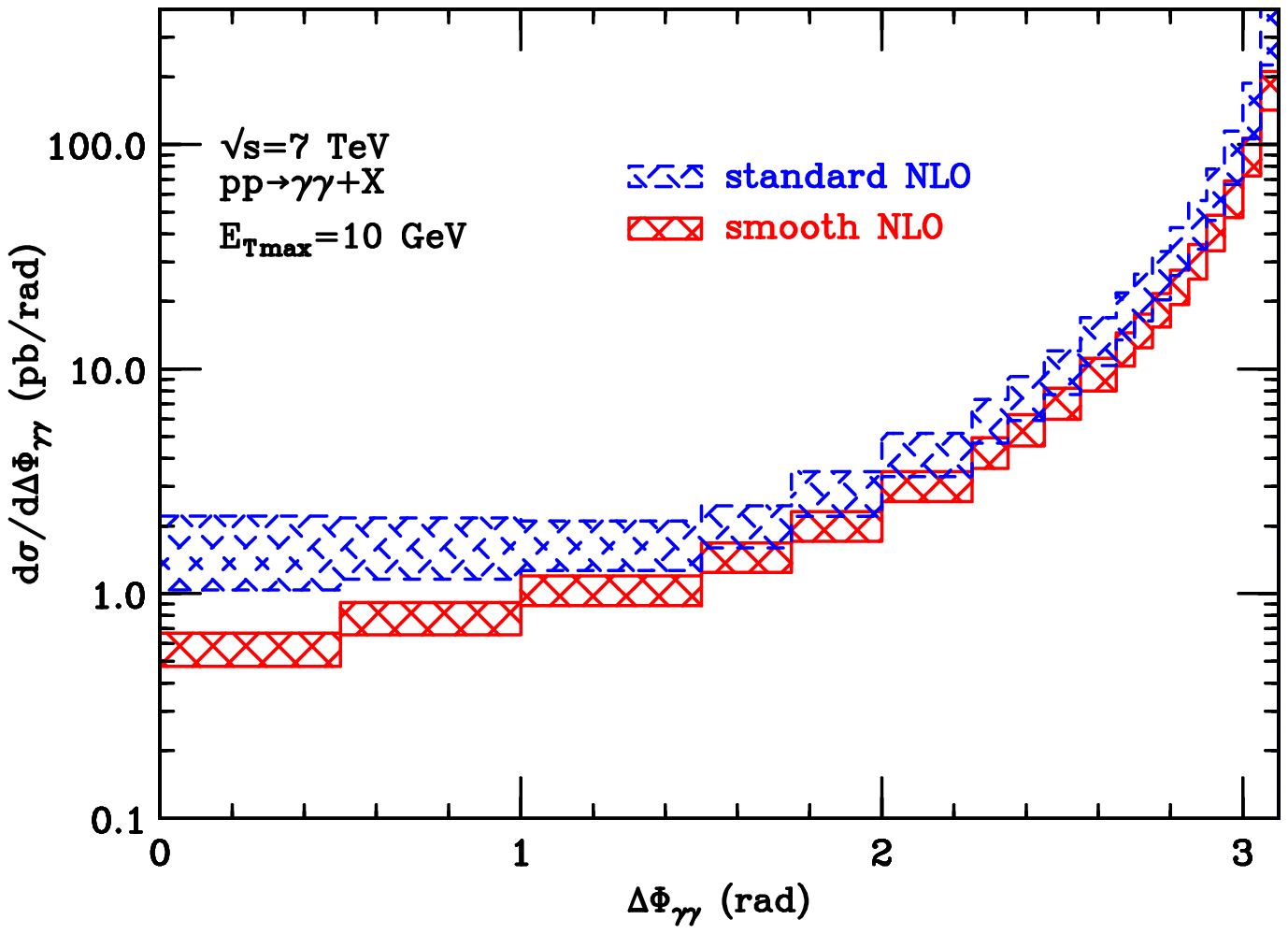}
\\
\end{tabular}
\end{center}
\caption{\label{fig:dphi}
{\em 
The NLO results (scale variation bands)
for the $\Dpgg$ differential cross section that are obtained 
by using the smooth (red solid band) and standard (blue dashed band) cone isolation criteria
with $\ETmax=2$~GeV (left panel) and $\ETmax=10$~GeV (right panel). 
The photon kinematical cuts are the same as in Fig.~\ref{fig:dm}.
}}
\end{figure}

\begin{figure}[htb]
\begin{center}
\begin{tabular}{cc}
\hspace*{-4.5mm}
\includegraphics[width=0.48\textwidth]{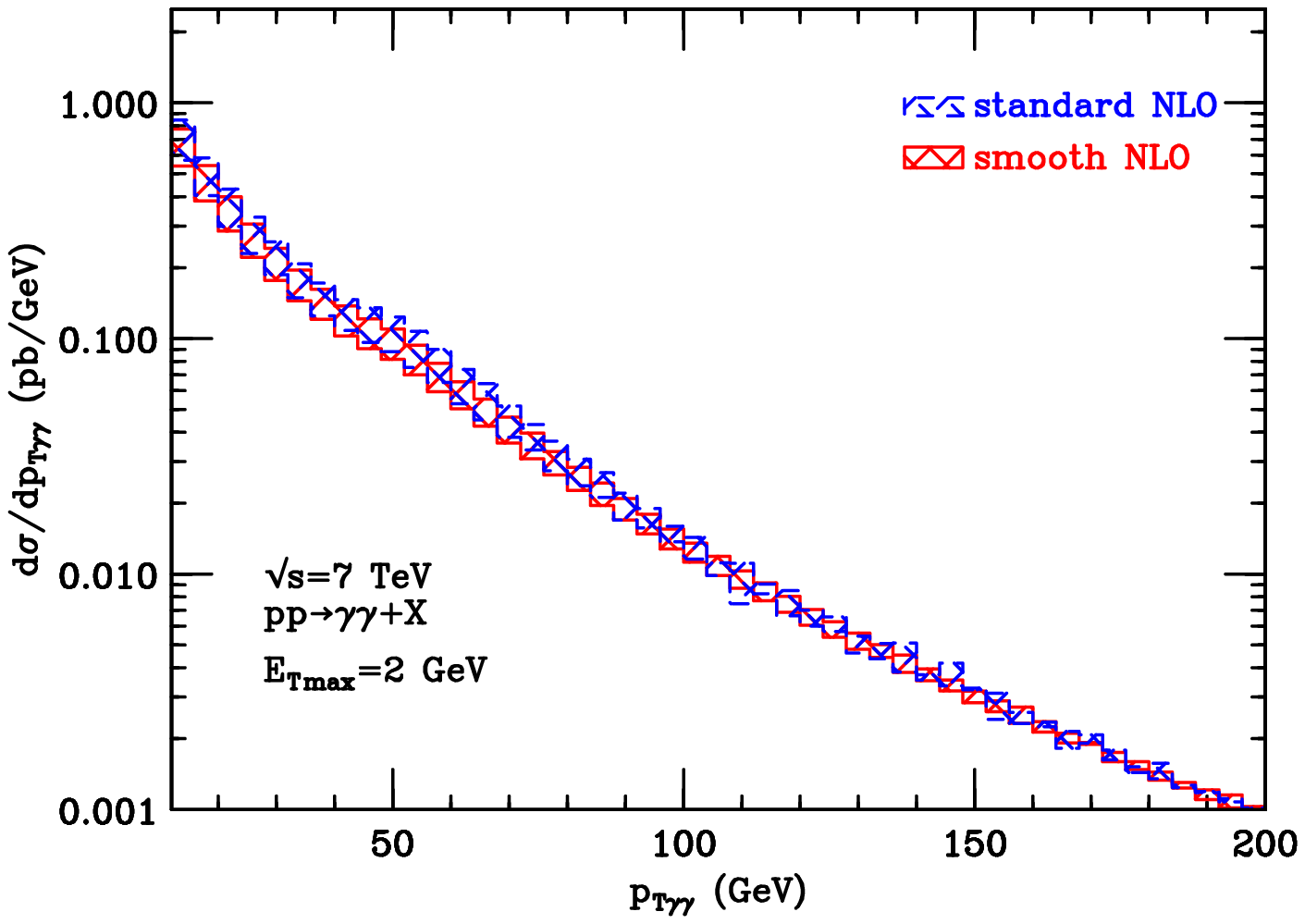}
& \includegraphics[width=0.48\textwidth]{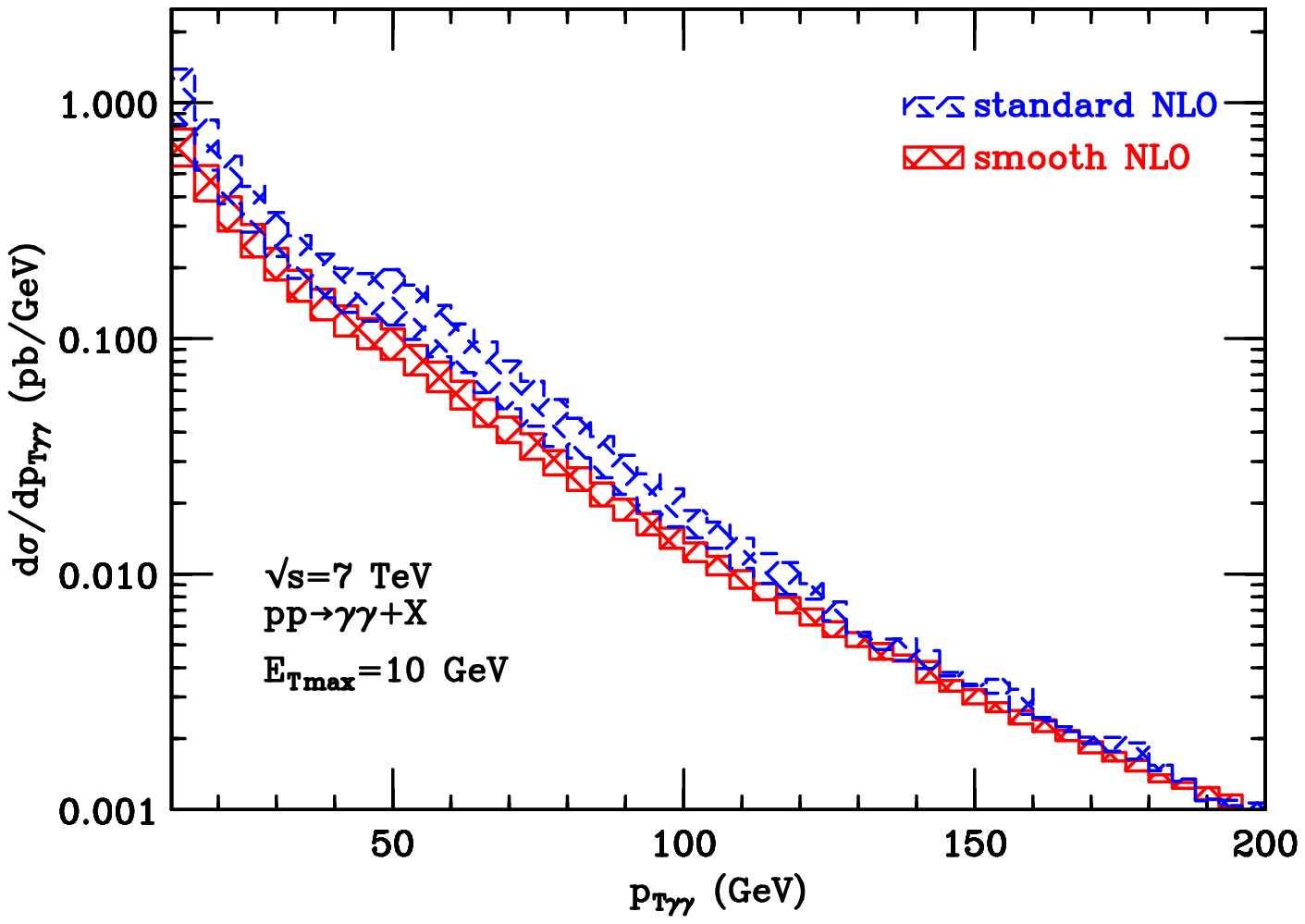}
\\
\end{tabular}
\end{center}
\caption{\label{fig:dpt}
{\em 
The NLO results (scale variation bands)
for the $p_{T \gamma \gamma}$ differential cross section that are obtained 
by using the smooth (red solid band) and standard (blue dashed band) cone isolation criteria
with $\ETmax=2$~GeV (left panel) and $\ETmax=10$~GeV (right panel). 
The photon kinematical cuts are the same as in Fig.~\ref{fig:dm}.
}}
\end{figure}

We now move to discuss the NLO results for the differential cross sections.
In addition to the $M_{\gamma \gamma}$ and $\cos \theta^{*}$ distributions,
we present the NLO results for the differential cross sections with respect
to the diphoton azimuthal separation $\Dpgg$ 
(Fig.~\ref{fig:dphi}) and to the transverse momentum $p_{T \gamma \gamma}$
of the photon pair (Fig.~\ref{fig:dpt}). The NLO results in 
Figs.~\ref{fig:dphi} and \ref{fig:dpt} are obtained by using the reference
kinematical cuts described at the beginning of Sect.~\ref{subsec:res}
(as in the case of Figs.~\ref{fig:dm} and \ref{fig:dcos}). As we have previously
noticed, the LO calculation leads to non-vanishing differential cross section
only in specific LO kinematical subregions. Therefore, outside these
LO  kinematical subregions (i.e., if $M_{\gamma \gamma}< M^{LO}$, 
$\Dpgg \neq \pi$ or $p_{T \gamma \gamma} > \ETmax$),
the NLO results presented in Figs.~\ref{fig:dm}, \ref{fig:dmnew}, 
\ref{fig:dphi} and \ref{fig:dpt} actually represent `effective' LO predictions
for the corresponding differential cross sections. We also note that, dealing
with `effective' LO predictions, the distinction between direct and fragmentation
components is unambiguous in the context of standard cone isolation.

We first discuss the results for the invariant-mass distribution 
(Fig.~\ref{fig:dm}). It is convenient to consider three different 
regions: the region of intermediate values of $M_{\gamma \gamma}$
(say, 45~GeV$< M_{\gamma \gamma} < 65$~GeV) around the LO kinematical threshold
at $M_{\gamma \gamma} \sim M^{LO}$,
and the regions of higher and lower values of $M_{\gamma \gamma}$.
For the purposes of the subsequent discussions, we also define
\beq
\label{lothres}
M^{LO} \equiv 2 \sqrt{ p_H p_S} \;\;.
\eeq
Note that $M^{LO}$ is equal to (or smaller than) the  minimum between 
the thresholds $M^{LO}_{\rm frag}$ and $M^{LO}_{\rm dir}$.

In the high-mass region where $M_{\gamma \gamma} > 65$~GeV
(Fig.~\ref{fig:dm}), the NLO results for smooth and standard isolation are
quantitatively very similar, with a scale dependence that is comparable to that
of the corresponding NLO total cross sections.
The NLO corrections are large for both isolation criteria and for both values of 
$\ETmax=2$~GeV and 10~GeV considered in Fig.~\ref{fig:dm}.
All these features are similar (both qualitatively and quantitatively) to those
of the NLO and LO total cross sections and they have exactly the same origin,
which we have already remarked in our discussion on the total cross sections.
We do not repeat such a discussion on the role of the 
$qg$ initial-state channel and of the corresponding PDF luminosity at different
perturbative orders and within the two different isolation 
criteria.

The NLO total cross section receives a negligible contribution from 
$d\sigma^{NLO}/dM_{\gamma \gamma}$ in the low-mass region 
($M_{\gamma \gamma} < 45$~GeV): the contribution is smaller than the scale
dependence of the total cross section.
The LO kinematical boundary on $M_{\gamma \gamma}$ is unphysical: it is due to
the $p_T$ cuts on the photons but it is an artifact of the LO kinematics, which
implies $\Dpgg = \pi$. Physical diphoton events (and also
corresponding partonic contributions beyond the LO) can have 
$\Dpgg < \pi$: they produce non-vanishing values of 
$d\sigma/dM_{\gamma \gamma}$ at $M_{\gamma \gamma} < M^{LO}$, although this
kinematical region is strongly suppressed by the photon $p_T$ cuts.
Owing to energy conservation and the presence of the $p_T$ cuts,
the low-mass region selects diphoton events with small values of $\Dpgg$.
Owing to transverse-momentum conservation, in the low-mass region these 
$p_T$ cuts effectively act also as a lower limit 
on $p_{T \gamma \gamma}$ or, equivalently,
on the total transverse
momentum of the hadronic (partonic) final-state system.
Roughly speaking, low values of $M_{\gamma \gamma}$ imply small values of
$\Dpgg$ and, in turn, relatively-large values of $p_{T \gamma \gamma}$.

In general, kinematics
leads to the minimal constraint 
\beq
\label{lowm}
M_{\gamma \gamma} > M^{LO} \sin\left(\frac{\Dpgg}{2}\right) \;\;,
\eeq
which implies that decreasing values of $M_{\gamma \gamma}$ necessarily require
decreasing values of $\Dpgg$. The constraint in Eq.~(\ref{lowm})
is obtained by setting
$\dy=0$; larger values of $|\dy|$ further reduce the
value of $\Dpgg$ at fixed value of $M_{\gamma \gamma}$.
Eventually the kinematical lower limit\footnote{If 
$\Dpgg < R_{\gamma \gamma}^{\rm min}$, kinematics leads to the replacement
$\sin(\Dpgg/2) \to \sinh(\sqrt{(R_{\gamma \gamma}^{\rm min}/2)^2 - 
(\Dpgg/2)^2}\,)$ in the right-hand side of Eq.~(\ref{lowm}).}
on $M_{\gamma \gamma}$ 
is obtained by the combined effect of the $p_T$ cuts and the cut on the
minimum angular separation, $R_{\gamma \gamma} > R_{\gamma \gamma}^{\rm min}
= 0.4$, between the photons. 
Since the value of $R_{\gamma \gamma}^{\rm min}$ is small, the lower limit on 
$M_{\gamma \gamma}$ is 
$M_{\gamma \gamma} \gtap M^{LO} R_{\gamma \gamma}^{\rm min}/2 \simeq 10$~GeV.

As a consequence
of the kinematical constraint in Eq.~(\ref{lowm}), 
if $M_{\gamma \gamma} \ltap 41$~GeV ($M_{\gamma \gamma} \ltap 33$~GeV)
we have $\Dpgg \ltap 2.1 \simeq 2\pi/3$ ($\Dpgg \ltap 1.6 \simeq \pi/2$):
therefore, due to transverse-momentum conservation, the total transverse
momentum of the partonic (hadronic) final state is necessarily larger than 
$p_{T \gamma}^{soft}$ ($p_{T \gamma}^{hard}$). 
At smaller values of $M_{\gamma \gamma}$ we have
$\Dpgg \ltap 1.4$ if $M_{\gamma \gamma} \simeq 30$~GeV and
$\Dpgg \ltap 0.9$ if $M_{\gamma \gamma} \simeq 20$~GeV: therefore,
we are dealing with a relatively collimated diphoton system that recoils against
a high-$p_T$ hadronic (partonic) jet in the transverse-momentum plane.

Using kinematical considerations, the region of small values of $\Dpgg$ can also
be more directly related to $p_{T \gamma \gamma}$. As a consequence of
transverse-momentum conservation  and of the photon $p_T$ cuts,
provided $\Dpgg < \pi/2$ we have
\beq
\label{ptvsphi}
p_{T \gamma \gamma} > \sqrt{p_H^2 + p_S^2 + 2p_H p_S \,\cos\,(\Dpgg)} \;\;\;, 
\quad \quad \quad
(\Dpgg < \pi/2) \;\;.
\eeq
This relation shows that small values of $\Dpgg$ necessarily imply
relatively-large values of $p_{T \gamma \gamma}$. 
For instance, if $\Dpgg = \pi/2$ we have 
$p_{T \gamma \gamma} > \sqrt{p_H^2 + p_S^2}$ 
(i.e., $p_{T \gamma \gamma} > 34$~GeV if $p_H=25$~GeV and $p_S=22$~GeV),
whereas at very small values of $\Dpgg$ we have
\beq
\label{ptphi0}
p_{T \gamma \gamma} > p_H + p_S \;\;,
\quad \quad \quad
(\Dpgg \simeq 0) \;\;.
\eeq
Therefore, if $p_H=25$~GeV and $p_S=22$~GeV, the region where 
$\Dpgg \simeq 0$ does not contribute to the $p_{T \gamma \gamma}$ spectrum
unless $p_{T \gamma \gamma} \gtap 47$~GeV.

In the low-mass region (Fig.~\ref{fig:dm}), the scale dependence of the NLO
differential cross section is larger than the corresponding dependence of the NLO
total cross section, as expected from an effective LO prediction.
In the case of smooth isolation, the scale dependence slightly increases by
decreasing $M_{\gamma \gamma}$; at $M_{\gamma \gamma} \sim 20$~GeV the
scale dependence is roughly a factor of 2 larger than the scale dependence of
the NLO total cross section. The scale dependence of the standard isolation
result is larger, and it increases by either decreasing $M_{\gamma \gamma}$
or increasing $\ETmax$; at $M_{\gamma \gamma} \sim 20$~GeV and
$\ETmax=10$~GeV the scale dependence of the NLO differential cross section is
roughly a factor of 3.6 larger than the scale dependence of
the NLO total cross section. At the lower value of $\ETmax$ (2~GeV), smooth
and standard isolations give similar NLO differential cross sections, within the
corresponding scale variation uncertainties. At the higher value of 
$\ETmax$ (10~GeV), the smooth isolation result is systematically smaller than
the standard isolation result and the relative difference increases by decreasing
$M_{\gamma \gamma}$: the NLO results for standard isolation is roughly a factor
of 3.8 (2.5) larger than the corresponding result for smooth isolation at
$M_{\gamma \gamma} \sim 20$~GeV ($M_{\gamma \gamma} \sim 30$~GeV).

The observed NLO differences between smooth and standard isolation in the
low-mass region (analogously to the corresponding LO differences in the
high-mass region and to the LO differences of the total cross section) 
deserve specific comments.
The NLO calculation for smooth isolation has two photons and one parton in the
final state. Owing to transverse-momentum conservation, the parton can be inside
the photon isolation cones only if $\Dpgg > \pi -R \simeq 2.7$, 
which corresponds to $M_{\gamma \gamma} \gtap 46$~GeV in view of the constraint
in Eq.~(\ref{lowm}).
Therefore, in the entire
low-mass region the NLO result for smooth isolation is exactly independent of 
$\ETmax$, and that represents a much simplified approximation of the
expected physical behaviour. The independence of  $\ETmax$ also implies that
the smooth isolation result is exactly equal to the result of the direct
component for standard isolation. Therefore, the observed differences between the
two isolation criteria are entirely due to the fragmentation component of the
standard isolation calculation. The NLO result for smooth isolation (or,
equivalently, for the direct component) is due to the partonic processes
$q{\bar q} \to g \gamma \gamma$ and $qg \to q \gamma \gamma$ (or 
${\bar q} g \to{\bar q} \gamma \gamma$), where the collimated diphoton system
recoils against the final-state parton, and the initial-state $qg$ process gives
the dominant contribution because of the larger ${\cal L}_{qg}$ PDF luminosity.
The large NLO effect of the fragmentation component (especially at the higher
value of $\ETmax$, which leads to a smaller suppression effect from isolation)
is due to its numerous partonic processes, which, moreover, also include the
$gg$ and $qq$ initial states: the corresponding partonic cross sections
(although they are suppressed by isolation) can be enhanced by the size of the
PDF luminosity (${\cal L}_{gg}$ and ${\cal L}_{qg}$ have a comparable size). In
particular, two of these partonic processes are 
\beq
\label{qgnlofrag}
qg \to g q \gamma \quad\quad (q \to \gamma +X) \;\;
\eeq
and
\beq
\label{ggnlofrag}
gg \to {\bar q} q \gamma \quad\quad (q \to \gamma +X, \;\; 
{\rm or} \;\; {\bar q}  \to \gamma +X) \;\;,
\eeq
where the final-state 
quark (or the antiquark in the case of the $gg$ channel of 
Eq.~(\ref{ggnlofrag})) 
is collimated with the
$\gamma$ and fragments into a second photon.
At low values of $M_{\gamma \gamma}$ these two partonic processes are enhanced
by the relative factor $(M^{LO}/M_{\gamma \gamma})^2$ (its value is about 2.4
and 5.5 at $M_{\gamma \gamma} \sim 30$~GeV and $M_{\gamma \gamma} \sim 20$~GeV,
respectively), which originates from the final-state perturbative singularity 
in the $q \gamma$ collinear limit
(or ${\bar q} \gamma$ collinear limit in the case of the $gg$ channel of
Eq.~(\ref{ggnlofrag})).
Analogous processes, namely,
\beq
\label{qgnlosmooth}
qg \to g q \gamma \gamma
\eeq
and 
\beq
\label{ggnlosmooth}
gg \to {\bar q} q \gamma \gamma \;\;,
\eeq
where a
final-state quark (or antiquark) is inside the isolation cone of one photon, 
contribute
(and their contribution depends on $\ETmax$) to the NNLO calculation for
smooth isolation. We thus expect that these processes lead to large radiative
corrections\footnote{At the strictly formal level, we note that the enhancing
factor $(M^{LO}/M_{\gamma \gamma})^2$ in the NLO calculation for standard
isolation is `unphysical' in the limit $M_{\gamma \gamma} \to 0$, and the
$(M_{\gamma \gamma})^{-2}$ behaviour is softened by higher-order radiative
corrections for both standard and smooth isolation.}
(this expectation is confirmed by the NNLO results presented and discussed in
Sect.~\ref{sec:diffNNLO}; see, in particular, Fig.~\ref{fig:mlhc}-left)
and that the corresponding NNLO result removes the large differences with
respect to the standard isolation result.

In summary, according to our discussion, the sizeable NLO differences between
standard and smooth isolation results that are observed in the low-mass region
(Fig.~\ref{fig:dm}) are more an artifact of the NLO calculation than a physical
effect due to the two different isolation criteria. Certainly, the achievement
of high-precision QCD predictions in the low-mass region is challenging,
basically because of the relatively-low value of the characteristic
hard-scattering scale $M_{\gamma \gamma}$.

The region of intermediate values of $M_{\gamma \gamma}$
(45~GeV $< M_{\gamma \gamma} < 65$~GeV in Fig.~\ref{fig:dm}) is the region where
the shape of $d\sigma/dM_{\gamma \gamma}$ is directly most sensitive to the
$p_T$ cuts. The NLO results for smooth and standard isolation are quite similar
in this region. Independently of the isolation criterion, the differential cross
section starts to rapidly increase at  $M_{\gamma \gamma} \simeq M^{LO} \simeq
47$~GeV and the position of the peak (the maximum of 
$d\sigma/dM_{\gamma \gamma}$) is basically unchanged with respect to that of the
LO result for standard isolation. It is noticeable that 
$d\sigma/dM_{\gamma \gamma}$ starts to increase at 
$M_{\gamma \gamma} \simeq M^{LO}$ ($M^{LO} \simeq M^{LO}_{\rm frag}$)
despite the presence of two different LO threshold, $M^{LO}_{\rm frag}$ and
$M^{LO}_{\rm dir}$, that are displaced (the LO threshold at $M^{LO}_{\rm dir}$
is somehow more `unphysical': it is insensitive to the value of $p_S$ since
absolutely no partonic transverse momentum is allowed in the final state of the
corresponding LO calculation). The NLO corrections are obviously large for 
$M_{\gamma \gamma} < M^{LO}_{\rm dir} =50$~GeV, and they are sizeable also at 
$M_{\gamma \gamma} > 50$~GeV (for the same reason as for the high-mass region).
In the case with $\ETmax=10$~GeV we note that the NLO result for smooth
isolation tends to be larger than the corresponding result for standard
isolation, in disagreement with the physical constraint in Eq.~(\ref{eq:a}).
We do not regard this behaviour as particularly worrisome since the values of 
$d\sigma/dM_{\gamma \gamma}$ for the two isolation criteria are basically
consistent with each other within the computed scale variation 
uncertainties\footnote{We also note that in this $M_{\gamma \gamma}$ region
the scale dependence of the NLO standard isolation result at 
$\ETmax=10$~GeV is quite small, in particular, if it is compared with the
corresponding scale dependence at  $\ETmax=2$~GeV. Such a small NLO scale
dependence seems accidental.}.
Moreover, these scale variation effects are expected to underestimate 
higher-order perturbative uncertainties. Indeed,
perturbative calculations in regions
around unphysical fixed-order thresholds are known
\cite{Catani:1997xc}
to be generally affected by perturbative instabilities at higher orders
(further comments on this point are postponed to the end of this section).
In our specific
case, $d\sigma/dM_{\gamma \gamma}$ is very steep 
in the region around the LO kinematical threshold 
and even the effect of little
instabilities is amplified by the large slope of $d\sigma/dM_{\gamma \gamma}$.
The ensuing effect on the comparison between smooth and standard isolation
results can be relevant since the slopes of the two results are both large and
they are different.
We also add that the standard isolation results do not include the uncertainty
(which is difficult to be quantified) that is due to the limited knowledge
of the photon fragmentation functions (an increased value of the fragmentation
functions can reduce the differences between smooth and standard isolation results
in this $M_{\gamma \gamma}$ region).

\setcounter{footnote}{2}

We briefly comment on the results in Fig.~\ref{fig:dmnew}, which are analogous
to those in Fig.~\ref{fig:dm}-right apart from having more asymmetric $p_T$
cuts. The value of $p_H$ increases in going from Fig.~\ref{fig:dm}-right
to Fig.~\ref{fig:dmnew}, so that the size of the unbalance between the minimum
values of the photon transverse momenta increases from $p_H - p_S=3$~GeV
to $p_H - p_S=10$~GeV. The main features of the results in 
Figs.~\ref{fig:dm}-right and \ref{fig:dmnew} are very similar. In particular, we
can comment on the behaviour of $d\sigma/dM_{\gamma \gamma}$ in the region of
intermediate values of $M_{\gamma \gamma}$, just above $M_{\gamma \gamma}= 
M^{LO}$ ($M^{LO}\simeq 47$~GeV and $M^{LO}\simeq 53$~GeV in 
Figs.~\ref{fig:dm} and \ref{fig:dmnew}, respectively). The NLO results for
smooth and standard isolation are very similar in this region despite the fact
that the difference between the LO thresholds $M^{LO}_{\rm frag}$
($M^{LO}_{\rm frag}=M^{LO}$) and
$M^{LO}_{\rm dir}$ is larger in the case of more asymmetric cuts
($M^{LO}_{\rm dir} - M^{LO} \simeq 11$~GeV in Fig.~\ref{fig:dmnew}, while
$M^{LO}_{\rm dir} - M^{LO} \simeq 3$~GeV in Fig.~\ref{fig:dm}). 
This similarity between the NLO results also confirms our observation in the
previous paragraph that the `threshold' at $M_{\gamma \gamma} \sim M^{LO}$
is somehow more `physical'. Even in the case
of more asymmetric cuts, the NLO corrections are able to remove the 
relative `deficit'
of the LO results for smooth isolation in the vicinity of the LO threshold.
Comparing Figs.~\ref{fig:dm}-right and \ref{fig:dmnew} we see that 
the slope of $d\sigma/dM_{\gamma \gamma}$ is smaller in the case
of more asymmetric cuts,  and the NLO results for smooth and standard isolation
are more similar (in particular, the smooth isolation result does not tend to be
larger than the standard isolation result, consistently with our previous
comment on the behaviour observed in Fig.~\ref{fig:dm}). Additional comments on
symmetric and asymmetric $p_T$ cuts are presented in Sect.~\ref{sec:results}.

In Fig.~\ref{fig:dphi} we present the NLO results for the differential
cross section with respect to the azimuthal angle separation 
$\Dpgg$ of the two photons. At the LO,  
$\Dpgg = \pi$ for both isolation criteria. 
As is well known \cite{Binoth:1999qq},
at higher perturbative orders the computation of the differential cross section
is affected by large logarithmic corrections in the region near 
$\Dpgg = \pi$. In this region, any fixed-order QCD result is 
physically not reliable, and reliable quantitative predictions for the detailed
shape of the $\Dpgg$ distribution can be recovered only through
all-order perturbative resummation \cite{Balazs:2007hr,Cieri:2015rqa}
of these large logarithmic contributions. Owing to this reason, in the results
of Fig.~\ref{fig:dphi} we have excluded the region around 
$\Dpgg = \pi$. This also implies that the NLO results in 
Fig.~\ref{fig:dphi} (with $\Dpgg < \pi$) actually represent
`effective' LO predictions for the $\Dpgg$ distribution.

Independently of the value of $\ETmax$, we note that the $\Dpgg$
distribution is
sharply peaked at large values of $\Dpgg$: the cross section decreases
by more than one order of magnitude by decreasing $\Dpgg$ from
$\Dpgg \sim 3$ to $\Dpgg \sim 2.5$, and it still decreases
by about one order of magnitude in going from $\Dpgg \sim 2.5$
to $\Dpgg \sim 0.5$.
Standard and smooth isolation results have a similar scale dependence that
increases from roughly $\pm 20$\% at $\Dpgg \sim 2.5$
to roughly $\pm 30$\% at $\Dpgg \sim 0.5$ 
(in the case of standard isolation the scale dependence also
slightly increases by increasing the value of $\ETmax$). 
As expected from an effective LO prediction,
this scale dependence
is larger (by about a factor of 2--3) than the scale dependence of the NLO total
cross section (the bulk of the NLO total cross section
is due to the region near $\Dpgg \sim \pi$, where the scale dependence
of the $\Dpgg$ differential cross section is reduced).

The main quantitative differences between
standard and smooth isolation results appear by examining the dependence on 
$\ETmax$ of the $\Dpgg$ distribution.
At very small values of $\ETmax$ ($\ETmax=2$~GeV) the two
isolation criteria lead to very similar results, within the corresponding scale
variation uncertainties. At larger values of $\ETmax$ ($\ETmax=10$~GeV),
smooth cone results tend to be smaller than standard cone results, and the
differences increase by decreasing the value of $\Dpgg$. At very small
values of $\Dpgg$, the differences are sizeable; for instance,
standard NLO results are roughly a factor of 2.1 larger than smooth NLO results 
at $\Dpgg \sim 0.5$. In the case of smooth isolation, the $qg$
initial-state partonic channel contributes more than the $q{\bar q}$ channel
(because of the larger ${\cal L}_{qg}$ PDF luminosity).
At high values of $\Dpgg$ the smooth isolation result depends very
weakly on $\ETmax$ and it is very similar (though it is not exactly equal)
to the direct component of the standard isolation result. As previously
mentioned, if $\Dpgg < \pi -R \simeq 2.7$
the smooth isolation result is exactly independent of $\ETmax$
and it coincides with the direct
component of the standard isolation cross section.
Therefore, at low values of $\Dpgg$ the NLO differences between the
two isolation criteria are entirely due to the fragmentation component of the
standard isolation calculation.

In Fig.~\ref{fig:dpt} we present
the NLO results for the differential
cross section
with respect to the transverse momentum $p_{T \gamma \gamma}$
of the photon pair. At LO, as already mentioned, $p_{T \gamma \gamma}=0$
for smooth isolation  while $p_{T \gamma \gamma} < \ETmax$ for standard
isolation. Analogously to the case of $d\sigma/d\Dpgg$ at
$\Dpgg \simeq \pi$, the perturbative computation of 
$d\sigma/dp_{T \gamma \gamma}$ at small values of $p_{T \gamma \gamma}$
is affected by large logarithmic corrections
\cite{Binoth:1999qq}
that have to be treated by an all-order resummation procedure
\cite{Balazs:2007hr,Cieri:2015rqa}
to obtain reliable QCD predictions. Therefore, in Fig.~\ref{fig:dpt}
we do not show the NLO results 
at small values of 
$p_{T \gamma \gamma}$.  Actually, we consider only the region with 
$p_{T \gamma \gamma} > \ETmax$, where the results have to be regarded as
`effective' LO predictions.

The scale dependence of the NLO result for $d\sigma/dp_{T \gamma \gamma}$
in the case of smooth isolation is similar to that of the NLO total cross
section, while the scale dependence in the case of standard isolation
is larger. At small values of $\ETmax$ (e.g., $\ETmax=2$~GeV) the NLO
results for smooth and standard isolation are very similar, within the
corresponding scale dependence. At larger values of 
$\ETmax$ ($\ETmax=10$~GeV) the standard isolation result tends to
be larger than the smooth isolation result: however, the ratio of the two
results is always smaller than approximately 1.8. In the case of
smooth isolation, the $qg$ initial-state channel gives the dominant contribution.
The partonic final state of the NLO calculation for smooth isolation includes
the two photons and a parton.
Owing to transverse-momentum conservation, the transverse momentum of the parton
is equal to $p_{T \gamma \gamma}$ and, therefore, if 
$p_{T \gamma \gamma} > \ETmax$ (as in Fig.~\ref{fig:dpt}) the parton is not
allowed (by isolation) to be inside the photon isolation cones. Consequently, if
$p_{T \gamma \gamma} > \ETmax$ the NLO smooth isolation result is 
independent\footnote{Parton radiation inside the isolation cone is kinematically
allowed. If $p_{T \gamma \gamma} > \ETmax$ the smooth isolation result
is independent of $\ETmax$ but it still depends on the photon isolation
criterion: events with $p_{T \gamma \gamma} > \ETmax$ are kinematically
allowed but they are rejected by the isolation requirement.}
of $\ETmax$ and it exactly coincides with the result of the direct component
for standard isolation. Therefore, the differences between the smooth and
standard isolation results in Fig.~\ref{fig:dpt} are entirely due to the
fragmentation component of the standard isolation result.

We comment on the comparison between the smooth and standard isolation results in
Figs.~\ref{fig:dphi} and \ref{fig:dpt} at large values of $\ETmax$
(at small values of $\ETmax$, the results are very similar since the
fragmentation component is much suppressed).
As previously discussed, the fragmentation component of the NLO calculation is
dynamically enhanced at small values of $M_{\gamma \gamma}$
(mainly because of the presence of the fragmentation processes in 
Eqs.~(\ref{qgnlofrag}) and (\ref{ggnlofrag})). 
The low-$\Dpgg$ region in Fig.~\ref{fig:dphi}
receives contributions from both small and large values of $M_{\gamma \gamma}$.
The same happens in the large-$p_{T \gamma \gamma}$ region considered in
Fig.~\ref{fig:dpt} and, moreover, at large values of $p_{T \gamma \gamma}$
even diphoton events with small $\Dpgg$ tend to have larger values
of $M_{\gamma \gamma}$ (at fixed $\Dpgg$, $M_{\gamma \gamma}$
kinematically increases by increasing $p_{T \gamma \gamma}$). 
Therefore, the relative enhancement
effect of the NLO fragmentation component at low values of $M_{\gamma \gamma}$
continuously decreases in going to the low-$\Dpgg$ region and to the
large-$p_{T \gamma \gamma}$ region. As a consequence of this reasoning,
the relative
difference between standard and smooth isolation results correspondingly and
continuously decreases in going from $d\sigma/dM_{\gamma \gamma}$ at small
$M_{\gamma \gamma}$ (Fig.~\ref{fig:dm}-right), to 
$d\sigma/d\Dpgg$ (Fig.~\ref{fig:dphi}-right) and to
$d\sigma/dp_{T \gamma \gamma}$ (Fig.~\ref{fig:dpt}-right): this behaviour is
indeed observed in the NLO results of the corresponding figures. Since the
partonic processes that enhance the NLO fragmentation component contribute to
the smooth isolation result at NNLO (as discussed
in the accompanying comments to Eqs.~(\ref{qgnlofrag})--(\ref{ggnlosmooth})), 
we expect NNLO
corrections for smooth isolation that are large in the 
low-$M_{\gamma \gamma}$ region and that continuously decrease in going to small
values of $\Dpgg$, to large values of $p_{T \gamma \gamma}$
and to high values of $M_{\gamma \gamma}$. The actual NNLO results that are
presented in Sect.~\ref{sec:diffNNLO} 
basically confirm the expectation from our discussion.

In the context of standard cone
isolation, the relevance of 
the fragmentation processes in Eqs.~(\ref{qgnlofrag}) and (\ref{ggnlofrag})
for the differential
cross sections $d\sigma/d\Dpgg$ and $d\sigma/dp_{T \gamma \gamma}$
was already remarked in Ref.~\cite{Binoth:2000zt}.
In the case of $d\sigma/d\Dpgg$ at small values of $\Dpgg$, the authors of 
Ref.~\cite{Binoth:2000zt} noticed that these fragmentation processes give a 
NLO contribution that is sizeably larger than that of the direct component 
of the NLO cross section, as we have also remarked. The authors of 
Ref.~\cite{Binoth:2000zt} also nicely discussed how these fragmentation
processes have impact on $d\sigma/dp_{T \gamma \gamma}$ at relatively-large
values of $p_{T \gamma \gamma}$, thus producing a shoulder-type shape of the 
$p_{T \gamma \gamma}$ distribution. The $p_{T \gamma \gamma}$ shoulder
is clearly visible in the standard isolation results with $\ETmax=10$~GeV
in Fig.~\ref{fig:dpt}-right (the $p_{T \gamma \gamma}$ shoulder is much less
visible in Fig.~\ref{fig:dpt}-left, since at very small values of $\ETmax$,
such as $\ETmax=2$~GeV, the fragmentation component is highly suppressed).
The NLO shape of $d\sigma/dp_{T \gamma \gamma}$ for standard isolation 
changes (it flattens out) in the region where 
$40~{\rm GeV} \ltap p_{T \gamma \gamma} \ltap 50~{\rm GeV}$, and 
$d\sigma/dp_{T \gamma \gamma}$ is 
larger than the corresponding
NLO smooth isolation result (which is exactly equal to the direct component of
the standard isolation result at NLO) for increasing values of 
$p_{T \gamma \gamma}$. According to our previous discussion, the smooth
isolation NNLO processes in Eqs.~(\ref{qgnlosmooth}) and (\ref{ggnlosmooth})
have a dynamical role which is analogous to that of the NLO fragmentation
processes in Eqs.~(\ref{qgnlofrag}) and (\ref{ggnlofrag}) (and to part of their
NNLO radiative corrections). Therefore, we expect a pronounced 
$p_{T \gamma \gamma}$ shoulder in the smooth isolation results at NNLO. This
expectation is confirmed by the results that we present and further discuss in
Sect.~\ref{sec:diffNNLO}.

\setcounter{footnote}{2}

We now turn to consider the differential
cross section with respect to the polar angle $\theta^{*}$.
Figure~\ref{fig:dcos} presents
the NLO results by using the customary
kinematical cuts
of this subsection. Since the LO result covers the entire range of 
$\cos \theta^{*}$, the corresponding NLO results are effectively NLO QCD
predictions, and the main kinematical effects that we have discussed at the LO
level are unchanged at the NLO level. 

The main features of the comparison between
the LO and NLO results in Fig.~\ref{fig:dcos} are very similar to those of the
comparison between the corresponding total cross sections. At variance with other
kinematical distributions, the shape of the NLO results for 
$d\sigma/d\cos \theta^{*}$ has very little dependence on $\ETmax$: the increase
of $\ETmax$ from 2~GeV to 10~GeV has mainly an effect on the overall
normalization, whose size increases analogously to the value of the corresponding
NLO total cross section. The scale dependence of the NLO results is very similar
for the two isolation criteria: its value is approximately $\pm 15$\% at 
$\cos \theta^{*}=0$, and it is slightly smaller at larger values of 
$|\!\cos \theta^{*}|$. Smooth and standard isolation results at NLO are very
similar (with overlapping scale variation bands) for 
$|\!\cos \theta^{*}| \ltap 0.7$. At larger values of $|\!\cos \theta^{*}|$
the NLO result for standard isolation is systematically larger than the
corresponding result for smooth isolation: the ratio between the
standard and smooth  results is approximately 1.3 at 
$|\!\cos \theta^{*}| \sim 0.9$. In the region of central values of 
$\cos \theta^{*}$ ($|\!\cos \theta^{*}| \ltap 0.5$), we note that the NLO result
for smooth isolation tends to be larger than the corresponding result for standard
isolation, although the two NLO results are consistent with each other within the
computed scale variation dependence. A similar tendency has been already noticed
in the case of the differential cross section 
$d\sigma/dM_{\gamma \gamma}$ in the region where 
$M_{\gamma \gamma} \sim M^{LO}$ (Fig.~\ref{fig:dm}).
The two effects, for $d\sigma/d\cos \theta^{*}$ and $d\sigma/dM_{\gamma \gamma}$,
are certainly related since, as a consequence of the photon $p_T$ cuts
(as previously discussed in Sect.~\ref{sec:diffLO}), 
the differential cross section  
$d\sigma/d\cos \theta^{*}$ in the region where $|\!\cos \theta^{*}| \ltap 0.5$
is very sensitive to the $M_{\gamma \gamma}$ dependence of
$d\sigma/dM_{\gamma \gamma}d\cos \theta^{*}$ in the region 
50~GeV$\ltap M_{\gamma \gamma} \ltap 60$~GeV. 

We also comment on the 
$\cos \theta^{*}$ distribution in the region of large values of 
$|\!\cos \theta^{*}|$. Comparing the NLO calculations for the standard and smooth
isolation criteria,
standard isolation involves many more partonic processes. In particular, the 
single-fragmentation component receives contribution from 
the partonic processes 
\beq
\label{qqbartchanfrag}
q{\bar q} \to q{\bar q} \gamma \quad \quad 
(q \to \gamma + X \;\;{\rm or} \;\;{\bar q} \to \gamma + X)
\eeq
and
\beq
\label{qqtchanfrag}
q q \to q q \gamma \quad \quad (q \to \gamma + X) \;\;,
\eeq
in which the photon and
a final-state fermion can have a small relative angle and  the other final-state
fermion fragments into a second photon. 
At large values of 
$|\!\cos \theta^{*}|$ (which roughly correspond to large rapidity separations
$|\dy|$), these two processes are dominated by the effect of the
exchange of one gluon in the $t$-channel of the $2 \to 2$ fermion scattering
subprocess: the gluon exchange leads to an angular distribution that is
proportional to $(1-\cos^2 \theta^{*} )^{-2}$. 
Therefore, the gluon exchange subprocesses are
dynamically enhanced by the relative factor $(1-\cos^2 \theta^{*} )^{-1}$
with respect to the fermion exchange subprocesses (see, e.g., Eqs.~(\ref{cosdir}) 
and (\ref{cosfrag})) that contribute to the NLO calculation for smooth isolation.
Although the gluon exchange processes are suppressed by the isolation requirements,
their effect on the NLO results for standard isolation is not negligible at large
values of $|\!\cos \theta^{*}|$. 
We also note that corresponding gluon exchange
processes enter the calculation for smooth isolation at the NNLO. Such processes
are
\beq
\label{qqbartchansmooth}
q{\bar q} \to q{\bar q} \gamma \gamma
\eeq
and
\beq
\label{qqtchansmooth}
q q \to q q \gamma \gamma \;\;,
\eeq
in which the two photons are produced at large
rapidity separation and each photon is at small relative angle 
with respect to one final-state fermion.
These
processes can enhance the size of the NNLO correction within smooth isolation at
large values of $|\!\cos \theta^{*}|$.

We briefly comment on the NLO results for $d\sigma/d\cos \theta^{*}$ in 
Fig.~\ref{fig:dcosnew}, which are obtained by applying the additional kinematical
cut 200~GeV$ < M_{\gamma \gamma} < 800$~GeV. For the sake of simplicity, in 
Fig.~\ref{fig:dcosnew} we present LO and NLO results without considering scale
variation dependence and we simply use $\mu_R=M_{\gamma \gamma}/2$ and
$\mu_F=\mu_{frag}= 2M_{\gamma \gamma}$. As already remarked in our discussion of
the LO results, the constraint $M_{\gamma \gamma} > 200$~GeV has a major effect on
the shape of $d\sigma/d\cos \theta^{*}$, which qualitatively follows the shape of
the angular distribution of the underlying partonic processes. This behaviour
is not affected by including NLO corrections. Comparing the results in 
Fig.~\ref{fig:dcos}-right and \ref{fig:dcosnew}, we see that the high-mass
constraint $M_{\gamma \gamma} > 200$~GeV sizeably (and obviously) reduces the
values of the cross sections and it also modifies the size of the NLO corrections. 
The NLO results for smooth and standard isolation are very similar also at large
values of $|\!\cos \theta^{*}|$ (e.g., $|\!\cos \theta^{*}| \sim 0.9$). At central
values of $|\!\cos \theta^{*}|$, the NLO result for standard isolation is
(slightly) higher than the NLO result for smooth isolation: this behaviour is
different from that in the NLO results of Fig.~\ref{fig:dcos} and it is consistent
with our discussion on the relevance of the region of 
intermediate values of $M_{\gamma \gamma}$ for the behaviour of the NLO
results in Fig.~\ref{fig:dcos}.
 
As shown in Table~\ref{Table:Isol1} and Figs.~\ref{fig:Xtot}--\ref{fig:dcos},
standard and smooth isolations lead to QCD results in good quantitative
agreement for physical observables that are effectively computed up to the NLO.
In view of this agreement we present some additional investigations on the role
of the fragmentation component in the QCD computation for the standard isolation
criterion. A perturbative scheme (approximation) that is sometimes used in
photon isolation computations at the NLO
(see, e.g., Refs.~\cite{Campbell:2011bn, Gehrmann:2013aga, Campbell:2014yka})
consists in combining the evaluation of the direct component at the NLO with
that of the fragmentation component at the LO. This has the practical advantage
of avoiding the computation of the more cumbersome photon fragmentation
subprocesses at the NLO. Within this context, one can `equivalently' 
\cite{Campbell:2014yka}
use LO and NLO parton-to-photon fragmentation functions.
Such a scheme is applied in the diphoton production 
calculation of Ref.~\cite{Campbell:2011bn},
which we use (as implemented in the 
\texttt{MCFM} program) in our subsequent numerical investigation 
(we do not use the \texttt{DIPHOX} program, by removing the NLO corrections to
the fragmentation component, because LO fragmentation functions are not readily
available in the default setup of the \texttt{DIPHOX} code).

\begin{figure}[htb]
\begin{center}
\begin{tabular}{cc}
\hspace*{-4.5mm}
\includegraphics[width=0.48\textwidth]{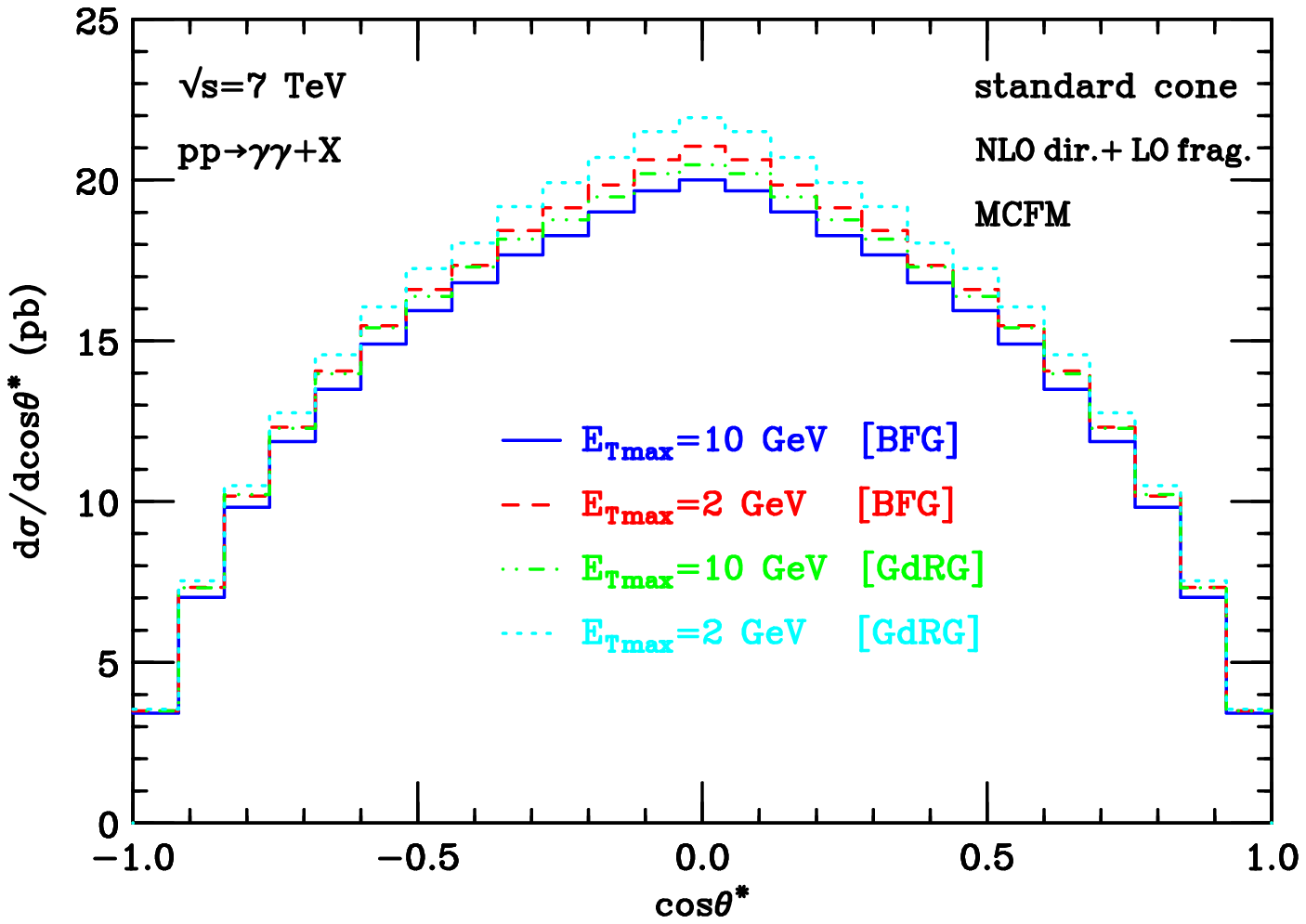}
& \includegraphics[width=0.48\textwidth]{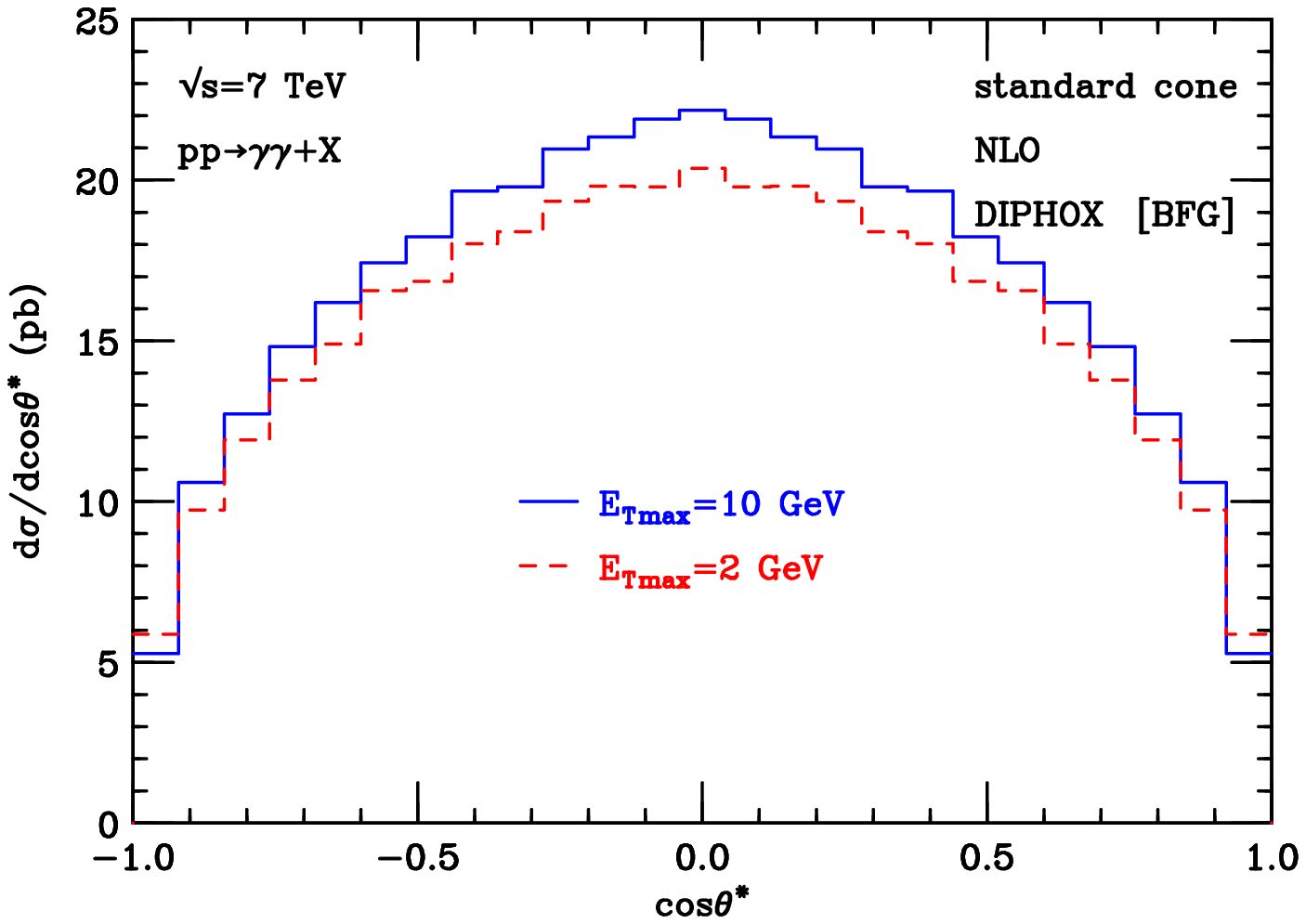}
\\
\end{tabular}
\end{center}
\caption{\label{fig:frag_cos}
{\em The $\cos~\theta^{*}$  differential cross section for standard cone
isolation with two different values of  $\ETmax$  (2~GeV and 10~GeV) and the same photon
kinematical cuts as in Fig.~\ref{fig:dcos}.
The QCD results are obtained at the central value of the scales
($\mu_F=\mu_R=\mu_{frag}=\mu_0 \equiv \mgg$). The results with NLO direct + LO
fragmentation components (left panel) use BFG and GdRG\_LO fragmentation
functions. The NLO results (right panel) use BFG fragmentation functions.
}}
\end{figure}

We consider standard cone isolation and the photon kinematical cuts used in 
Table~\ref{Table:Isol1} and Figs.~\ref{fig:Xtot}--\ref{fig:dcos}.
The QCD calculation that includes NLO~direct+LO~fragmentation
components is carried out by using the \texttt{MCFM} program
\cite{Campbell:2011bn}, and we use either the NLO fragmentation functions of the 
BFG set II \cite{Bourhis:1997yu} (as in our customary NLO calculations)
or the LO fragmentation functions of
the GdRG\_LO set \cite{GehrmannDeRidder:1998ba}.
At the central value of the scales and using BFG fragmentation functions,
we obtain the following values of total cross sections:  
$\sigma^{MCFM}_{\rm standard}=29.66$~pb with $\ETmax=2$~GeV, and
$\sigma^{MCFM}_{\rm standard}=28.46$~pb with $\ETmax=10$~GeV
(the GdRG\_LO set of fragmentation functions leads to results that are larger by
approximately 1~pb). The corresponding results for
$d\sigma/d\cos \theta^{*}$ are presented in Fig.~\ref{fig:frag_cos}-left.
We note that these results are similar and consistent with the complete NLO
results, within the corresponding scale variation dependence. This implies that
the NLO corrections to the fragmentation component are not particularly
sizeable. However, we also note that $\sigma^{MCFM}_{\rm standard}$ and
$\sigma^{NLO}_{\rm standard}$ differ by about 15\% at $\ETmax=10$~GeV and,
moreover, $\sigma^{MCFM}_{\rm standard}$ decreases by increasing the value of
$\ETmax$ from 2~GeV to 10~GeV (we have checked that the value of 
$\sigma^{MCFM}_{\rm standard}$ at $\ETmax=4$~GeV is intermediate between the
values at $\ETmax=2$~GeV and $\ETmax=10$~GeV). This $\ETmax$
dependence of the MCFM results, which occurs for both the total cross section
and $d\sigma/d\cos \theta^{*}$ at fixed values of $\cos \theta^{*}$
(see Fig.~\ref{fig:frag_cos}-left), violates the expected physical behaviour
(see Eq.~(\ref{eq:b})). The use of BFG or GdRG\_LO fragmentation functions
does not change the qualitative dependence on $\ETmax$.
In contrast, the complete NLO result for standard isolation (see 
Table~\ref{Table:Isol1}, Fig.~\ref{fig:dcos} and Fig.~\ref{fig:frag_cos}-right)
and also the NLO result for smooth isolation (see Table~\ref{Table:Isol1} and
Fig.~\ref{fig:dcos}) have the expected physical dependence on $\ETmax$
(total and differential cross sections increase by increasing $\ETmax$).

We interpret the unphysical dependence on $\ETmax$ of the MCFM results as the
effect of a mismatch between the perturbative orders in the direct and
fragmentation components. Indeed the $\ETmax$ dependence of the MCFM results
is mostly produced by the NLO corrections to the direct component through the
partonic subprocess $qg \to q \gamma \gamma$ at the tree level.
The tree-level contribution of this partonic subprocess is formally positive
definite and, as such, it would lead to an increasing (physical) dependence on 
$\ETmax$ by increasing $\ETmax$. However, this contribution is divergent
in the phase space region where the final-state $q$ is collinear to one of the
produced photons. The collinear divergence (which is in turn absorbed and
factorized in the non-perturbative definition of the quark-to-photon
fragmentation function) is removed through its subtraction (which is actually
performed in the ${\overline {\rm MS}}$ factorization scheme 
\cite{Campbell:2011bn}) from the direct component contribution at NLO.
Since the collinear-divergent term that is subtracted is formally positive definite, the
final NLO correction to the direct component is finite but it is not
necessarily positive definite and, consequently, it can have an unphysical
dependence on $\ETmax$ (this unphysical dependence is actually also visible
in the direct component of the NLO results in Table~\ref{Table:Isol1}). 
It turns out that the result of the calculation with NLO direct + LO
fragmentation components has an unphysical dependence on $\ETmax$.

In the case of the complete NLO result for standard
isolation, the NLO corrections to the fragmentation component 
(corrections to both the partonic cross sections and the scale evolution of the
fragmentation functions) are consistently (at the formal level) included within
the ${\overline {\rm MS}}$ factorization scheme. Although these corrections are
not particularly sizeable in absolute terms, their inclusion leads to NLO
results with a qualitative dependence on $\ETmax$ that agrees with physical
expectation. 

In the context of smooth isolation, the entire $\ETmax$
dependence at NLO is due to partonic subprocesses with two photons and a parton
in the final state, which are evaluated at the tree level. The contribution of
these tree-level subprocesses is  positive definite (with no collinear
divergences to be subtracted) and it cannot produce any unphysical dependence on
$\ETmax$: NLO smooth isolation cross sections cannot decrease by increasing 
$\ETmax$. Incidentally, we note that this elementary reasoning, based on
positivity, cannot be applied at the NNLO, since the $\ETmax$ dependence at
NNLO receives contributions from both tree-level and loop-level partonic
processes. Nonetheless, as shown in Sect.~\ref{sec:ispar}
(see Fig.~\ref{fig:nnEtdep} and the accompanying comments) 
the NNLO results for smooth isolation do not
show unphysical dependence on $\ETmax$.

\begin{figure}[htb]
\begin{center}
\begin{tabular}{cc}
\includegraphics[width=0.485\textwidth]{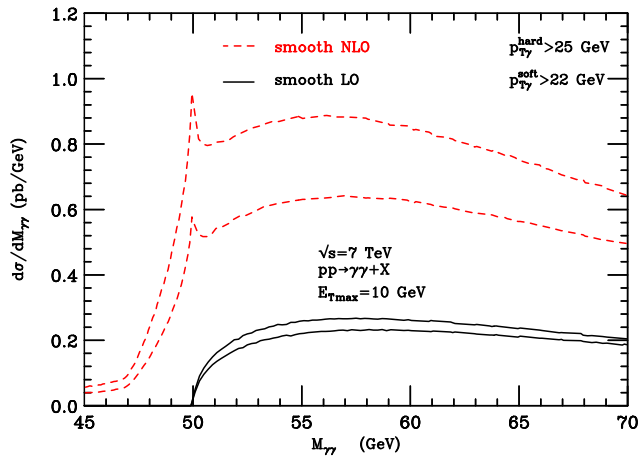}
\\
\end{tabular}
\end{center}
\caption{\label{fig:dmth}
{\em  The differential cross section
$d\sigma/dM_{\gamma \gamma}$ for 
smooth isolation with $\ETmax=10$~GeV and the same kinematical cuts as in
Fig.~\ref{fig:dm}. The LO (black solid) and NLO (red dashed) numerical results
use $M_{\gamma \gamma}$ bins with constant size of 0.1~GeV. At both perturbative
orders, the maximum and minimum values of $d\sigma/dM_{\gamma \gamma}$ 
correspond to the scale choices 
$\{ \mu_R= M_{\gamma \gamma}/2, \mu_F=2 M_{\gamma \gamma} \}$ and
$\{ \mu_R= 2 M_{\gamma \gamma}, \mu_F=M_{\gamma \gamma}/2 \}$, 
respectively. 
}}
\end{figure}

As anticipated in our comments on the results in Fig.~\ref{fig:dm}, we present a
more detailed discussion on the differential cross section 
$d\sigma/dM_{\gamma \gamma}$ in the region where $M_{\gamma \gamma}$ is close to
the LO threshold ($M_{\gamma \gamma} \sim M^{LO}$). The numerical results in 
Fig.~\ref{fig:dm} are obtained by using a (constant) bin size of 2~GeV in 
$M_{\gamma \gamma}$. To quantitatively examine the detailed shape of 
$d\sigma/dM_{\gamma \gamma}$, we perform the numerical calculation with a finer
resolution in $M_{\gamma \gamma}$ and we use $M_{\gamma \gamma}$ bins with a
constant size of 0.1~GeV (which is 20 times smaller than that used in 
Fig.~\ref{fig:dm}). The results of the calculation for smooth isolation 
at both LO and NLO are presented in Fig.~\ref{fig:dmth}. Specifically, we use 
$\ETmax=10$~GeV and, at each perturbative order, we report the two results
with the scale choices 
$\{ \mu_R= M_{\gamma \gamma}/2, \mu_F=2 M_{\gamma \gamma} \}$ and
$\{ \mu_R= 2 M_{\gamma \gamma}, \mu_F=M_{\gamma \gamma}/2 \}$
(the region enclosed by these two scale-dependent results corresponds to the
scale variation band in Fig.~\ref{fig:dm}).
We remark that the smooth isolation results in Fig.~\ref{fig:dmth} exactly refer
to the same quantity (and to the same theoretical setup) as the corresponding
results in Fig.~\ref{fig:dm}-right, the only difference being the much smaller
$M_{\gamma \gamma}$ bin size used in Fig.~\ref{fig:dmth}. Various shape details
that are visible in Fig.~\ref{fig:dmth} disappear in Fig.~\ref{fig:dm} since
they are smeared by the larger bin size.

We note that the qualitative shape of  $d\sigma/dM_{\gamma \gamma}$ in 
Fig.~\ref{fig:dmth} is independent of the scale choice, which only affects the
size of $d\sigma/dM_{\gamma \gamma}$. The behaviour of the LO result has been
discussed in the paragraph below Eq.~(\ref{ptcut}). The LO result has a
threshold at $M_{\gamma \gamma}=M^{LO}_{\rm dir}=50$~GeV and, close to the
threshold, it vanishes as
\beq
\label{loth}
\frac{d\sigma^{LO}_{\rm smooth}}{dM_{\gamma \gamma}}
\propto {\sqrt {\epsilon_M}} \;\; \Theta(M_{\gamma \gamma} - M^{LO}_{\rm dir})
\;\;, \quad \quad \quad (M_{\gamma \gamma} \sim M^{LO}_{\rm dir}) \;,
\eeq
where $\epsilon_M = | 1 - (M_{\gamma \gamma}/M^{LO}_{\rm dir})|$.
The threshold and the square-root behaviours near threshold are visible in the LO
results of Fig.~\ref{fig:dmth}. At the NLO, $d\sigma/dM_{\gamma \gamma}$
is not vanishing at $M_{\gamma \gamma}=M^{LO}_{\rm dir}$ (the threshold
disappears) and it has a {\em cusp} behaviour in the vicinity of the LO 
threshold. 
Following the general analysis in Ref.~\cite{Catani:1997xc}, 
we have examined the NLO shape of 
$d\sigma/dM_{\gamma \gamma}$ at $M_{\gamma \gamma} \sim M^{LO}_{\rm dir}$
in analytic form
and we find the 
dominant behaviour
\beq
\label{nloth}
\frac{d\sigma^{NLO}_{\rm smooth}}{dM_{\gamma \gamma}} = a_0 - 
{\sqrt {\epsilon_M}} \left[ a_{(-)} \,\ln\left( \frac{1}{\epsilon_M}\right) 
\Theta(M^{LO}_{\rm dir} - M_{\gamma \gamma})
+ a_{(+)} \,\ln^2\left( \frac{1}{\epsilon_M}\right)
\Theta(M_{\gamma \gamma} - M^{LO}_{\rm dir}) \right] + \;\; \dots
\;\;, 
\eeq
where $a_0, a_{(-)}$ and $a_{(+)}$ are positive constants (i.e., they do not
depend on $M_{\gamma \gamma}$) and the dots on the right-hand side denote
subdominant contributions (terms that are relatively suppressed by powers of 
$(\ln \epsilon_M)^{-1}$ or ${\sqrt {\epsilon_M}}\,$) in the limit 
$M_{\gamma \gamma} \to M^{LO}_{\rm dir}$.
In particular, Eq.~(\ref{nloth}) implies that the first derivative of 
$d\sigma/dM_{\gamma \gamma}$ with respect to $M_{\gamma \gamma}$
(i.e., the slope of $d\sigma/dM_{\gamma \gamma}$) diverges to 
$+\infty$
($-\infty$) if $M_{\gamma \gamma} \to M^{LO}_{\rm dir}$ from below
(above) the LO threshold. This slope leads to the {\em double-side} cusp
of Fig.~\ref{fig:dmth}.

At low values of $M_{\gamma \gamma}$ in Fig.~\ref{fig:dmth}, 
$d\sigma^{NLO}/dM_{\gamma \gamma}$ is very small. Its value starts to rapidly 
increase
at $M_{\gamma \gamma} \sim 47$~GeV. This confirms the observation in our 
previous comments on the NLO result in Figs.~\ref{fig:dm} and \ref{fig:dmnew}:
although $d\sigma/dM_{\gamma \gamma}$ has no physical threshold, it displays an
`approximate' threshold behaviour at 
$M_{\gamma \gamma} \sim M^{LO} \simeq 47$~GeV (see Eq.~(\ref{lothres}))
since the kinematical region where $M_{\gamma \gamma} \ltap M^{LO}$ receives
contributions only from physical events in which the diphoton pair is
accompanied by hard (high transverse momentum) parton radiation in the final
state (see Eq.~(\ref{lowm}) and related comments).
At high values of $M_{\gamma \gamma}$ in Fig.~\ref{fig:dmth}, 
the shape of the LO and NLO results is quite similar. In particular, the
position of the (broad) peak of $d\sigma/dM_{\gamma \gamma}$ (at 
$M_{\gamma \gamma} \sim 57$~GeV) does not substantially vary in going from the
LO to the NLO results.

Considering the NLO result with the scale configuration
$\{ \mu_R= M_{\gamma \gamma}/2, \mu_F=2 M_{\gamma \gamma} \}$
in Fig.~\ref{fig:dmth}, we also notice
that the value of $d\sigma/dM_{\gamma \gamma}$ at 
$M_{\gamma \gamma} = M^{LO}_{\rm dir}$ (i.e., the height of the cusp) is
quite large and, in particular, it is larger than the value in the peak region at 
$M_{\gamma \gamma} \sim 57$~GeV. Independently of the scale configuration, we
observe that the cusp behaviour is located in a tiny region of 
$M_{\gamma \gamma}$ around $M_{\gamma \gamma}=M^{LO}_{\rm dir}$ and,
consequently, such behaviour quantitatively disappears by increasing 
the bin size in $M_{\gamma \gamma}$ (see Fig.~\ref{fig:dm}).

The shape of $d\sigma^{NLO}/dM_{\gamma \gamma}$ at 
$M_{\gamma \gamma} \sim M^{LO}_{\rm dir}$ 
(Fig.~\ref{fig:dmth} and Eq.~(\ref{nloth})) is definitely unphysical and it
deserves additional comments. This shape (and the expression 
in Eq.~(\ref{nloth})) follows from the general discussion and results of
Ref.~\cite{Catani:1997xc}. According to Ref.~\cite{Catani:1997xc}, an observable
that has a discontinuity at some point $x=x_0$ (inside the physical region) at
the LO necessarily has a logarithmic divergent (though integrable) discontinuity
at the same point at the NLO. In our specific case, the slope of 
$d\sigma^{LO}/dM_{\gamma \gamma}$ has a discontinuity at 
$M_{\gamma \gamma}=M^{LO}_{\rm dir}$ (though $d\sigma^{LO}/dM_{\gamma \gamma}$
is continuous at the same point): therefore, the slope of 
$d\sigma^{NLO}/dM_{\gamma \gamma}$ has a logarithmically-enhanced discontinuity
(see Eqs.~(\ref{loth}) and (\ref{nloth})) at $M_{\gamma \gamma}=M^{LO}_{\rm dir}$
(though $d\sigma^{NLO}/dM_{\gamma \gamma}$ remains continuous at 
the same point). 

The logarithmic enhancement of the discontinuity is due to
soft-gluon radiation at the NLO \cite{Catani:1997xc}.
In our specific case, the relevant NLO partonic processes are the real emission
contribution 
\beq
\label{softtree}
q {\bar q} \to \gamma \gamma g
\eeq
at the tree level (here, the final-state gluon is soft and collinear to one of
the initial-state partons), and the virtual contribution to 
$q {\bar q} \to \gamma \gamma$ at the one-loop level.
A non-smooth kinematical mismatch (such as that produced by a discontinuous
observable) between the real and virtual contributions produces the 
logarithmic enhancement \cite{Catani:1997xc}.
In our specific case, the NLO virtual process $q {\bar q} \to \gamma \gamma$
still fulfils the same kinematical constraints as at the LO (in particular,
transverse-momentum conservation implies  
$p_{T \gamma}^{soft} = p_{T \gamma}^{hard} \geq p_H=25$~GeV,
independently of the value of $p_S$) and, consequently, it only contributes at
$M_{\gamma \gamma} \geq M^{LO}_{\rm dir}$.
In contrast, the soft gluon radiated in the process of Eq.~(\ref{softtree})
produces diphoton events with $M_{\gamma \gamma} < M^{LO}_{\rm dir}$, since the
softer photon absorbs the soft-gluon momentum recoil\footnote{Note that such
recoil is forbidden in the case of symmetric $p_T$ cuts with $p_H=p_S$
(see a related discussion in Sect.~\ref{sec:ptcut}).} 
thus decreasing its
transverse momentum $p_{T \gamma}^{soft}$ below its LO kinematical limit (i.e.,
we have $p_{T \gamma}^{soft} < p_H \leq p_{T \gamma}^{hard}$, although 
$p_{T \gamma}^{soft} \geq p_S$).
It follows that real soft-gluon radiation is
completely unbalanced by virtual radiation in the region just below the LO
threshold, and this produces the corresponding logarithmically-enhanced cusp at 
$M_{\gamma \gamma} < M^{LO}_{\rm dir}$ in 
Fig.~\ref{fig:dmth} and Eq.~(\ref{nloth}).
Therefore, the shape of $d\sigma^{NLO}/dM_{\gamma \gamma}$ just below the LO
threshold is exactly a consequence of the reasoning and results of 
Ref.~\cite{Catani:1997xc}. Instead, the logarithmic enhancement of the slope 
of $d\sigma^{NLO}/dM_{\gamma \gamma}$ just above the LO threshold 
($M_{\gamma \gamma} > M^{LO}_{\rm dir}$) can be somehow regarded as a {\em
corollary} to the reasoning in Ref.~\cite{Catani:1997xc}. Both real and virtual
terms contribute above the threshold, but here the slope of 
$d\sigma/dM_{\gamma \gamma}$ is already divergent (see Eq.~(\ref{loth})),
and not only discontinuous, at the LO: this LO divergent behaviour produces a
strong real--virtual mismatch and an ensuing logarithmic enhancement at the NLO
(see Fig.~\ref{fig:dmth} and Eq.~(\ref{nloth})).

The unphysical cusp behaviour of $d\sigma^{NLO}/dM_{\gamma \gamma}$
at $M_{\gamma \gamma} \sim M^{LO}_{\rm dir}$ occurs for {\em both} smooth and
standard isolation (although Fig.~\ref{fig:dmth} only shows smooth isolation
results). Indeed, the LO direct component of the standard isolation result
exactly coincides with the LO smooth isolation result and, therefore, the NLO
soft-gluon radiative correction to the direct component for standard isolation
exactly behaves in the same way as we have just described for the smooth isolation
criterion. In contrast, the NLO radiative corrections to the fragmentation
component do not produce a cusp behaviour in the vicinity of the corresponding LO
threshold at $M_{\gamma \gamma} \sim M^{LO}_{\rm frag}$
($M^{LO}_{\rm frag}=M^{LO} \simeq 47$~GeV in Fig.~\ref{fig:dmth}).
This follows from the
fact that $d\sigma^{LO}_{\rm frag}/dM_{\gamma \gamma}$ is sufficiently smooth at
$M_{\gamma \gamma} \sim M^{LO}_{\rm frag}$ (see the discussion at the end of
Sect.~\ref{sec:diffLO}) and, in particular, the slope of 
$d\sigma^{LO}_{\rm frag}/dM_{\gamma \gamma}$ is not discontinuous (it actually
vanishes) at $M_{\gamma \gamma} = M^{LO}_{\rm frag}$. Obviously, the NLO
fragmentation component contributes to $d\sigma^{NLO}/dM_{\gamma \gamma}$
at $M_{\gamma \gamma} \sim M^{LO}_{\rm dir}$, but its contribution is smooth
and it only produces a finite vertical displacement of the cusp of 
the NLO direct component.

The unphysical shape (i.e., the cusp behaviour) of $d\sigma/dM_{\gamma \gamma}$
at NLO persists at each subsequent perturbative order. Such an 
unphysical fixed-order behaviour can be removed by a proper all-order resummation
of soft-gluon effects \cite{Catani:1997xc}. Resummation leads to a smooth behavior
(Sudakov shoulder) \cite{Catani:1997xc}
of both $d\sigma/dM_{\gamma \gamma}$ and its slope from the peak region at
$M_{\gamma \gamma} \sim 57$~GeV to the `approximate' threshold at 
$M_{\gamma \gamma} \sim 47$~GeV. 

In the context of fixed-order computations,
the unphysical behaviour produces perturbative instabilities that can be reduced
only by considering observables that are sensitive to a `sufficiently-smeared'
region in the vicinity of $M_{\gamma \gamma} \sim M^{LO}_{\rm dir}$. The degree
of `sufficient' smearing or insensitivity depends on various factors, such as the
type of observable, the bin size and the kinematical cuts. We briefly comments on
these factors.
As for the dependence on the observable, the total cross section has little
sensitivity to these instabilities, the differential cross section 
$d\sigma/d\cos \theta^{*}$ at central values of $cos \theta^{*}$ has some
sensitivity, and the differential cross section $d\sigma/dM_{\gamma \gamma}$
in the region close to $M^{LO}_{\rm dir}$ is certainly sensitive. The bin size
dependence is obvious at the qualitative level, but it is less obvious at the
quantitative level. A sufficiently-large $M_{\gamma \gamma}$-bin at 
$M_{\gamma \gamma} \sim M^{LO}_{\rm dir}$ removes the cusp behaviour, but it leads
to a binned value of $d\sigma/dM_{\gamma \gamma}$ that depends on the bin size and
also on the average slope of $d\sigma/dM_{\gamma \gamma}$ in the region close to
the LO threshold (a larger value of the average slope increases the sensitivity to the bin
size). In our discussion on the results in Fig.~\ref{fig:dm}, we have argued that 
$d\sigma^{NLO}/dM_{\gamma \gamma}$ is possibly sensitive to the perturbative
instabilities even if the $M_{\gamma \gamma}$-bin size is 2~GeV.
The dependence on the $p_T$ cuts has been pointed out in our comments on the
results in Fig.~\ref{fig:dmnew}, where we have observed that the size of the
perturbative instabilities can be affected by the average slope of 
$d\sigma/dM_{\gamma \gamma}$ at $M_{\gamma \gamma} \sim M^{LO}_{\rm dir}$,
which directly depends on the difference between $M^{LO}_{\rm dir}$ and
$M^{LO}$ (i.e., on the values of the $p_T$ cuts $p_H$ and $p_S$).
Certainly, the relevance of these perturbative instabilities also depends on the
required theoretical accuracy of the QCD calculation.
In Sect.~\ref{sec:ptcut}
we present additional results and comments on this kind \cite{Catani:1997xc} of 
instabilities and related observables.

\setcounter{footnote}{2}

\section{Diphoton production at the LHC and NNLO results}
\label{sec:results}

In this section we consider diphoton production in $pp$ collisions at LHC
energies and we present perturbative QCD results at the NNLO. 
We use smooth cone isolation since the NNLO calculation for standard cone
isolation has not yet been performed.
Within smooth isolation, we also present corresponding results at LO and NLO to
directly comment on features of the perturbative QCD expansion.

In Ref.~\cite{Catani:2011qz}
we presented NNLO results for diphoton production at the LHC in diphoton
kinematical configurations that are typically used in the context of Higgs boson
searches and studies. In particular, we considered the region 
$\mgg < 200$~GeV and highly-asymmetric cuts on the photon transverse momenta
(i.e., 
$p_T^{hard} > 40$~GeV, $p_T^{soft} > 25$~GeV). 
In the following we consider different kinematical configurations and we discuss
various aspects of the NNLO results. Diphoton production results at NNLO
and comparisons with LHC data are also presented in 
Refs.~\cite{Aad:2012tba,Chatrchyan:2014fsa,Aaboud:2017vol,Campbell:2016yrh}.
NNLO results for diphoton production in $pp$ collisions at ${\sqrt s}=100$~TeV
are presented in Ref.~\cite{Mangano:2016jyj} (see Sect.~8.3 therein).

In our computation
the radius of the photon isolation cone is set at the value
$R=0.4$. We use
the smooth isolation function $\chi(r;R)$ in Eq.~(\ref{Eq:Isol_chinormal}) 
(the same form of the isolation function is used in the NNLO results reported in
Refs.~\cite{Aad:2012tba,Chatrchyan:2014fsa,Aaboud:2017vol})
and the
value of the power $n$ is set to $n=1$ 
for most of the results, although we comment on the $n$ dependence of 
total cross sections and of some differential cross sections.

The QCD results are obtained by using the programs
\texttt{2$\gamma$NNLO} and \texttt{MATRIX}.
The theoretical setup of our calculation is the same
as described at the beginning of Sect.~\ref{subsec:res}
(in particular, we use the value $\alpha=1/137$ of the QED coupling constant and
the MMHT 2014 sets \cite{Harland-Lang:2014zoa} of PDFs).
The only difference with respect to Sect.~\ref{subsec:res}
regards the central value $\mu_0$ of the renormalization ($\mu_R$) and factorization 
($\mu_F$) scales. Unlike the case of Sect.~\ref{subsec:res}
(where $\mu_0=\mgg$), throughout this section we use the {\em dynamical}
central value 
$\mu_0 = \sqrt{{\mgg}^2+p^2_{T\gamma \gamma}}=M_{T\gamma \gamma}$
($M_{T\gamma \gamma}$ 
is the transverse mass of the
diphoton system),
which also depends on the transverse
momentum $p_{T\gamma \gamma}$ of the photon pair.
We consider independent scale variations of $\mu_R$ and $\mu_F$ within the ranges
$0.5 \leq \mu_R/\mu_0 \leq 2$ and $0.5 \leq \mu_F/\mu_0 \leq 2$
around the central value  $\mu_0$. Practically, we obtain the results for nine
scale configurations (we independently combine $\mu_R/\mu_0= \{ 0.5,1,2\}$
and $\mu_F/\mu_0= \{ 0.5,1,2\}$) and we evaluate scale uncertainties by
considering the maximum value and minimum value among these results.
We have checked that, for most
of the computed quantities (including total cross
sections), the maximum and minimum values correspond to the scale configurations
$\{\mu_R=\mu_0/2$, $\mu_F=2\mu_0\}$ and $\{\mu_R=2\mu_0$ , $\mu_F=\mu_0/2\}$,
respectively.

The bulk of the diphoton cross section is produced at small values of
$p_{T\gamma \gamma}$ ($p_{T\gamma \gamma} \ll \mgg$). Therefore, to the purpose of computing 
the total
cross section,  the choice of the dynamical scale 
$\mu_0 = M_{T\gamma \gamma}$
leads to results that are basically
similar to those obtained with the scale choice $\mu_0 \sim \mgg$.
The dynamical scale 
$\mu_0 = M_{T\gamma \gamma}$
sizeably differs
from the diphoton invariant mass $\mgg$ only at high values of 
$p_{T\gamma \gamma}$ (i.e., in the kinematical regions where 
$\mgg \ll p_{T\gamma \gamma}$).
High values of $p_{T\gamma \gamma}$ can be reached either in the highly unbalanced regime
where $p_{T \gamma}^{soft} \ll p_{T \gamma}^{hard}$ or in the highly boosted
regime where the two photons have comparable transverse momenta
($p_{T \gamma}^{soft} \sim p_{T \gamma}^{hard}$)  and small values of both the
azimuthal angle separation $\Delta \Phi_{\gamma \gamma}$ and the rapidity
separation $\Delta y_{\gamma \gamma}$.
In both these regimes, $p_{T\gamma \gamma}$ is balanced by a recoiling 
high-$p_T$ jet,
and the dynamical scale 
$\mu_0 = M_{T\gamma \gamma} \sim p_{T\gamma \gamma}$ 
parametrically mimics the scale
of the invariant mass $M_{\gamma \gamma jet}$
($ p_{T\gamma \gamma} \sim {\cal O}(M_{\gamma \gamma jet})$) of the diphoton+jet
final-state system, which is the characteristic scale of the underlying
hard-scattering subprocesses.

In this paper we do not evaluate PDF uncertainties. NLO PDF uncertainties at the
LHC are computed in the diphoton studies of
Refs.~\cite{Aad:2012tba,Chatrchyan:2014fsa,Aaboud:2017vol}
and combined with scale variation uncertainties (PDF uncertainties are found to
be typically smaller than scale uncertainties). 
In Ref.~\cite{Chatrchyan:2014fsa}
the PDF uncertainty on the NLO total cross section is explicitly quoted and it
amounts to about $\pm 5$\%. The PDF uncertainties that are 
computed
in Ref.~\cite{Aaboud:2017vol} are at the level of $\pm 2$\%.

As main reference kinematical configuration we consider
LHC collisions at the centre--of--mass
energy $\sqrt{s}=7$~TeV and we use the kinematical acceptance cuts 
implemented by the ATLAS Collaboration in the analysis of
Ref.~\cite{Aad:2012tba}.
We require $p_T^{hard} \geq 25$~GeV and $p_T^{soft}\geq 22$~GeV, 
we restrict the rapidity of both photons to the regions
$|y_\gamma|<1.37$ 
and \mbox{$1.52<|y_\gamma| \leq 2.37$},
and the minimum angular separation between the two 
photons is $R_{\gamma \gamma}^{\rm min}=0.4$. 
These acceptance cuts coincide with the main reference cuts in 
Sect.~\ref{subsec:res}, apart from the exclusion of the rapidity interval
$1.37 < |y_\gamma|< 1.52$.
The ATLAS data are selected by using standard cone
isolation with the isolation parameters $R=0.4$ and $\ETmax=4~$GeV.
We use smooth cone isolation with $R=0.4$ and $\ETmax=4~$GeV.
All the results in this section refer to this configuration and to this value of
$\ETmax$,
unless otherwise explicitly stated.

\subsection{Total cross sections}
\label{sec:totalNNLO}

The results of the LO, NLO and NNLO total cross sections are reported 
in Table~\ref{Table:total}.

At the LO the value of the total cross section is
$\sigma^{LO}= 9.293~{\rm pb}~^{+10.9\, \percent }_{-11.9\,\percent }$~(scale),
and it does not depend on the power $n$ of the isolation function in
Eq.~(\ref{Eq:Isol_chinormal}). At the NLO with $n=1$ we have
$\sigma^{NLO}= 28.55~{\rm pb}~^{+12.5\, \percent }_{-12.2\,\percent }$~(scale).
The features of these LO and NLO results are very similar to those of the smooth
isolation results presented in Sect.~\ref{sec:sigmatot} 
(see Table~\ref{Table:Isol1}),
apart from an overall reduction (by roughly 10\%) of the cross section values,
which is due to the exclusion of the photon rapidity region
$1.37 < |y_\gamma|< 1.52\,$.

\begin{table}

\begin{center}

\renewcommand{\arraystretch}{1.5}

\begin{tabular}{ |c||c|c|c| }

\hline


\cline{2-4}

& $\sigma^{{\rm LO}}$~~(pb) & $\sigma^{{\rm NLO}}$~~(pb) & $\sigma^{{\rm
NNLO}}$~~(pb)    \\\hline

\multirow{1}{*}{$n$~{\rm ind.}} & $9.293~^{+10.9\, \percent }_{-11.9\,
\percent }$

 & & \\

\multirow{1}{*}{$n=0.5$} &   
& $29.40~^{+12.8\, \percent }_{-12.4\,
\percent }$ & $40.98(68)~^{+8.3\, \percent }_{-8.7\, \percent }$ \\

\multirow{1}{*}{$n=1$} &  
& $28.55~^{+12.5\, \percent }_{-12.2\,
\percent }$ & $39.50(50)~^{+7.9\, \percent }_{-8.4\, \percent }$ \\

\multirow{1}{*}{$n=2$} &  
& $27.98~^{+12.3\, \percent }_{-11.9\,
\percent }$ & $37.53(52)~^{+7.0\, \percent }_{-7.8\, \percent }$  \\

\hline

\end{tabular}


\caption{\label{Table:total}
{\em Results for LO, NLO and NNLO total cross sections with
the photon kinematical cuts described in the text
(beginning of Sect.~\ref{sec:results}). The results are obtained by using smooth
isolation and $n$ denotes the power in the isolation function of 
Eq.~(\ref{Eq:Isol_chinormal}).
The central values of $\sigma$ are obtained with the
scale choice $\mu_R=\mu_F=\mu_0=M_{T\gamma \gamma}$, 
and the scale dependence corresponds to independent variations of $\mu_R$ and
$\mu_F$ between $\mu_0/2$ and $2\mu_0$.
The last digit
of the LO and NLO cross section values has an error of one unit
from the statistical uncertainty of the numerical calculation.
The systematical error on $\sigma^{NNLO}$ is explicitly given in round brackets.
}}
\renewcommand{\arraystretch}{1}

\end{center}

\end{table}

The value of the NNLO total cross section with $n=1$ is
$\sigma^{NNLO}= 39.50~{\rm pb}~^{+7.0\, \percent }_{-7.8\, \percent }$~(scale).
This result is obtained  by using \texttt{MATRIX} with the default setup~\cite{matrix}.
The value of the NNLO cross section
is obtained by the code through an $r_{\rm cut}\to 0$ extrapolation of the $r_{\rm cut}$ dependence
for $r_{\rm cut}>0.15\%$. The systematic uncertainty of the extrapolated result, as explicitly reported in Table~\ref{Table:total}, is at the $\pm {\cal O}(1\%)$ level.
Such uncertainty is much smaller than the NNLO perturbative uncertainties, and thus still acceptable to the purpose of the present paper.
More accurate results could in principle be obtained with \texttt{MATRIX}
by using the option \texttt{switch\_qt\_accuracy=1}, which lowers the value of
the minimum $r_{\rm cut}$ down to $0.05\%$.
The \texttt{MATRIX} result is in agreement with the corresponding result obtained by using \texttt{2$\gamma$NNLO}
within the systematical uncertainties.

We have computed the $n$ dependence of the total cross section in the range
$0.5 \leq n \leq 2$ (see Table~\ref{Table:total}).
At the NNLO (NLO) we find that the result with $n=1$ increases by about 
4\% (3\%) with $n=0.5$ and decreases by about 5\% (2\%) with $n=2$.
At both NLO and NNLO we note that this $n$ dependence of the total cross
section is monotonic and in qualitative agreement with the physical expectation
in Eq.~(\ref{eq:d}).
In particular, we point out that the $n$ dependence of the cross section increases
(especially at $n=2$) in going from NLO to NNLO: the NNLO results are more
sensitive to 
the value
of $n$.
We also note that variations of $n$ in the interval $0.5 \leq n \leq 2$
produce (at both NLO and NNLO) variations of the total cross section that are
smaller than those produced by the scale dependence at fixed $n$.
In view of this, 
we limit ourselves to using $n=1$ (unless otherwise stated) for most of our
subsequent studies.

Throughout this section we present explicit results on the contribution of the
different initial-state partonic channels to various observables at the NNLO.
To simplify the presentation we consider 
the partition of the total result in three contributions:
the contribution of the $gg$ partonic
channel (and its partial component from the box contribution $gg \to \gamma
\gamma$)\footnote{The $gg$ partonic channel contributes at NNLO through two
partonic subprocesses: the box $gg \to \gamma\gamma$ subprocess (which is
positive definite) and the real radiation subprocess $gg \to \gamma\gamma q
{\bar q}$ at the tree level. The real radiation subprocess (which would be
positive definite at the formal level) leads to initial-state collinear
singularities that have to be factorized in the PDFs of the colliding protons.
After factorization, the contribution of the real radiation subprocess turns out
to be negative for most of the phase space region.}, 
the contribution of the $qg+{\bar q}g$ partonic channel 
(it is dominated by its $qg$
component and it is briefly labelled as $qg$ channel) and the contribution
of the remaining
partonic channels (this contribution is dominated by the $q{\bar q}$ initial
state and it is briefly labelled as $q{\bar q}$ channel).
This decomposition in partonic channels has a mild scale dependence and we 
always present results at the central values
$\mu_R=\mu_F=\mu_0=M_{T\gamma \gamma}$ of the scales.

About 9\% of $\sigma^{NNLO}$ is due to the $gg$ channel.
Therefore, the contribution of the $gg$ channel at NNLO is not sizeable
(it is quantitatively comparable to the size of the scale dependence of 
$\sigma^{NNLO}$), despite the fact that the contribution of the 
box $gg \to \gamma \gamma$ is approximately equal to one half of the LO total 
cross
section. We also note that the total $gg$ contribution at NNLO is partly smaller
than its box component: the additional 
NNLO contributions from $gg$ collisions turn out to be negative and have an
absolute size that is approximately one quarter of the box contribution. 
The NNLO result for the total cross section is dominated by the $qg$ and 
$q{\bar q}$ channels, which approximately equally contribute to $\sigma^{NNLO}$.
The NNLO cross section receives contributions of about 48\% from the
$q{\bar q}$ channel and of about 43\% from the
$qg$ channel. 

The NLO total cross section is roughly 3 times larger than $\sigma^{LO}$.
The NNLO $K$ factor, $K^{NNLO}= \sigma^{NNLO}/\sigma^{NLO}$, at central values
of the scales is $K^{NNLO} \simeq 1.4$. We see that both NLO and NNLO
corrections are sizeable. The large size of the QCD radiative corrections up to
NNLO is justifiable and understandable following and extending the reasoning
that we have presented in Sect.~\ref{sec:sigmatot} 
(see Eqs.~(\ref{qqbarlo}) and (\ref{qgnlo}) and accompanying comments).
At the LO only the $q{\bar q}$ channel contributes. At the NLO, 
$\sigma^{NLO}$ receives contribution also from the $qg$ channel. The relative
NLO correction from the $qg$ channel is of the order of 
$\as \,{\cal L}_{q g}/{\cal L}_{q {\bar q}}$ and the large value of the PDF
luminosity ratio ${\cal L}_{q g}/{\cal L}_{q {\bar q}}$ compensates the
suppression factor produced by $\as$. Therefore, the presence of a new NLO
partonic channel, the $qg$ channel, with a large PDF luminosity implies that,
at the quantitative level, there is no parametric hierarchy of 
${\cal O}(\as)$ between NLO corrections and LO result.
As a consequence, the NLO
result has to be quantitatively regarded as an effective lowest-order estimate
of the cross section. The next-order corrections to this result are 
parametrically of ${\cal O}(\as)$ (the contribution of the new $gg$ channel
at NNLO is not particularly sizeable) and they turn out to have a `moderate'
size. Indeed, the value $K^{NNLO} \simeq 1.4$ has a size that is not much
different  from the typical (and expected) size of NLO $K$ factors for various
hard-scattering processes at hadron colliders.
We also note that the scale dependence of the total cross section is partly
reduced in going from NLO to NNLO (it is not so in going from LO to NLO).
However, we remark that the values of $\sigma^{NNLO}$ and $\sigma^{NLO}$
do not overlap by including their corresponding scale dependence. This implies
that the computed scale dependence of $\sigma^{NNLO}$ cannot be consistently
regarded as a reliable estimate of uncalculated radiative corrections to the
total cross section. The `true' theoretical uncertainty of $\sigma^{NNLO}$
due to higher-order corrections is certainly larger than the NNLO scale
dependence that we have computed.

\subsection{Differential cross sections}
\label{sec:diffNNLO}

We now move to consider kinematical distributions. In particular, we consider
the differential cross sections $d\sigma/dM_{\gamma \gamma},
d\sigma/d\cos \theta^{*}, d\sigma/d\Dpgg$ and $d\sigma/dp_{T \gamma \gamma}$,
which are also considered in Ref.~\cite{Aad:2012tba},
and we use the same kinematical bins as used for the corresponding experimental
data in Tables~2--5 of Ref.~\cite{Aad:2012tba}.

\begin{figure}[htb]
\begin{center}
\begin{tabular}{cc}
\hspace*{-4.5mm}
\includegraphics[width=0.48\textwidth]{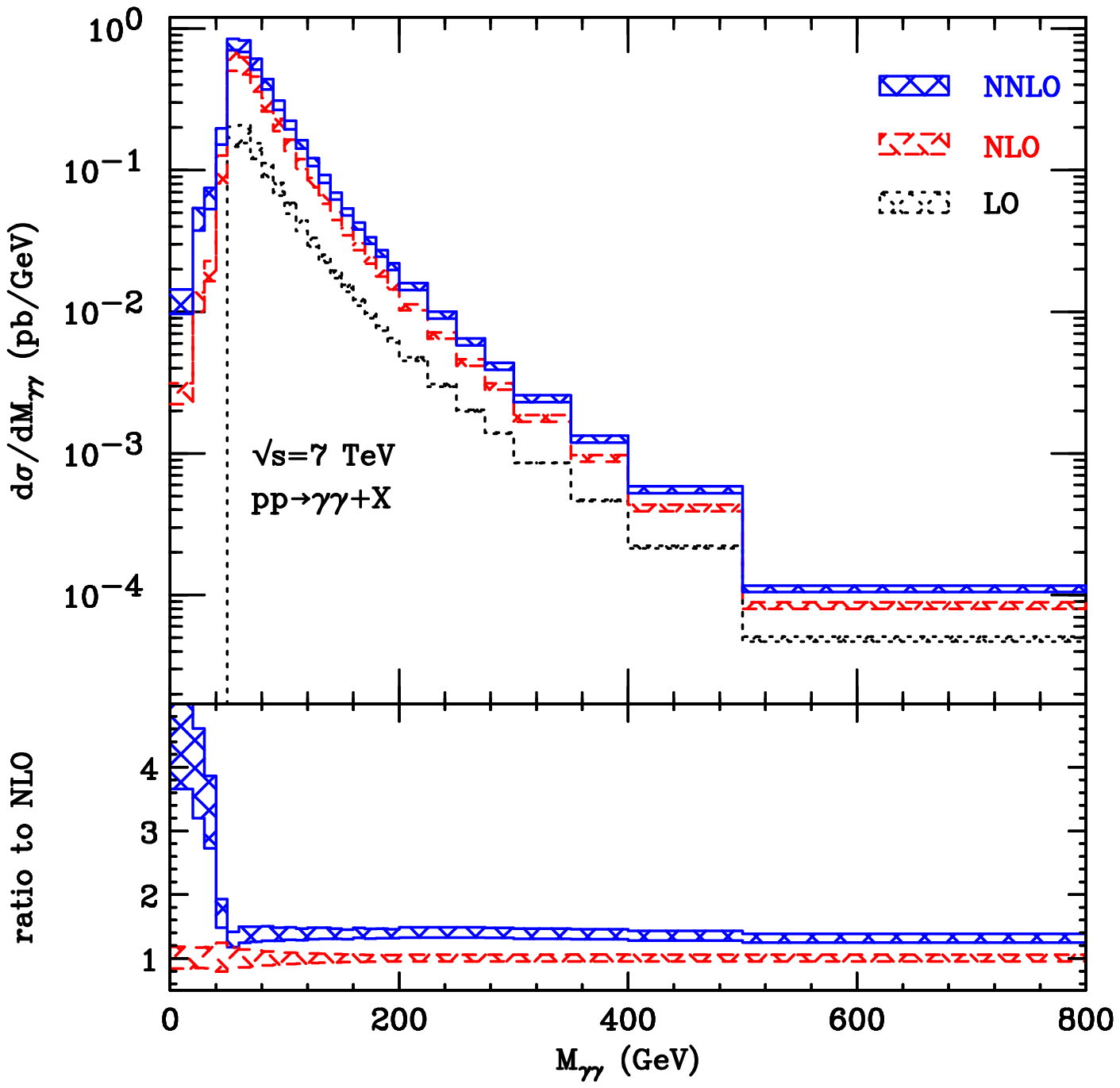}
& 
\includegraphics[width=0.48\textwidth]{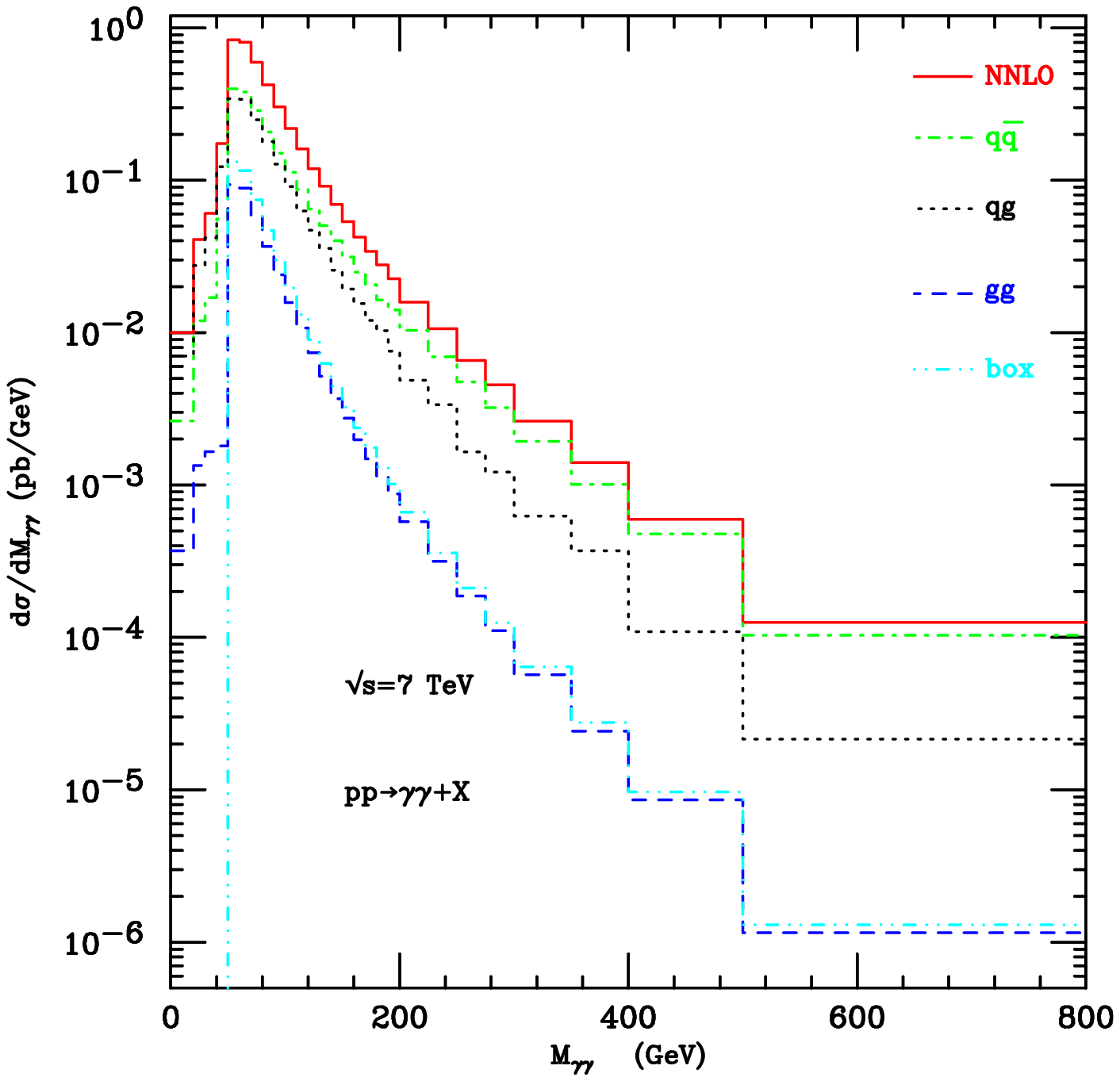}
\\
\end{tabular}
\end{center}
\caption{\label{fig:mlhc}
{\em The differential cross section $d\sigma/dM_{\gamma \gamma}$
with the same photon kinematical cuts as in Table~\ref{Table:total}.
The main panel in the left-hand side shows the LO (black dotted), NLO (red dashed) and NNLO
(blue solid) results, with the corresponding scale dependence. The lower subpanel
presents the NNLO $K$ factor (including its scale dependence) and the relative
scale dependence at NLO. The panel in the right-hand side shows the NNLO result
at central scales and its decomposition in the contributions of different
initial-state partonic channels: $q{\bar q}$ (green dot-dashed), $qg$ (black dotted) and $gg$ (blue dashed). The separate
contribution of the
box $gg \to \gamma \gamma$ squared amplitude is also shown (light-blue dot-dot-dashed).
}}
\end{figure}

The LO, NLO and NNLO results (including their scale dependence) for 
$d\sigma/dM_{\gamma \gamma}$ are presented in the main panel of
Fig.~\ref{fig:mlhc}-left. In the lower panel of Fig.~\ref{fig:mlhc}-left
we present the NNLO $K$ factor and the relative scale dependence at NLO.

The NNLO $K$ factor is defined as
\beq
\label{knnlo}
K^{NNLO}(x) = \frac{d\sigma^{NNLO}(\mu)/dx}{d\sigma^{NLO}(\mu_0)/dx} \;\;,
\eeq
where $d\sigma^{NNLO}(\mu)/dx$ is the scale dependent NNLO result for the
differential cross section $d\sigma/dx$ with respect to the kinematical 
variable $x$, and $d\sigma^{NLO}(\mu_0)/dx$
is the corresponding NLO result at central scales 
($\mu_R=\mu_F=\mu_0=M_{T\gamma \gamma}$). The relative scale dependence at NLO
is defined as
\beq
\label{rnlo}
\frac{d\sigma^{NLO}(\mu)/dx}{d\sigma^{NLO}(\mu_0)/dx} \;\;,
\eeq
where $d\sigma^{NLO}(\mu)/dx$ is the scale dependent NLO result.

The comparisons between the numerical values of the quantities in 
Eqs.~(\ref{knnlo}) and (\ref{rnlo}) gives a direct quantitative illustration of
the degree of overlap (including scale dependencies) of the NNLO and NLO
results.
By inspection of Fig.~\ref{fig:mlhc}-left we see that the two bands in the lower
panel do not overlap. This lack of overlap has already been observed for the case
of the NNLO and NLO total cross sections and the same qualitative features will
be observed for the differential cross sections 
$d\sigma/d\cos \theta^{*}, d\sigma/d\Dpgg$ and $d\sigma/dp_{T \gamma \gamma}$
that we present in the following. 

As we have already remarked, this implies that
the `true' perturbative uncertainty of the NNLO result for these observables
is larger than the corresponding NNLO scale dependence that we explicitly
compute.  

A minor comment on the results in Fig.~\ref{fig:mlhc} regards the first bin,
where $0 \leq M_{\gamma \gamma} \leq 20$~GeV. As discussed in
Sect.~\ref{sec:diffNLO} 
(see, in particular the comments that accompany Eq.~(\ref{lowm})), the
differential cross section $d\sigma/dM_{\gamma \gamma}$ in Fig.~\ref{fig:mlhc}
has the lower kinematical limit $M_{\gamma \gamma} \gtap 10$~GeV.
This implies that we do not actually compute $d\sigma/dM_{\gamma \gamma}$
with $M_{\gamma \gamma} \to 0$.
It also implies
that the rapid decrease of $d\sigma/dM_{\gamma \gamma}$ from the second to the
first bin is partly an artifact of the vanishing of 
$d\sigma/dM_{\gamma \gamma}$ in the region where 
$M_{\gamma \gamma} \ltap 10$~GeV (this region covers about one half of the first
bin). Obviously the same kinematical artifact equally affects the experimental
data in the first bin (the artifact has no effect on the comparison between data
and theory and on the quantities in the lower panel of Fig.~\ref{fig:mlhc}-left). 

The main features of $d\sigma/dM_{\gamma \gamma}$ and of the corresponding LO and
NLO results are discussed in Sects.~\ref{sec:diffLO} and \ref{sec:diffNLO}
(see, in particular, Fig.~\ref{fig:dm} and related comments). That discussion also
includes some comments on our expectations about the NNLO results. As noticed
below, those expectations are confirmed by the NNLO results in 
Fig.~\ref{fig:mlhc}.

The presence of the (unphysical) LO threshold at $M_{\gamma \gamma}=50$~GeV
is responsible for the shape of $d\sigma/dM_{\gamma \gamma}$
in Fig.~\ref{fig:mlhc}. The two invariant-mass bins 
($40~{\rm GeV} \leq M_{\gamma \gamma} \leq 50~{\rm GeV}$ and
$50~{\rm GeV} \leq M_{\gamma \gamma} \leq 60~{\rm GeV}$) that are closest 
to the LO threshold have a relatively-large size, which does not offer enough
resolution to examine the detailed shape of $d\sigma/dM_{\gamma \gamma}$ at
$M_{\gamma \gamma} \sim 50$~GeV. Therefore, we simply comment on the high-mass
($M_{\gamma \gamma} \gtap 50$~GeV) and low-mass 
($M_{\gamma \gamma} \ltap 50$~GeV) regions
(comments on the region that is very close to $M_{\gamma \gamma}= 50$~GeV
are postponed to Sect.~\ref{sec:ptcut}).

In the high-mass region, the NLO and NNLO corrections to 
$d\sigma/dM_{\gamma \gamma}$ have a size that is similar to that of the
corresponding results for the total cross section. The value of 
$K^{NNLO}(M_{\gamma \gamma})$ (including its scale dependence)
for $d\sigma/dM_{\gamma \gamma}$ is very similar 
to the corresponding NNLO $K$ factor for the total cross section.
In particular,  $K^{NNLO}(M_{\gamma \gamma})$ is remarkably independent of 
$M_{\gamma \gamma}$: we have $K^{NNLO}(M_{\gamma \gamma}) \simeq 1.4$ for 
$50~{\rm GeV} \ltap M_{\gamma \gamma} \ltap 350$~GeV; at higher values of
$M_{\gamma \gamma}$,  $K^{NNLO}(M_{\gamma \gamma})$ slightly decreases and 
$K^{NNLO}(M_{\gamma \gamma}) \simeq 1.3$ in the highest-mass bin 
($500~{\rm GeV} \leq M_{\gamma \gamma} \leq 800$~GeV) of Fig.~\ref{fig:mlhc}.
The NNLO contributions of the various initial-state partonic channels 
to $d\sigma/dM_{\gamma \gamma}$ are shown in Fig.~\ref{fig:mlhc}-right.
The $gg$ channel gives a little NNLO contribution to 
$d\sigma/dM_{\gamma \gamma}$ (analogously to the case of the total 
cross section). In particular, the results in the high-mass region of 
Fig.~\ref{fig:mlhc} explicitly show that the total contribution of the 
$gg$ channel is partly smaller than the sole component of the box contribution
$gg \to \gamma \gamma$. In the region where 
$50~{\rm GeV} \ltap M_{\gamma \gamma} \ltap 150$~GeV (which gives the bulk of the
total cross section), the $q{\bar q}$ and $qg$ channels give comparable
contributions to $d\sigma/dM_{\gamma \gamma}$ (the $q{\bar q}$ contribution is
slightly larger). At larger values of $M_{\gamma \gamma}$, the relative
contribution of the $q{\bar q}$ channel increases. 

In the low-mass region ($M_{\gamma \gamma} \ltap 50$~GeV) the NNLO (NLO) result
represents, at the formal level, an effective NLO (LO) perturbative prediction.
In view of that and of the discussion in Sect.~\ref{sec:diffNLO}
(see, in particular, Eqs.~(\ref{qgnlosmooth}) and (\ref{ggnlosmooth}) and
accompanying comments), the NNLO corrections are expected to be large. Indeed,
the NNLO corrections sizeably increase by decreasing $M_{\gamma \gamma}$.
The NNLO $K$ factor has the value $K^{NNLO}(M_{\gamma \gamma}) \sim 3.8$
in the second bin ($20~{\rm GeV} \leq M_{\gamma \gamma} \leq 30~{\rm GeV}$), and
it increases to $K^{NNLO}(M_{\gamma \gamma}) \sim 4.4$ in the first bin. The
scale dependence of the NNLO result also increases by decreasing 
$M_{\gamma \gamma}$, and it reaches the size of about $\pm 20$\% in the first
bin. As shown in Fig.~\ref{fig:mlhc}-right, the contribution of the $gg$ partonic
channel to the NNLO result remains small also in the low-mass region (note that
the box contribution $gg \to \gamma \gamma$ is absent in this region).
The major contribution to the NNLO result is due to the $qg$ channel, because 
both the
$qg$ and $q{\bar q}$ channels already contribute at the lowest order 
in the low-mass region
and the PDF luminosity ${\cal L}_{qg}$ is larger than
${\cal L}_{q{\bar q}}$.

\begin{figure}[htb]
\begin{center}
\begin{tabular}{cc}
\hspace*{-4.5mm}
\includegraphics[width=0.48\textwidth]{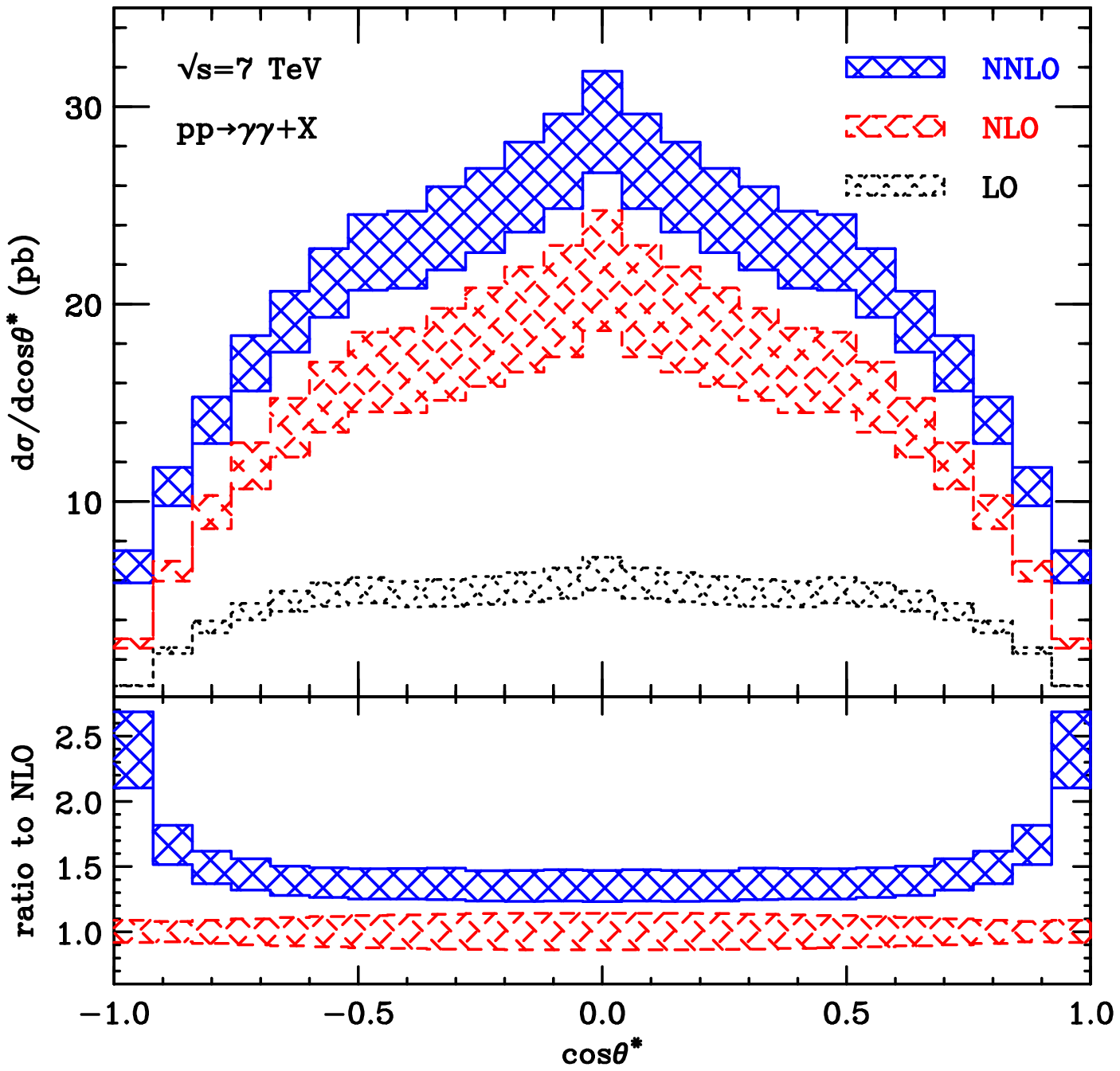}
& \includegraphics[width=0.48\textwidth]{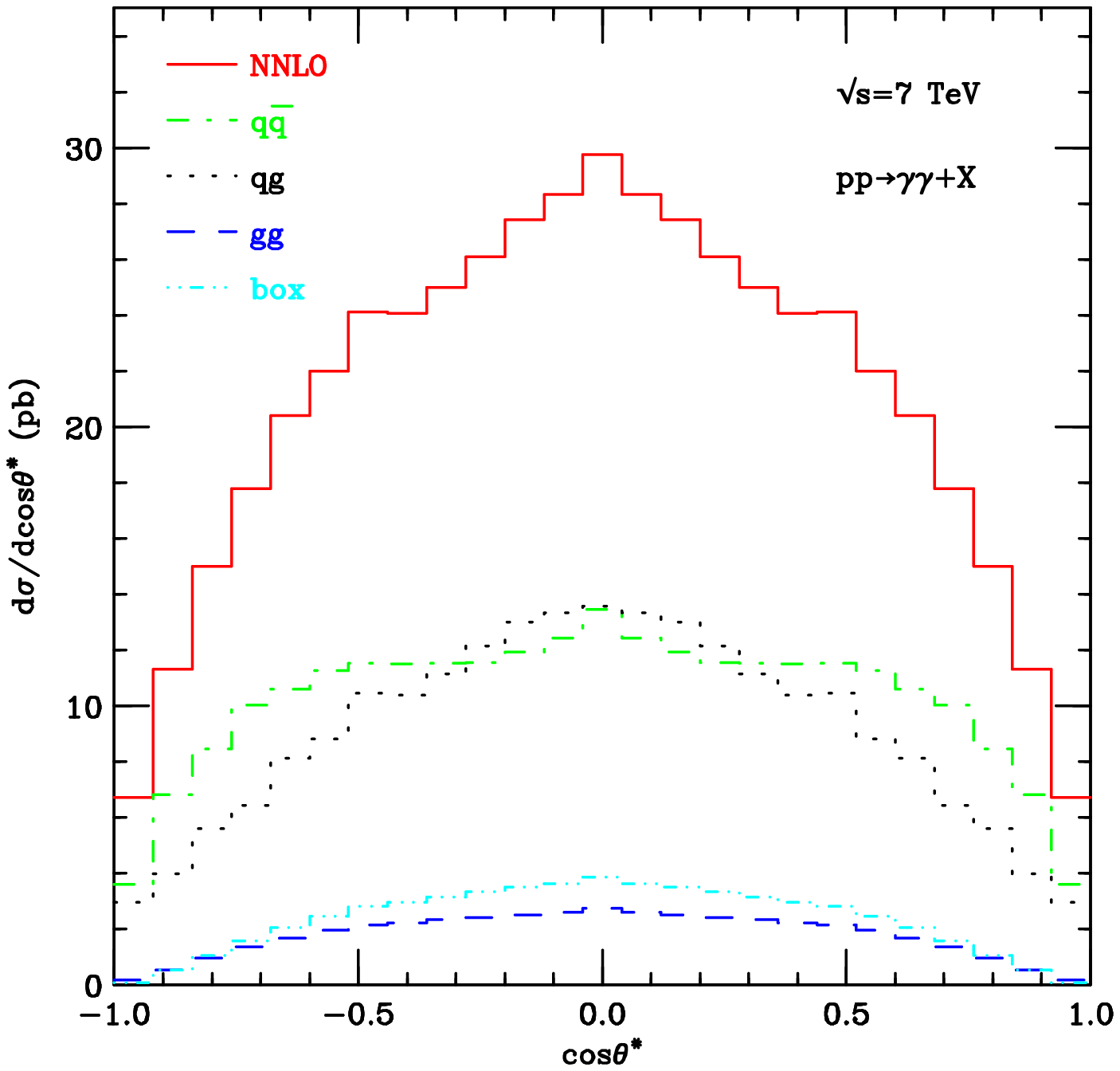}
\\
\end{tabular}
\end{center}
\caption{\label{fig:coslhc}
{\em The differential cross section $d\sigma/d\cos \theta^{*}$.
The results are analogous to those in Fig.~\ref{fig:mlhc}.
}}
\end{figure}

The NNLO results for $d\sigma/d\cos \theta^{*}$ (and the corresponding results at
LO and NLO) are presented in Fig.~\ref{fig:coslhc}.
The main features of the shape of $d\sigma/d\cos \theta^{*}$ are discussed in
detail in Sects.~\ref{sec:diffLO} and \ref{sec:diffNLO}. By direct inspection of
the NLO (and LO) results in Fig.~\ref{fig:dcos} 
and Fig.~\ref{fig:coslhc}-left, we note that the shape of 
$d\sigma/d\cos \theta^{*}$ in Fig.~\ref{fig:coslhc}-left is slightly different
in the central region ($|\cos \theta^{*}| \ltap 0.5$), since it tends to be more
sharpened as $\cos \theta^{*} \to 0$. This shape distortion between the results
in Fig.~\ref{fig:dcos} and Fig.~\ref{fig:coslhc} is basically produced by
the exclusion of the photon rapidity region
$1.37 < |y_\gamma|< 1.52\,$ in the case of Fig.~\ref{fig:coslhc}.

In the central region ($|\cos \theta^{*}| \ltap 0.5$) we have 
$K^{NNLO}(\cos \theta^{*}) \simeq 1.4\,$. This value of the NNLO $K$ factor
and its scale dependence are remarkably similar to those of the corresponding 
$K$ factors for both the differential cross section
$d\sigma/dM_{\gamma \gamma}$ in the high-mass region and the
total cross section. In this  $\cos \theta^{*}$ region the $q{\bar q}$ and
$qg$ channels give comparable NNLO contribution to $d\sigma/d\cos \theta^{*}$
(the NNLO contribution of the $gg$ channel is small, at the level of about
10\%), as shown in Fig.~\ref{fig:coslhc}-right. All these features are perfectly
consistent with the fact (as discussed in detail in Sect.~\ref{sec:diffLO};
see, in particular, the second paragraph below Eq.~(\ref{ptcut})) that the
central $\cos \theta^{*}$ region is kinematically strongly correlated (through
the photon $p_T$ cuts) to the $M_{\gamma \gamma}$ region that gives the bulk of
the total cross section.

At large values of $|\cos \theta^{*}|$ ($|\cos \theta^{*}| \gtap 0.5$), the size
of the NNLO corrections to $d\sigma/d\cos \theta^{*}$ increases by increasing 
$|\cos \theta^{*}|$ (see Fig.~\ref{fig:coslhc}-left). In the second bin
($0.84 < |\cos \theta^{*}| < 0.92$) we have 
$K^{NNLO}(\cos \theta^{*}) \simeq 1.7$, and in the first bin 
($0.92 < |\cos \theta^{*}| < 1$) we have 
$K^{NNLO}(\cos \theta^{*}) \simeq 2.4$ 
(the NNLO scale dependence also increases in the first bin).
As discussed in Sect.~\ref{sec:diffLO}, the large $|\cos \theta^{*}|$ region
is kinematically more sensitive to high values of $M_{\gamma \gamma}$
(values of $M_{\gamma \gamma}$ higher than those that mostly contribute to the
total cross section), and it is also relatively more sensitive to the angular
distribution of the underlying partonic hard-scattering processes.
In particular, this region can receive enhanced NNLO corrections from the
$q{\bar q}$ and $qq$ initiated processes in Eqs.~(\ref{qqbartchansmooth}) and 
(\ref{qqtchansmooth}). The increasing value of $K^{NNLO}(\cos \theta^{*})$
at large $|\cos \theta^{*}|$ is consistent with our comments that accompany
Eqs.~(\ref{qqbartchanfrag})--(\ref{qqtchansmooth}) in Sect.~\ref{sec:diffNLO}\,.
As shown in Fig.~\ref{fig:coslhc}-right, the relative NNLO contribution to
$d\sigma/d\cos \theta^{*}$ of the $q{\bar q}$ channel increases in the region
where $|\cos \theta^{*}| \gtap 0.5$. This increasing behaviour
of the $q{\bar q}$ channel is consistent
with that observed at high values of $M_{\gamma \gamma}$ 
(see Fig.~\ref{fig:mlhc}-right).

\begin{figure}[htb]
\begin{center}
\begin{tabular}{cc}
\hspace*{-4.5mm}
\includegraphics[width=0.475\textwidth]{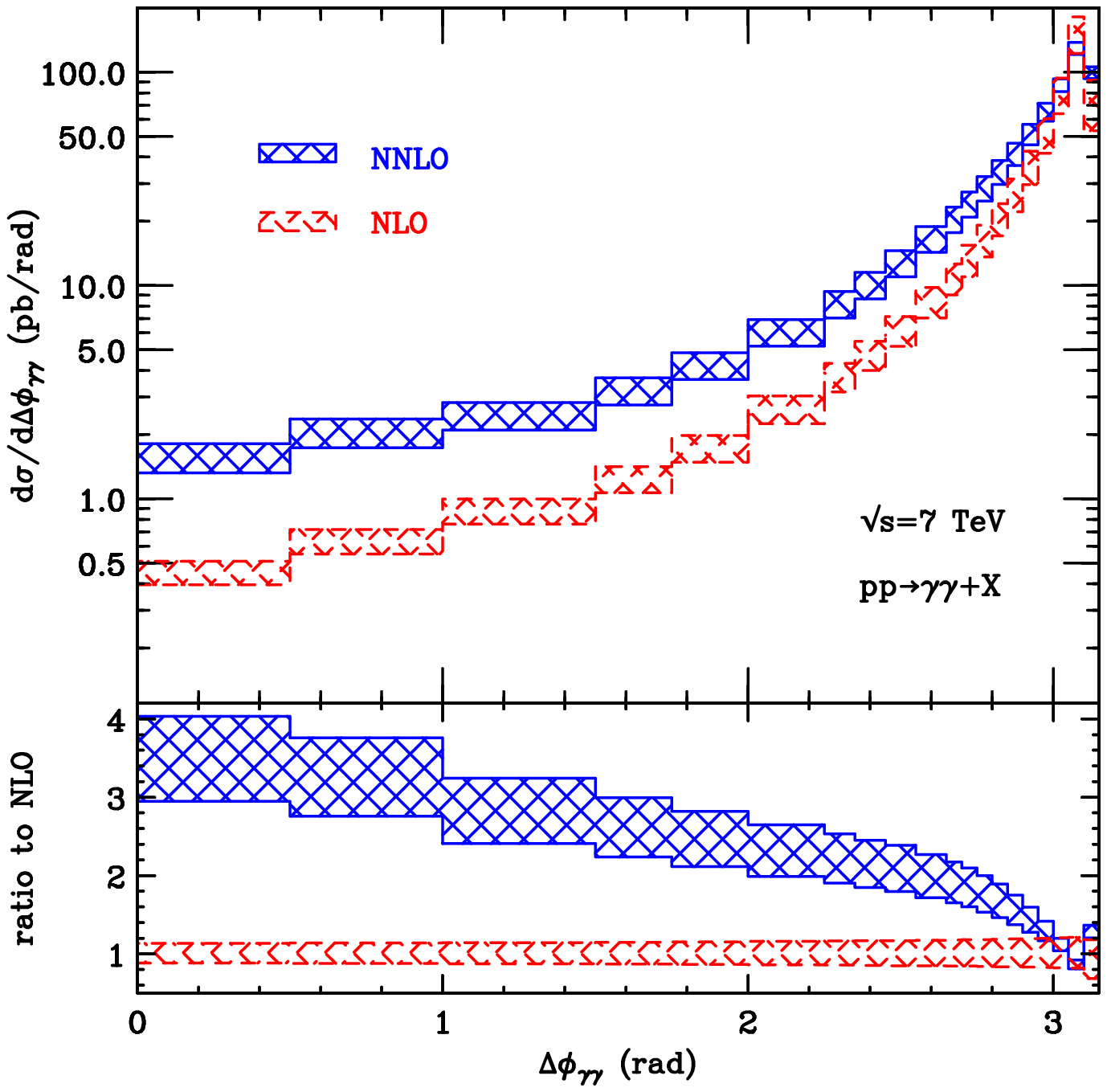}
& \includegraphics[width=0.485\textwidth]{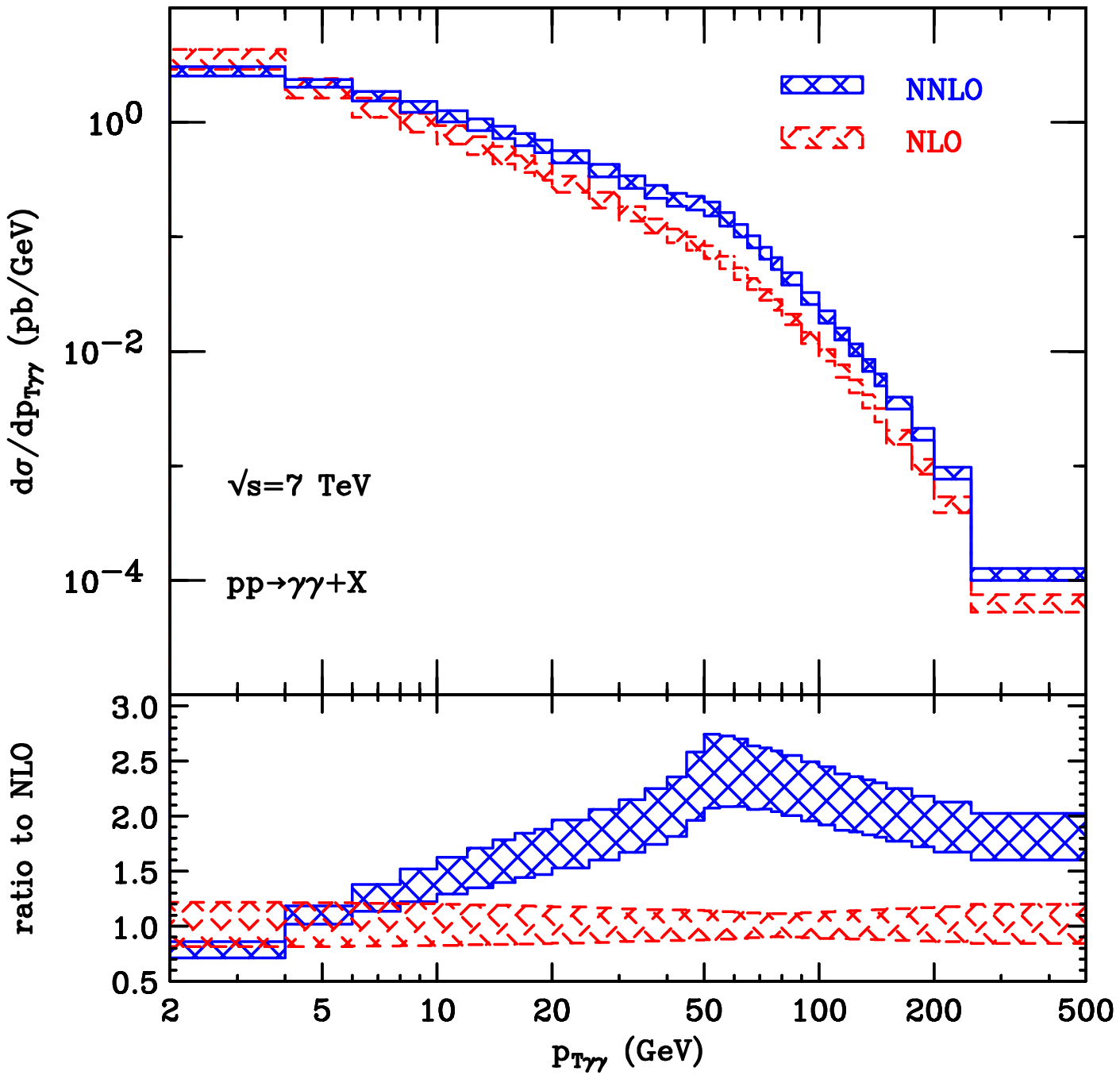}
\\
\end{tabular}
\end{center}
\caption{\label{fig:phiptlhc}
{\em The differential cross sections $d\sigma/d\Dpgg$ (left) and 
$d\sigma/dp_{T \gamma \gamma}$ (right). The NLO and NNLO results are analogous
to those in Fig.~\ref{fig:mlhc}.
}}
\end{figure}

The NNLO (and NLO) results for the differential cross sections
$d\sigma/d\Dpgg$ and $d\sigma/dp_{T \gamma \gamma}$ are presented in 
Fig.~\ref{fig:phiptlhc}. These differential cross sections have some
similarities and some differences.

The similarities certainly regard the Sudakov sensitive region, namely the
region that is close to the exclusive boundary of the phase space (either 
$\Dpgg \sim \pi$ or $p_{T \gamma \gamma} \sim 0$).
In this region the LO result (which is not shown in Fig.~\ref{fig:phiptlhc})
is non-vanishing only in the most extreme bin (the bin that includes either
$\Dpgg = \pi$ or $p_{T \gamma \gamma} = 0$).
As already recalled in Sect.~\ref{sec:diffNLO}, fixed-order QCD computations
of $d\sigma/dp_{T \gamma \gamma}$ at small values of $p_{T \gamma \gamma}$
are affected by large logarithmic contributions (powers of 
$\ln(M_{\gamma \gamma}/p_{T \gamma \gamma}) \sim 
\ln(M^{LO}/p_{T \gamma \gamma})$) that invalidate the physical predictivity of
the fixed-order result. Since $p_{T \gamma \gamma} \to 0$ implies 
$\Dpgg \to \pi$, ensuing large logarithms ($\ln(\pi - \Dpgg)$) appears in the 
fixed-order computation of $d\sigma/d\Dpgg$. These large logarithmic corrections
produce a rapid change of the shape of the differential cross section
order-by-order in QCD perturbation theory. Such a variation of shape is clearly
visible in the results of Fig.~\ref{fig:phiptlhc}-left for $\Dpgg \gtap 2.8$
and Fig.~\ref{fig:phiptlhc}-right for $p_{T \gamma \gamma} \ltap 15$~GeV
(see, in particular, the shape of $K^{NNLO}$ in the lower subpanels 
of Fig.~\ref{fig:phiptlhc}).
As a consequence of this shape variation, the NLO and NNLO results tend to
overlap (leading to $K^{NNLO} \sim 1$) in a tiny region. This overlap and the
much reduced scale dependence of the NNLO result in this region should not be
regarded as a signal of perturbative convergence: in contrast,
they are just a consequence and an artifact of the order-by-order perturbative
instability of the shape of the differential cross section (see 
Ref.~\cite{Bozzi:2008bb} and, in particular, Sect.~3 therein for a completely
related discussion in the context of $Z$ production).
In the Sudakov sensitive region, reliable QCD predictions requires all-order
resummation of the large logarithmic contributions 
\cite{Balazs:2007hr,Cieri:2015rqa}.

\begin{figure}[htb]
\begin{center}
\includegraphics[width=0.49\textwidth]{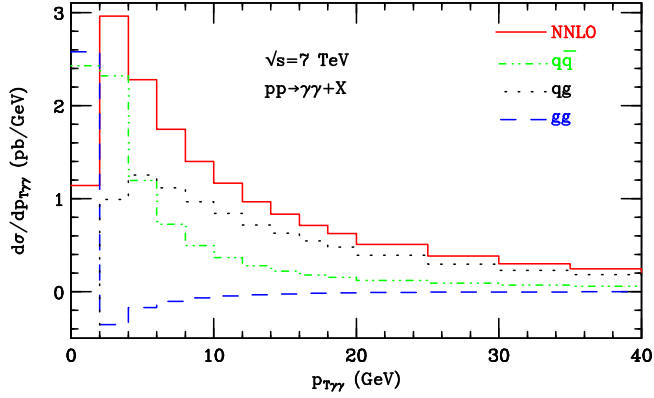}
\end{center}
\caption{\label{fig:ptchannel}
{\em The NNLO cross section $d\sigma/dp_{T \gamma \gamma}$ at central scales
in the low-$p_{T \gamma \gamma}$ region of Fig.~\ref{fig:phiptlhc}-right.
The NNLO result is decomposed in the contributions of different
initial-state partonic channels: $q{\bar q}$ (green dot-dashed), $qg$ (black dotted) and $gg$ (blue dashed). The box 
$gg \to \gamma \gamma$ squared amplitude only contributes in the first
(lowest-$p_{T \gamma \gamma}$) bin.
}}
\end{figure}

The NNLO contribution of the various initial-state partonic channels to
$d\sigma/dp_{T \gamma \gamma}$ in the region where 
$p_{T \gamma \gamma} < 40$~GeV is presented in Fig.~\ref{fig:ptchannel}.
Note that the partial
contribution of the box $gg \to \gamma \gamma$ amplitude is non-vanishing only 
in the lowest-$p_{T \gamma \gamma}$ bin ($p_{T \gamma \gamma}< 2$~GeV).
In this first bin the $q{\bar q}$ and $gg$ channels  give comparable and
positive contributions to the NNLO differential cross section, while the
contribution of the $qg$ channel is large and negative (it is outside the range
of the vertical scale in Fig.~\ref{fig:ptchannel}).
If $p_{T \gamma \gamma}> 2$~GeV the $gg$ channel always gives a minor
contribution: in particular, the contribution is 
negative
in the entire 
$p_{T \gamma \gamma}$ region 
(excluding the lowest-$p_{T \gamma \gamma}$ bin)
of Fig.~\ref{fig:ptchannel}, and it becomes
positive at larger values of $p_{T \gamma \gamma}$.
At very low values 
of $p_{T \gamma \gamma}$, the $q{\bar q}$ channel gives the largest NNLO
contribution 
(this channel is responsible
for the dominant logarithmic enhancement at small values of 
$p_{T \gamma \gamma}$).
By increasing the value of $p_{T \gamma \gamma}$, the $qg$ channel tends to give
the largest NNLO contribution because of its larger PDF luminosity.
The contribution of the $qg$ channel remains the largest also in the region
where $p_{T \gamma \gamma} > 40$~GeV.

We do not explicitly report the decomposition of $d\sigma^{NNLO}/d\Dpgg$
in initial-state partonic channels, since it is qualitatively similar to the
decomposition in Fig.~\ref{fig:phiptlhc} through the 
correspondence (low~$p_{T \gamma \gamma}) \leftrightarrow$~(large~$\Dpgg$)
and (large~$p_{T \gamma \gamma}) \leftrightarrow$~(small~$\Dpgg$).
In particular, the $q{\bar q}$ channel gives the largest NNLO
contribution at $\Dpgg \gtap 3$, whereas the $qg$ channel gives the largest
contribution at moderate and small values of $\Dpgg$. The contribution of the
$gg$ channel is always small and, in particular, it is negative at 
$\Dpgg \gtap 2.7\,$.

Outside the Sudakov sensitive region the LO result vanishes and, therefore,
the NNLO (NLO) results in Fig.\ref{fig:phiptlhc} are `effective' NLO (LO) QCD
predictions. We comment on the results for $d\sigma/d\Dpgg$ and 
$d\sigma/dp_{T \gamma \gamma}$ in turn.

In the region from moderate to small values of $\Dpgg$ 
(Fig.~\ref{fig:phiptlhc}-left) the NNLO corrections monotonically increase by
decreasing $\Dpgg$. We have $K^{NNLO}(\Dpgg) \sim 2$ at $\Dpgg \sim 2.5$,
and $K^{NNLO}(\Dpgg) \simeq 3.4$ in the lowest-$\Dpgg$ bin
($0< \Dpgg <0.5$). 
The large size of the NNLO corrections at small $\Dpgg$ was pointed out in
Ref.~\cite{Cieri:2012tj}.
The scale dependence of the NNLO result
throughout this region is at the level of about $\pm 15$\%, and it is basically
unchanged with respect to the scale dependence of the NLO result.

In the region from moderate to large values of $p_{T \gamma \gamma}$ 
the size of the NNLO corrections does not have a monotonic dependence on 
$p_{T \gamma \gamma}$ (Fig.~\ref{fig:phiptlhc}-right). We have 
$K^{NNLO}(p_{T \gamma \gamma}) \sim 1.7$ at $p_{T \gamma \gamma} \sim 20$~GeV
and $K^{NNLO}(p_{T \gamma \gamma}) \simeq 1.8$ 
in the highest-$p_{T \gamma \gamma}$ bin ($250~{\rm GeV} < p_{T \gamma \gamma}
< 500$~GeV);
at these values of $p_{T \gamma \gamma}$ the NNLO scale dependence is at the
level of about $\pm 12$\% (the corresponding NLO scale dependence is about 
$\pm 18$\%). In the intermediate $p_{T \gamma \gamma}$ region, the size of the
NNLO corrections has a maximal value at $p_{T \gamma \gamma} \sim 50$~GeV.
In the bin with $50~{\rm GeV} < p_{T \gamma \gamma} < 55$~GeV
we have $K^{NNLO}(p_{T \gamma \gamma}) \simeq 2.4$ with a scale dependence of
about $\pm 15$\% (it is basically the same scale dependence as at NLO,
since in the region where $50~{\rm GeV} \ltap p_{T \gamma \gamma} \ltap 100$~GeV
the NLO scale dependence tends to be minimal).

The NNLO results for $d\sigma/d\Dpgg$ and 
$d\sigma/dp_{T \gamma \gamma}$ deserve some overall comments. 

We first note that the size of the NNLO corrections to 
$d\sigma/d\Dpgg$ and $d\sigma/dp_{T \gamma \gamma}$ is typically larger
than that of the NNLO corrections to the total cross section, to
$d\sigma/dM_{\gamma \gamma}$ at high masses and to $d\sigma/d\cos \theta^{*}$.
This feature is consistent with the fact that we are dealing with `effective'
NLO (rather than NNLO) QCD predictions in the case of the 
$\Dpgg$ and $p_{T \gamma \gamma}$ differential cross sections.

Then, we also observe a hierarchy of the size of the NNLO corrections for
various differential cross sections that are computed at `effective' NLO:
$K^{NNLO}(M_{\gamma \gamma})$ at low $M_{\gamma \gamma}$ is larger than
$K^{NNLO}(\Dpgg)$ at small $\Dpgg$, which is in turn larger than
$K^{NNLO}(p_{T \gamma \gamma})$ at moderate and large values of 
$p_{T \gamma \gamma}$. This hierarchy is in agreement with the expectations of
our discussion in Sect.~\ref{sec:diffNLO}. In particular, it is qualitatively
consistent with the effect of the NNLO processes in 
Eqs.~(\ref{qgnlosmooth}) and (\ref{ggnlosmooth}): they are dynamically enhanced
at small values of $M_{\gamma \gamma}$ (see the accompanying comments to 
Eqs.~(\ref{qgnlosmooth}) and (\ref{ggnlosmooth})) and, because of kinematics,
they are still enhanced but with a decreasing relevance at small 
$\Dpgg$ and, in turn, at large $p_{T \gamma \gamma}$.

Finally, we note that the values of $K^{NNLO}$ for $d\sigma/d\Dpgg$ and 
$d\sigma/dp_{T \gamma \gamma}$ within smooth isolation are larger than the
corresponding differences between standard and smooth isolation
results at NLO (see Sect.~\ref{sec:diffNLO}). This fact and the discussion
throughout the paper is a strong indication of the presence of sizeable NNLO
radiative corrections to diphoton production also within the standard cone
isolation criterion.

\begin{figure}[htb]
\begin{center}
\begin{tabular}{cc}
\hspace*{-4.5mm}
\includegraphics[width=0.48\textwidth]{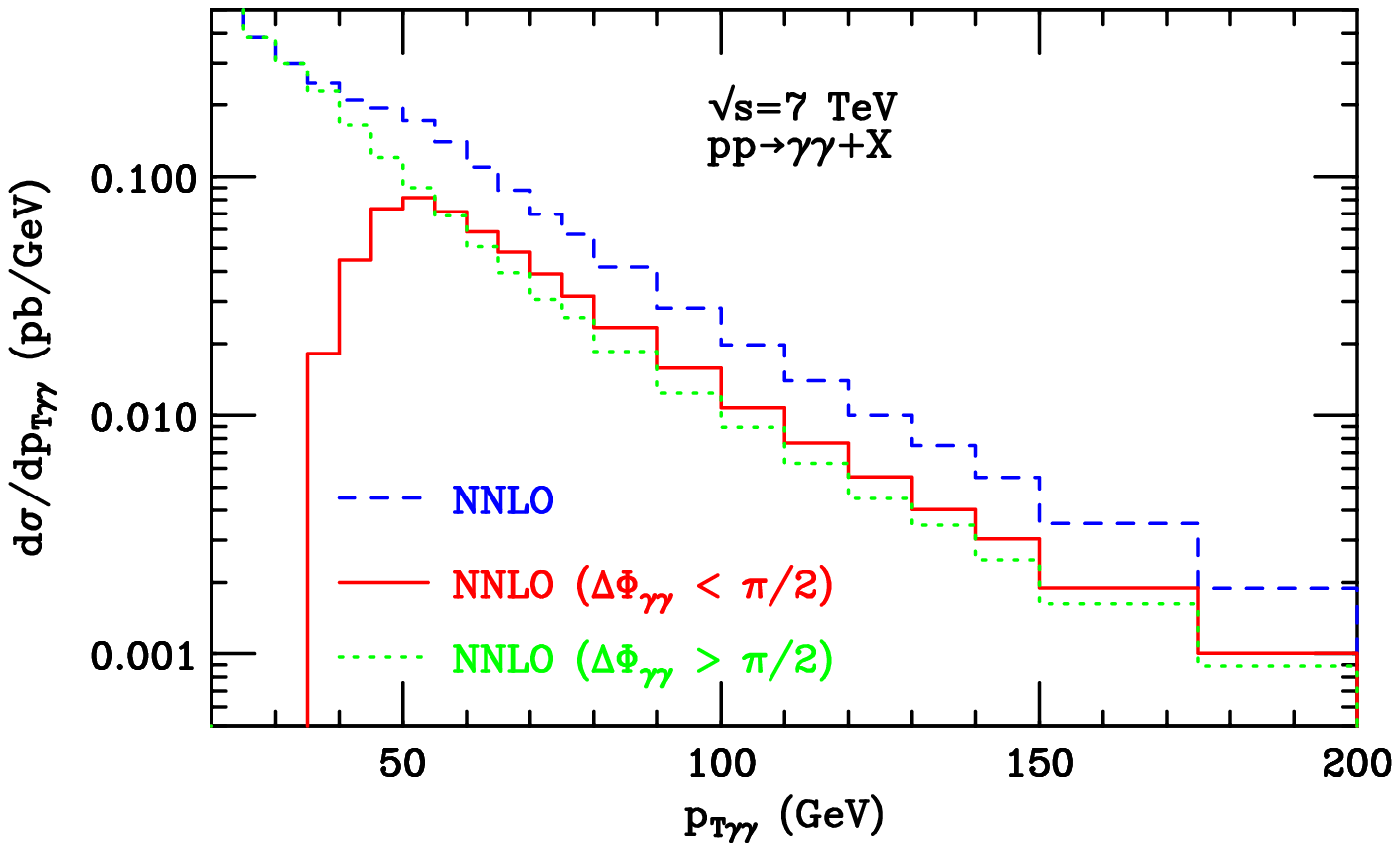}
& \includegraphics[width=0.48\textwidth]{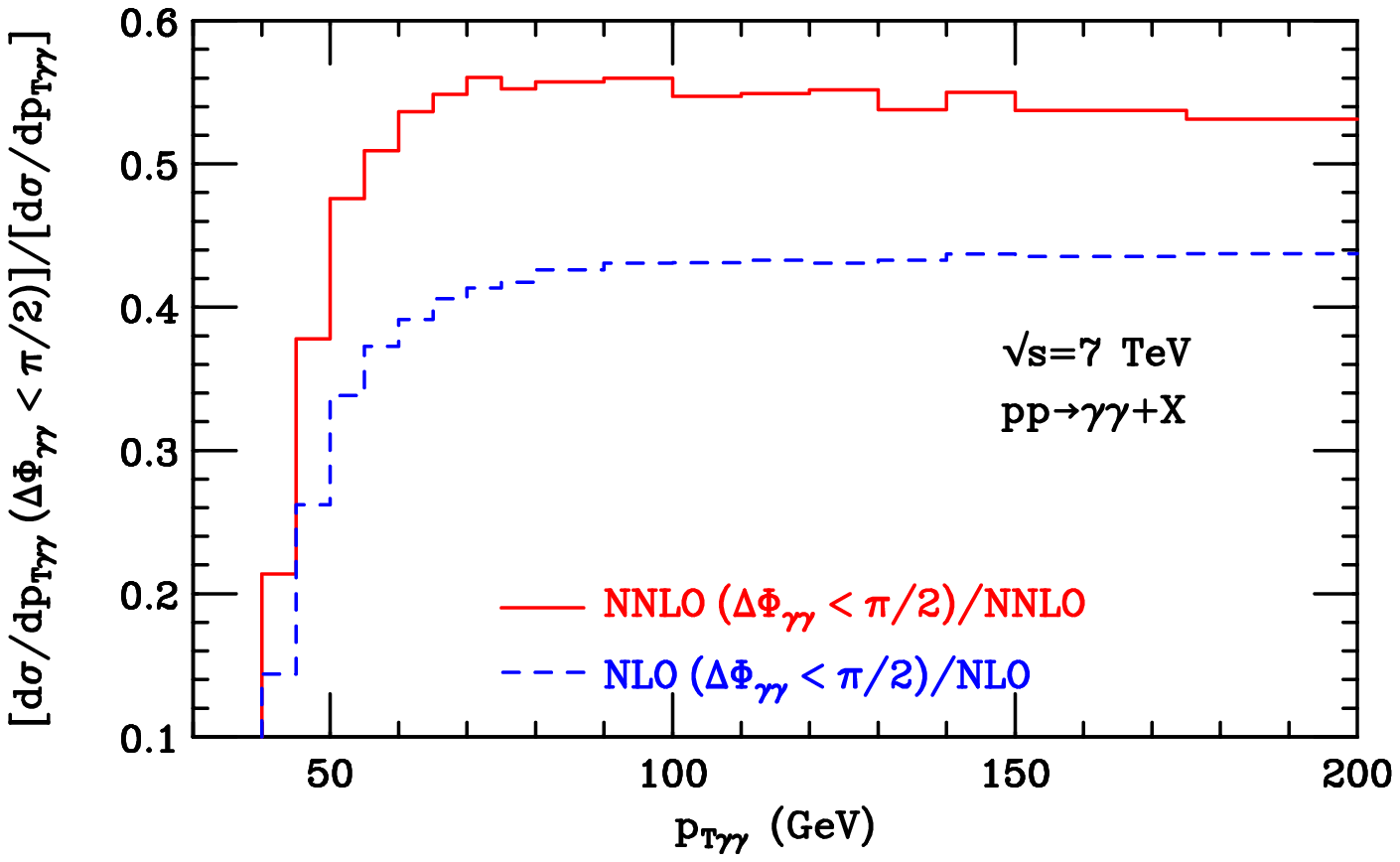}\\
\end{tabular}
\end{center}
\caption{\label{fig:ptshoulder}
{\em Left panel: the NNLO cross section  $d\sigma/dp_{T \gamma \gamma}$
(blue dashed) in Fig.~\ref{fig:phiptlhc}-right and its partition into the
contributions from the two complementary regions where 
$\Dpgg > \pi/2$ (green dotted)
and $\Dpgg<\pi/2$ (red solid).
Right panel: the fractional contribution to $d\sigma/dp_{T \gamma \gamma}$
from the region where $\Dpgg<\pi/2$ at NLO (blue dashed)
and NNLO (red solid).
}}
\end{figure}

We present some additional comments on $d\sigma/dp_{T \gamma \gamma}$
in the region of relatively-large values of $p_{T \gamma \gamma}$.
As we have noticed in Sect.~\ref{sec:diffNLO}, the NLO standard isolation 
results in Fig.~\ref{fig:dpt}-right have a shoulder-type behaviour at 
$p_{T \gamma \gamma} \sim 50$~GeV. The $p_{T \gamma \gamma}$ shoulder is also
visible in the NNLO smooth isolation results of Fig.~\ref{fig:phiptlhc}-right
(see also Figs.~\ref{fig:ptshoulder}-left and \ref{fig:dpdata}-right, which have
a linear scale in $p_{T \gamma \gamma}$), although it is less pronounced since
the results refer to $\ETmax=4$~GeV, which is smaller than the value
$\ETmax=10$~GeV of Fig.~\ref{fig:dpt}-right.
As previously recalled, in the context of standard isolation the $p_{T \gamma \gamma}$
shoulder was discussed in Ref.~\cite{Binoth:2000zt}. Here we present related
comments in the context of the smooth isolation criterion.

We follow Ref.~\cite{Binoth:2000zt} and we consider the partition of 
$d\sigma/dp_{T \gamma \gamma}$ in the contributions of the two complementary
kinematical regions with relatively-large ($\Dpgg > \pi/2$)
and relatively-small ($\Dpgg < \pi/2$)
values of $\Dpgg$. These contributions are denoted by
$\lpt$ (if $\Dpgg > \pi/2$)
and $\spt$ (if $\Dpgg < \pi/2$) in the following text.
The complete NNLO result for $d\sigma/dp_{T \gamma \gamma}$ and the corresponding
NNLO results from the two complementary regions of $\Dpgg$ are presented in 
Fig.~\ref{fig:ptshoulder}-left at values of $p_{T \gamma \gamma}$ around the
shoulder.
The component $\lpt$ has an approximately-constant slope at 
$p_{T \gamma \gamma} \gtap 30$~GeV, whereas the total contribution tends to
flatten out at $p_{T \gamma \gamma} \sim 50$~GeV and it remains systematically
higher than $\lpt$
at larger values of $p_{T \gamma \gamma}$. These differences are obviously
due to the component $\spt$, which exactly vanishes at 
$p_{T \gamma \gamma} \ltap 34$~GeV and has a maximum value at 
$p_{T \gamma \gamma} \sim 50$~GeV.

To comment on the behaviour of the results in Fig.~\ref{fig:ptshoulder}-left
we consider the effect of the kinematical constraint in Eq.~(\ref{ptvsphi})
(see also its accompanying comments), which is a consequence of the photon 
$p_T$ cuts. The component $\lpt$ is insensitive to the constraint.
Moving from large to smaller values of $p_{T \gamma \gamma}$, the size of this
component increases since the production of photons with smaller transverse
momenta is kinematically (because of energy conservation) and also dynamically
favoured. At $\Dpgg \sim 0$, the constraint 
in Eq.~(\ref{ptvsphi}) 
leads to
$p_{T \gamma \gamma} \gtap 47$~GeV (see Eq.~(\ref{ptphi0})). 
Therefore, provided 
$p_{T \gamma \gamma} \gtap 47$~GeV, the entire kinematical region of $\Dpgg$ is
kinematically allowed and 
$\spt$ is also insensitive to the constraint.
It follows that at large values of 
$p_{T \gamma \gamma}$ both components (and their total contribution) show a
similar increasing behaviour as $p_{T \gamma \gamma}$ decreases.
If $p_{T \gamma \gamma} \ltap 47$~GeV the constraint in Eq.~(\ref{ptvsphi})
is instead effective on $\spt$ and, consequently, on the total contribution to
$d\sigma/dp_{T \gamma \gamma}$. Going from $p_{T \gamma \gamma} \simeq 47$~GeV
to smaller values of $p_{T \gamma \gamma}$, the kinematical constraint forbids
the region of small values of $\Dpgg$: 
consequently, the increasing
behaviour of $d\sigma/dp_{T \gamma \gamma}$ flattens out since  
$\spt$ is increasingly suppressed. In particular, Eq.~(\ref{ptvsphi}) implies 
the restriction
$p_{T \gamma \gamma} \gtap 34$~GeV if $\Dpgg < \pi/2$ and, therefore, 
$\spt$ has a kinematical threshold at $p_{T \gamma \gamma} \simeq 34$~GeV.

From our discussion on the results in Fig.~\ref{fig:ptshoulder}-left, it follows
that the onset of the shoulder at $p_{T \gamma \gamma} \sim 50$~GeV is
kinematically driven (by the presence of the photon $p_T$ cuts).
However, it also has a dynamical component, since the NNLO corrections enhance
this effect (see the size of $K^{NNLO}(p_{T \gamma \gamma})$ at 
$p_{T \gamma \gamma} \gtap 50$~GeV in Fig.~\ref{fig:phiptlhc}-right).

To remark the NNLO enhancement we compute the ratio between $\spt$ and
$d\sigma/dp_{T \gamma \gamma}$, namely, the fractional contribution to
$d\sigma/dp_{T \gamma \gamma}$ from the region of relatively-small values of 
$\Dpgg$. In Fig.~\ref{fig:ptshoulder}-right we present the results for this
ratio at both NNLO and NLO (the shape of $\spt$ at NLO is qualitatively similar
to that of the corresponding NNLO result in Fig.~\ref{fig:ptshoulder}-left).
At both perturbative orders this ratio is sizeable and approximately independent
of $p_{T \gamma \gamma}$ in the region just above 
$p_{T \gamma \gamma} \sim 50$~GeV: the ratio is about 0.42 at NLO and 0.54 at
NNLO. Therefore, the region where $\Dpgg < \pi/2$ contributes to roughly half of
the complete differential cross section $d\sigma/dp_{T \gamma \gamma}$
at $p_{T \gamma \gamma} \gtap 50$~GeV, and this fact has a kinematical origin.
The increase of about 30\% of the ratio from the NLO to the NNLO result has
instead a dynamical origin (incidentally, this increase implies that the NNLO 
$K$ factor for $\spt$ is larger than that for $d\sigma/dp_{T \gamma \gamma}$
in Fig.\ref{fig:phiptlhc}-right).
It is due to the NNLO processes in Eqs.~(\ref{qgnlosmooth}) and 
(\ref{ggnlosmooth}) that are dynamically enhanced at small values of 
$M_{\gamma \gamma}$ and, consequently (see Eq.~(\ref{lowm}) and accompanying
comments), partly enhanced also at very small values of $\Dpgg$. 
There are dynamical similarities between these processes and
the NLO fragmentation processes 
in Eqs.~(\ref{qgnlofrag}) and (\ref{ggnlofrag}), which, according to 
Ref.~\cite{Binoth:2000zt}, originate the $p_{T\gamma \gamma}$ shoulder in the
NLO standard isolation results. Therefore, our discussion on the 
$p_{T\gamma \gamma}$ shoulder in the NNLO smooth isolation results is consistent
with the observations in Ref.~\cite{Binoth:2000zt}.

The partition of the phase space into the two regions with 
$\Dpgg > \pi/2$ and $\Dpgg<\pi/2$ was applied by the 
D0 Collaboration \cite{Abazov:2013pua} to diphoton production data in 
proton--antiproton collisions at the Tevatron.
The results in Ref.~\cite{Abazov:2013pua} show that the inclusion of the NNLO
corrections considerably improves the description of the data in both phase
space regions.

\subsection{Dependence on isolation parameters}
\label{sec:ispar}

We add some illustrative results and comments on differential cross sections
at NNLO and their dependence on isolation parameters.

\begin{figure}[htb]
\begin{center}
\begin{tabular}{cc}
\hspace*{-4.5mm}
\includegraphics[width=0.485\textwidth]{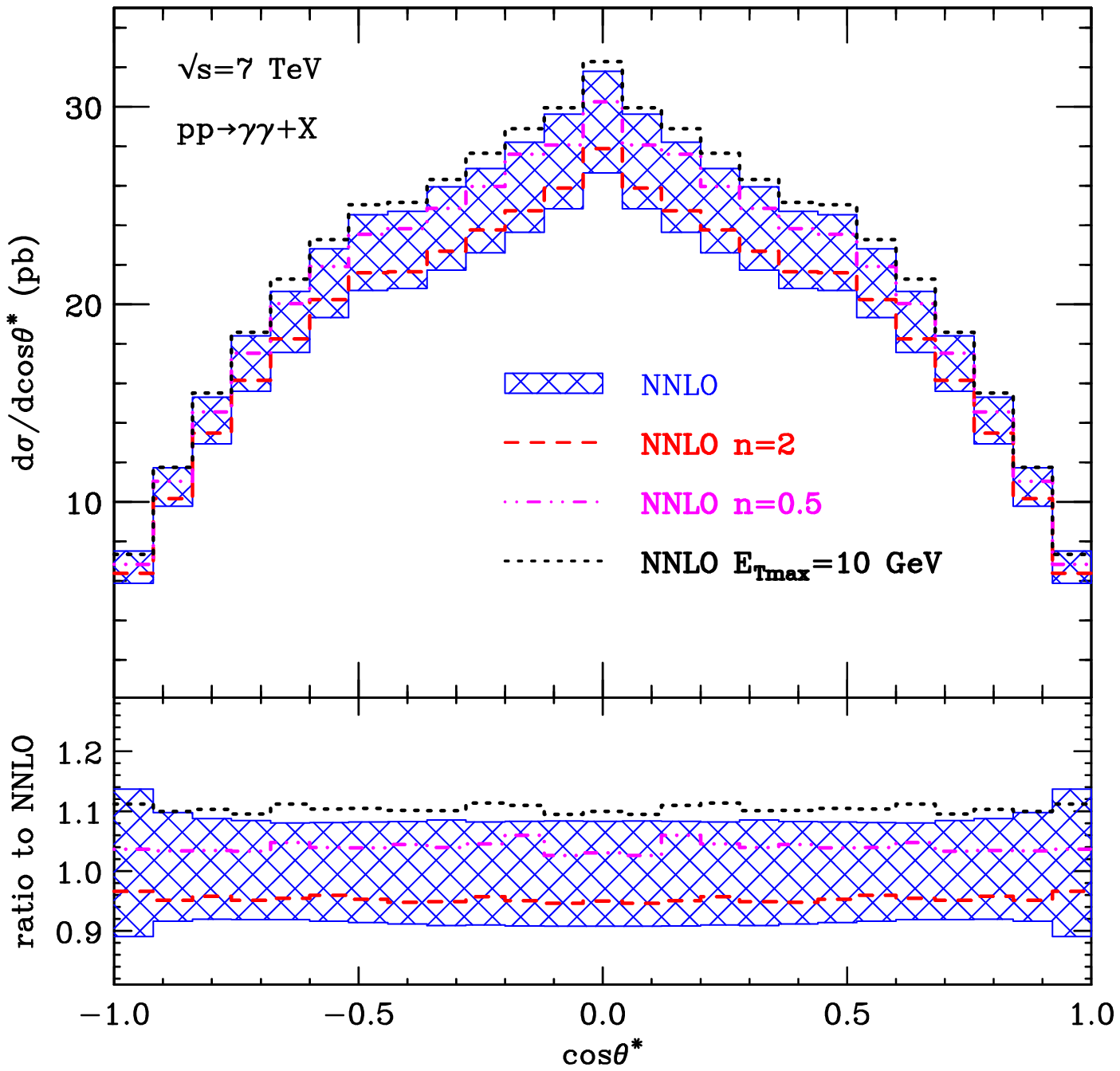} 
& \includegraphics[width=0.475\textwidth]{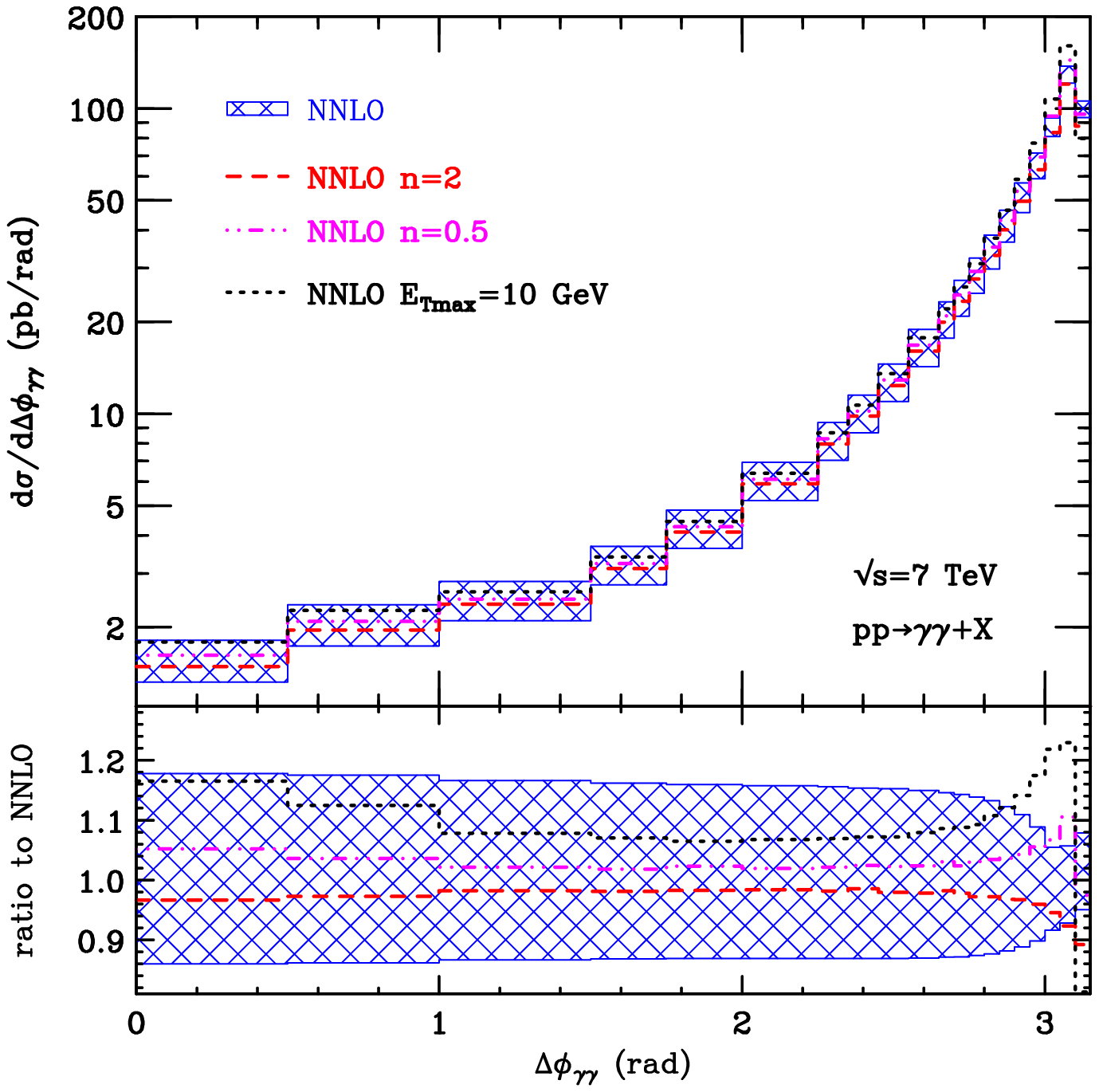}\\
\end{tabular}
\end{center}
\caption{\label{fig:nnEtdep}
{\em Isolation parameter dependence of $d\sigma/d\cos \theta^{*}$ (left)
and $d\sigma/d\Dpgg$ (right) at NNLO. The reference results (including their
scale dependence) with $n=1$ and $\ETmax=4$~GeV of Figs.\ref{fig:coslhc} 
and \ref{fig:phiptlhc}
are compared to corresponding results that are obtained by
varying either $n$ ($n=2$ and $n=0.5$) or $\ETmax$ ($\ETmax=10$~GeV).
Cross sections ratios with respect to the reference result are shown in the
lower subpanels.}
}
\end{figure}

We use $d\sigma/d\cos \theta^{*}$ and $d\sigma/d\Dpgg$ as representative cross
sections of typical features of the NNLO results. We consider the corresponding
reference NNLO results of this section 
(Figs.~\ref{fig:coslhc} and \ref{fig:phiptlhc}), which are obtained with 
$\ETmax=4$~GeV and the power $n=1$ in the isolation function $\chi(r;R)$,
and we perform variations of the isolation parameters $\ETmax$ and $n$.
Specifically, we use either $n=0.5$ or $n=2$ by keeping $\ETmax=4$~GeV fixed,
and we use $n=1$ by increasing $\ETmax$ to the value 
$\ETmax=10$~GeV. The ensuing NNLO results are presented in 
Fig.~\ref{fig:nnEtdep}. In the lower subpanels of Fig.~\ref{fig:nnEtdep}
we present the ratio of the results with different isolation parameters with
respect to the reference NNLO result ($n=1$, $\ETmax=4$~GeV).
All these results are obtained with central values of the scales
($\mu_R=\mu_F=\mu_0=M_{T\gamma \gamma}$).
The scale dependence of the reference NNLO result is also reported in
Fig.~\ref{fig:nnEtdep}
for comparison with isolation parameters effects.

We comment on the results in Fig.~\ref{fig:nnEtdep} by excluding those in the
region where $\Dpgg \gtap 2.8$ (comments on this region are postponed).
As a main overall comment we observe that the NNLO results have a mild
dependence on the isolation parameters. Indeed, the effects of the variations 
of the isolation parameters are not larger than the scale dependence at fixed
isolation parameters. Moreover, the qualitative dependence on both 
$\ETmax$ and $n$ is in agreement with physical expectation 
(see Eqs.~(\ref{eq:b}) and (\ref{eq:d}))
throughout the entire kinematical regions of $\cos \theta^{*}$
and $\Dpgg$: the value of the differential cross sections decreases
by requiring the photons to be more isolated. We also recall 
(see Sect.~\ref{sec:diffNLO}) that in the region where 
$\Dpgg \ltap \pi - R \simeq 2.7$ the NLO cross section $d\sigma/d\Dpgg$
(within smooth isolation) does not depend on the isolation parameters.
Therefore, the $\ETmax$ and $n$ dependence observed 
in the NNLO results 
for $d\sigma/d\cos \theta^{*}$ ($d\sigma/d\Dpgg$) is `effectively' an NLO (an LO)
QCD effect.

We note that the $n$ dependence (at fixed $\ETmax=4$~GeV)
is practically the same throughout the entire kinematical region of either 
$\cos \theta^{*}$ or $\Dpgg$, so that it does not produce any shape variations. 
The overall size of the $n$ dependence of the differential cross sections is
basically equal to that of the NNLO total cross section 
(see Sect.~\ref{sec:totalNNLO}).

The $\ETmax$ dependence in Fig.~\ref{fig:nnEtdep} is substantially larger
than the $n$ dependence, because $\ETmax=10$~GeV is sizeably different from
$\ETmax=4$~GeV. Nonetheless, also the $\ETmax$ dependence at NNLO is
quite moderate (as already noticed in Sect.~\ref{sec:sigmatot} for the case of
the NLO total cross section).
In the case of $d\sigma^{NNLO}/d\cos \theta^{*}$, the variation of 
$\ETmax$ does not produce any significant shape variation: the differential
cross section uniformly increases by approximately 10\% in going from
$\ETmax=4$~GeV to $\ETmax=10$~GeV. Consequently, an equal increase 
applies to the NNLO total cross section. We note that the size of the 
$\ETmax$ dependence is slightly larger at NNLO than at NLO (see 
Sect.~\ref{sec:sigmatot}); an analogous increased sensitivity to the value of
$n$ (at fixed $\ETmax$) has been already noticed in 
Sect.~\ref{sec:totalNNLO}\,.
In the case of $d\sigma^{NNLO}/d\Dpgg$, the variation of $\ETmax$
also produces a visible shape dependence. In going from 
$\ETmax=4$~GeV to $\ETmax=10$~GeV, $d\sigma^{NNLO}/d\Dpgg$ increases 
by approximately 8\% at $\Dpgg \sim 2$, and by approximately 
16\% at $\Dpgg \sim 0.5\,$. This shape dependent effect is consistent with
our previous discussions in Sect.~\ref{sec:diffNNLO} and in 
Sect.~\ref{sec:diffNLO}. At small values of $\Dpgg$ the NNLO cross section
$d\sigma^{NNLO}/d\Dpgg$ receives substantial contribution from configurations
in which the photons are accompanied by partonic (hadronic) transverse energy
inside the isolation cones 
(see Eqs.~(\ref{qgnlosmooth}) and (\ref{ggnlosmooth})):
by increasing $\ETmax$ these configurations are less suppressed by the
isolation requirement and the cross section increases (a qualitatively similar 
$\ETmax$ dependence at small $\Dpgg$ is observed in the NLO standard
isolation results of 
Fig.~\ref{fig:dphi}).

On the basis of our discussion about the results in Fig.~\ref{fig:nnEtdep}
and of our previous discussions about similarities between various NNLO
differential cross sections, we can argue about the $\ETmax$ dependence
of $d\sigma/dM_{\gamma \gamma}$ and $d\sigma/dp_{T \gamma \gamma}$
at NNLO. In the high-$M_{\gamma \gamma}$ region the $\ETmax$ dependence
of $d\sigma/dM_{\gamma \gamma}$ is similar to that of 
$d\sigma/d\cos \theta^{*}$. The $\ETmax$ sensitivity of 
$d\sigma/dM_{\gamma \gamma}$ at small values of $M_{\gamma \gamma}$ is partly
enhanced with respect to that of $d\sigma/d\Dpgg$ at small $\Dpgg$.
The $\ETmax$ sensitivity of $d\sigma/dp_{T \gamma \gamma}$ at moderate and
large values of $p_{T \gamma \gamma}$ is partly reduced with respect to that of 
$d\sigma/d\Dpgg$ at small $\Dpgg$. 

We come to comment about the results in the large-$\Dpgg$ region
($\Dpgg \gtap 2.8$) of Fig.~\ref{fig:nnEtdep}. This is definitely part of 
the Sudakov sensitive region. We point out that here the dependence on the
isolation parameters (both $\ETmax$ and $n$) is amplified with respect to the 
rest of the kinematical regions. We are dealing with diphoton production close
to the exclusive boundary of the phase space, where the photon pair is
accompanied by radiation of little transverse energy, also outside the photon
isolation cones. In such a configuration the effects of variations of the
isolation requirements are relatively enhanced at fixed order: 
therefore, an enhanced
sensitivity to isolation parameters can be expected 
in the results of Fig.~\ref{fig:nnEtdep}-right.
As recalled in 
Sect.~\ref{sec:diffNNLO}, the NNLO result (and, consequently, the ensuing
isolation parameter dependence) is not physically reliable in this Sudakov
sensitive region. Nonetheless,
the observed isolation parameter dependence at NNLO indicates that photon
isolation effects have to be carefully considered (and examined) in the context
of refined resummed calculations in the Sudakov sensitive region. 

\subsection{Comparison of NNLO results and data}
\label{sec:datavsth}

In Ref.~\cite{Aad:2012tba} the ATLAS data on diphoton production at the
centre--of--mass energy 
${\sqrt s}=7$~TeV were compared with the NNLO QCD results. As stated in the
conclusion of Ref.~\cite{Aad:2012tba}, the NNLO results are able to match the
data very closely within the uncertainties, except in limited regions.
In the following we present some comments on the comparison between 
NNLO results and data.

As a preliminary comment, we note that the NNLO results in 
Ref.~\cite{Aad:2012tba} and those in this paper are not exactly equal. The
differences have various origins. Part of the differences are due to the setup
of the NNLO calculations. The results of the NNLO calculation presented in
Ref.~\cite{Aad:2012tba} use the PDFs of the MSTW2008 set, the central scale
$\mu_0=M_{\gamma \gamma}$ and the scale dependence that is obtained by
considering the scale configurations with $\mu_R=\mu_F=M_{\gamma \gamma}/2$
and with 
$\mu_R=\mu_F=2 M_{\gamma \gamma}$. Here we use different PDFs, a different
central scale and independent scale variations of $\mu_R$ and $\mu_F$. The other
differences are due to the fact that the NNLO results of Ref.~\cite{Aad:2012tba} 
were obtained by using the first version
of the numerical program \texttt{2$\gamma$NNLO}, which had an implementation
error that 
was subsequently corrected \cite{Catani:2011qz}.
All these effects produce differences between the NNLO results used in 
Ref.~\cite{Aad:2012tba} and those reported here. Nonetheless the differences are
relatively small (in particular, they are smaller than the scale dependence
at NNLO) and they do not have a major effect on the comparison between NNLO
results and data.

\begin{figure}[htb]
\begin{center}
\begin{tabular}{cc}
\hspace*{-4.5mm}
\includegraphics[width=0.48\textwidth]{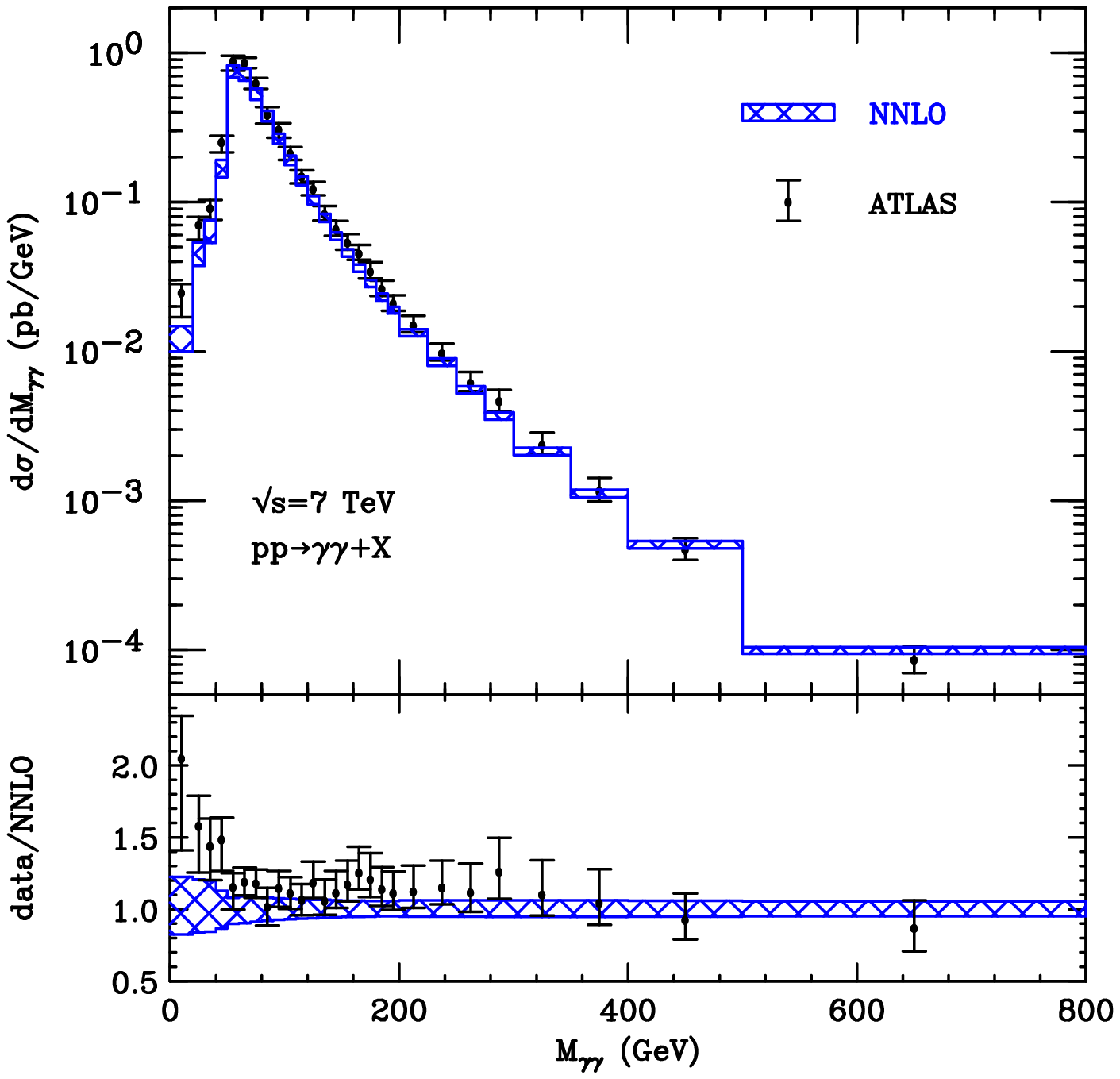} 
& \includegraphics[width=0.48\textwidth]{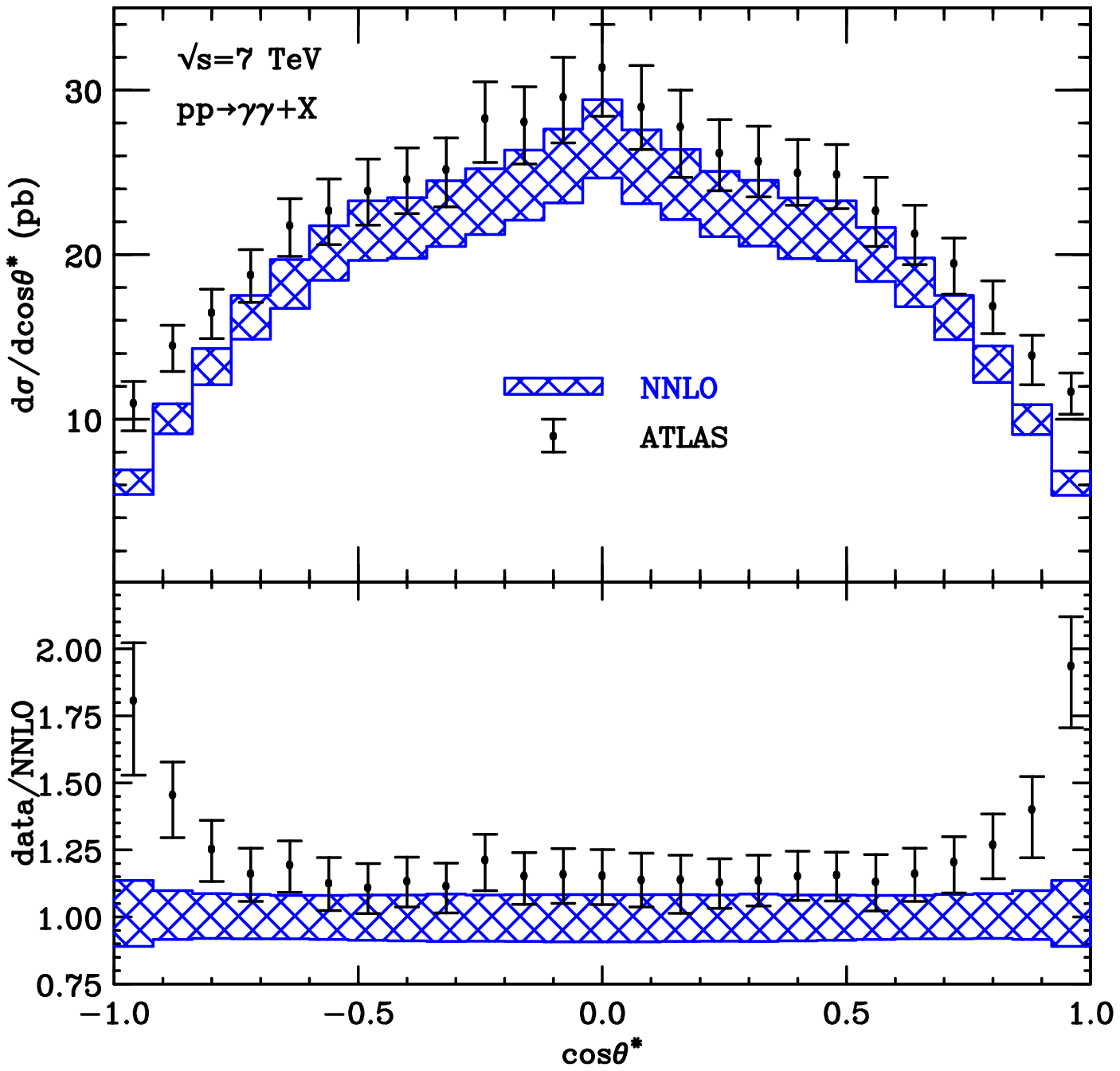}\\
\end{tabular}
\end{center}
\caption{\label{fig:mcdata}
{\em 
Comparison between ATLAS data at ${\sqrt s}=7$~TeV \cite{Aad:2012tba} 
and NNLO results 
(including their scale dependence) for 
$d\sigma/dM_{\gamma \gamma}$ (left) and $d\sigma/d\cos \theta^{*}$ (right).
The NNLO results are corrected for hadronization and underlying event effects
(see text).
}}
\end{figure}

\begin{figure}[htb]
\begin{center}
\begin{tabular}{cc}
\hspace*{-4.5mm}
\includegraphics[width=0.48\textwidth]{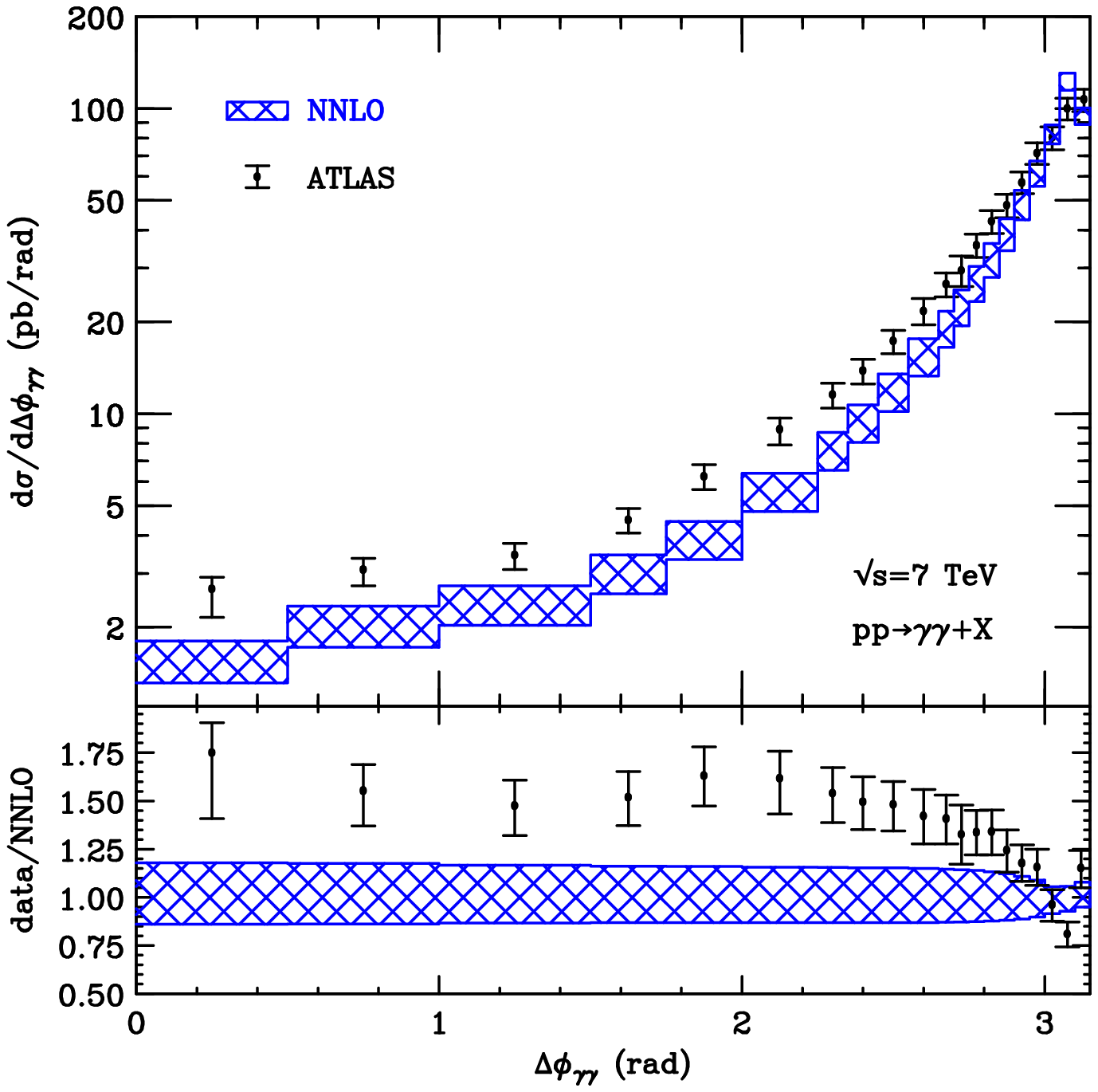} 
& \includegraphics[width=0.49\textwidth]{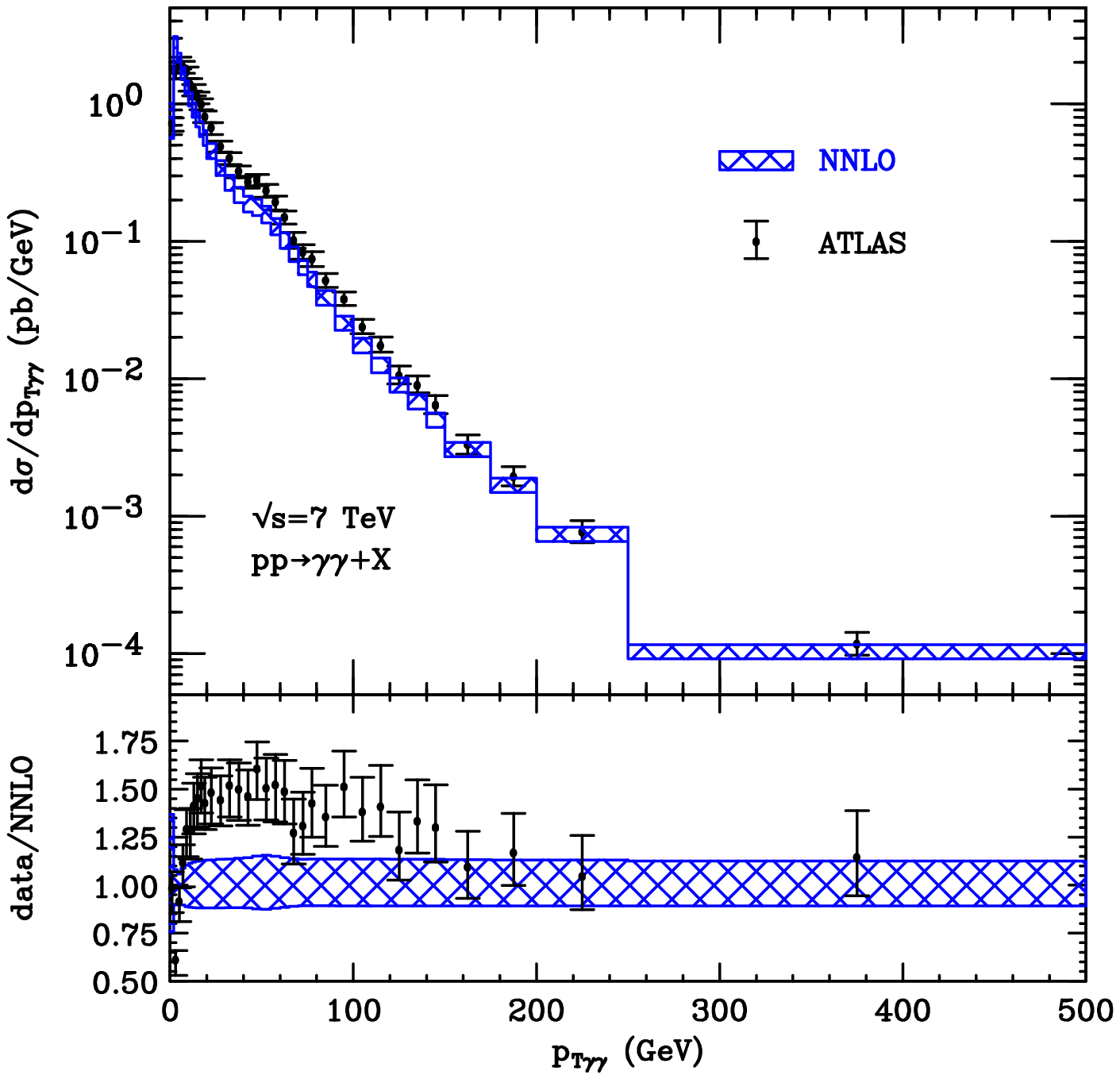}\\
\end{tabular}
\end{center}
\caption{\label{fig:dpdata}
{\em Comparison between ATLAS data at ${\sqrt s}=7$~TeV \cite{Aad:2012tba} 
and NNLO results 
(including their scale dependence) for 
$d\sigma/d\Dpgg$ (left) and $d\sigma/dp_{T \gamma \gamma}$ (right).
The NNLO results are corrected for hadronization and underlying event effects
(see text).}}
\end{figure}

The ATLAS Collaboration performs a quantitative estimate of underlying event,
pile-up and hadronization effects. These effects, which are not included in
fixed-order QCD calculations, are taken into account \cite{Aad:2012tba}
by applying a correction factor to the fixed-order QCD results. The typical size
of the correction factor is around 0.95 \cite{Aad:2012tba}. The bin-by-bin
correction factors for various differential cross sections are available in the
database of the Durham HepData Project \cite{Atlasdata}. We apply these
bin-by-bin correction factors to our NNLO calculations (for simplicity, we
simply rescale the NNLO central values by the bin-by-bin corrections, without
including the uncertainties of the correction factors themselves), and we
present the ensuing results in Figs.~\ref{fig:mcdata} and \ref{fig:dpdata}. 

A technical comment about Fig.~\ref{fig:mcdata} regards the differential cross
section $d\sigma/d\cos \theta^{*}$. Since the two photons of the diphoton pair
are identical particles, $d\sigma/d\cos \theta^{*}$ is symmetric with respect to
$\cos \theta^{*} \leftrightarrow -\cos \theta^{*}$. In our QCD results
throughout this paper we have always computed the differential cross section
with respect to $|\cos \theta^{*}|$ and then the results are presented in
symmetrized form. In the ATLAS data, $\cos \theta^{*}$ refers to the cosine of
the polar angle $\theta^{*}$ of the harder photon (the photon with  momentum
$p_{T \gamma}^{hard}$) and, consequently (because of experimental effects in the measurement, including statistical fluctuations), $d\sigma/d\cos \theta^{*}$ is not exactly symmetric under 
$\cos \theta^{*} \leftrightarrow -\cos \theta^{*}$ (although it is symmetric
within the experimental errors). The NNLO result in Fig.~\ref{fig:mcdata}-right
is not exactly symmetric because of the non-symmetric correction factors that
have been applied.

A main general comment about the comparison between data and NNLO results
regards the photon isolation criteria. The ATLAS data use standard isolation
(though the experimental details of the actual isolation procedure are quite
involved and differ from a plain implementation of standard isolation),
whereas the NNLO results use smooth isolation with the {\em same} values of the
isolation parameters $R$ and $\ETmax$. In our opinion the comparison is
meaningful despite the differences between the isolation criteria. This follows
from several observations that we list. Owing to the physical constraint in
Eq.~(\ref{eq:a}),
NNLO results for smooth isolation give a lower bound on the NNLO results for
standard isolation.
Smooth and standard isolation results are quantitatively quite similar starting
from computations at `effective' NLO accuracy (see Sects.~\ref{sec:sigmatot}
and \ref{sec:diffNLO}), and such similarity is expected to remain valid at NNLO
(see, e.g., the mild dependence on isolation parameters that we noticed 
in Sect.~\ref{sec:ispar}). Calculations within both isolation criteria are
affected by sizeable higher-order corrections, and at `effective' NLO accuracy
these corrections tend to be larger than the differences between the NLO 
results for smooth and standard isolation (see 
Sects.~\ref{sec:sigmatot}--\ref{sec:diffNLO}).
Therefore, neglecting NNLO effects can have more impact than using different
isolation criteria. Reference \cite{Aad:2012tba} compares the ATLAS data to both
NNLO results for smooth isolation and lower-order results for standard
isolation: the differences between these fixed-order results confirm that NNLO
corrections are sizeable and relevant. Obviously all these observations about
smooth and standard isolation results are valid within the corresponding
perturbative uncertainties.

We present more specific comments about total and differential cross sections.

The experimental value of the total cross section is \cite{Aad:2012tba}
$\sigma_{ATLAS}= 44.0~^{+3.2}_{-4.2}$~pb.
We compute the corresponding NNLO result from the bin-by-bin corrected
differential cross section $d\sigma/d\cos \theta^{*}$
(Fig.~\ref{fig:mcdata}-right), and we obtain
$\sigma_{ATLAS}^{NNLO}= 37.2~^{+3.2}_{-3.3} \rm{(scale)}$~pb.
The measured value and the NNLO result are consistent within the
corresponding experimental and scale dependence uncertainties.
Moreover, as remarked in Sect.~\ref{sec:totalNNLO},
the NNLO scale dependence cannot be regarded as a consistent estimate of the
perturbative uncertainty due to uncalculated higher-order terms. The `true'
perturbative uncertainty of $\sigma^{NNLO}$ is larger than the scale dependent
effect that we have computed. A better uncertainty estimate can be obtained, for
instance, by considering enlarged scale variations, such that the ensuing NNLO
and NLO results have some degree of overlap. An alternative, similar and simpler
procedure consists in comparing the NNLO and NLO results at central values of
the scales (incidentally, we have checked that NLO results for smooth and
standard isolation are quantitatively very similar) and using half of the
difference between them to assign the perturbative uncertainty of the NNLO
result. This procedure leads (see Table~\ref{Table:total})
to an NNLO uncertainty of about $\pm 14$\%, which amounts to almost double the
scale uncertainty of the NNLO result for the total cross section.

The ATLAS data \cite{Aad:2012tba} on differential cross sections are reported in
Figs.~\ref{fig:mcdata} and \ref{fig:dpdata} together with the corresponding NNLO
results. In the lower subpanels of these figures we present the ratio between
the data and the NNLO results at central scales, and the scale dependence of the
NNLO results.

At large values of $\Dpgg$ (e.g., $\Dpgg \gtap 2.8$) and small values of
$p_{T \gamma \gamma}$ (e.g., $p_{T \gamma \gamma} \ltap 20$~GeV) the shape of
the data (Fig.~\ref{fig:dpdata}) is definitely different from that of the NNLO
results. This is expected since, as discussed in Sect.~\ref{sec:diffNNLO}, the
shape of the NNLO results is not physically reliable in these Sudakov sensitive
regions. All-order resummed calculations 
\cite{Balazs:2007hr,Cieri:2015rqa,Aaboud:2017vol}
have to be used here. All-order resummation effects are implemented also in
parton shower event generators, which can 
lead to a consistent description
of the data \cite{Aad:2012tba,Aaboud:2017vol}.

The NNLO results in Fig.~\ref{fig:dpdata} (outside the Sudakov sensitive
regions) and those in the low-mass region ($M_{\gamma \gamma} < 50$~GeV)
of Fig.~\ref{fig:mcdata}-left
are perturbative results at `effective' NLO accuracy, while those in
the high-mass region and in the entire region of $\cos \theta^{*}$
(Fig.~\ref{fig:mcdata}) are results at `effective' NNLO accuracy.
In `effective' NNLO regions there is a good agreement between data and NNLO
results within the corresponding experimental and scale uncertainties.
In `effective' NLO regions the data tend to be systematically higher than the
NNLO results at central values of the scales. Note that the shape of the
differences between data and NNLO results tends to qualitatively follow the
shape of the NNLO $K$ factors (see the lower subpanels in 
Figs.~\ref{fig:mcdata} and \ref{fig:dpdata} and those in 
Figs.~\ref{fig:mlhc}, \ref{fig:coslhc} and \ref{fig:phiptlhc}).
This feature is consistent with our previous remarks on the fact
(see the differences
between NLO and NNLO results in Figs.~\ref{fig:mlhc}, \ref{fig:coslhc} 
and \ref{fig:phiptlhc}) that the computed scale dependence at NNLO does
underestimate the true perturbative uncertainties of the NNLO results.
If we proceed to assign a perturbative uncertainty on the basis of the
differences between NNLO and NLO results for differential cross sections
(analogously to the procedure that we have previously mentioned 
for the total cross section in this subsection), 
the NNLO uncertainty is almost doubled with respect to the NNLO scale
dependence in most of the regions of Figs.~\ref{fig:mcdata} and 
\ref{fig:dpdata} and it is further increased in limited extreme regions (such as
at very small values of $M_{\gamma \gamma}$, at very large values of 
$|\cos \theta^{*}|$ and at very low values of $\Dpgg$).
Such an NNLO uncertainty makes the ATLAS data consistent with the NNLO results in
both `effective' NNLO and `effective' NLO regions.

The tendency of the data to be systematically higher than the
NNLO results at central values of the scales is not inconsistent with the
expectation (see Eq.~(\ref{eq:a})) that the NNLO results with smooth isolation
should be  a lower bound on the corresponding results for standard isolation. One can
try to reduce the differences between the smooth and standard isolation criteria
by decreasing the isolation effects within smooth isolation. This can be done
by using different smooth isolation parameters such as, for instance, a smaller
value of the power $n$ of the isolation function $\chi(r;R)$ or a larger
value of $\ETmax$.
As shown by the results in Sect.~\ref{sec:ispar}, such a procedure can reduce
the systematic differences between ATLAS data and NNLO smooth isolation results.
Nonetheless, 
this `tuning' procedure
does not affect the overall features of the comparison
between data and NNLO results, because of the relatively-small dependence on the
isolation parameters (see Sect.~\ref{sec:ispar}) and of the substantial NNLO
theoretical uncertainties that we have previously discussed.

In this subsection we have considered the ATLAS diphoton data at ${\sqrt
s}=7$~TeV and we have presented comments on the main features of the data/NNLO
comparison. The same features and comments are equally valid for other LHC
diphoton measurements \cite{Chatrchyan:2014fsa,Aaboud:2017vol}
and related data/theory comparisons
\cite{Chatrchyan:2014fsa,Aaboud:2017vol,Campbell:2016yrh}.
In particular, the inclusion of the NNLO radiative corrections greatly improves
the description of the data \cite{Aad:2012tba,Chatrchyan:2014fsa,Aaboud:2017vol}
with respect to lower-order results. The effect of the NNLO corrections is from
moderate to sizeable in different kinematical regions. This leads to a good or
consistent (depending on kinematical regions) agreement with the data by taking
into account the perturbative uncertainty of the NNLO results.

\subsection{Asymmetric and symmetric photon $p_T$ cuts}
\label{sec:ptcut}

Throughout the paper we have remarked that the presence of the photon 
$p_T$ cuts ($p_{T \gamma}^{hard} \geq p_H, \;p_{T \gamma}^{soft} \geq p_S$)
has relevant effects on various features of diphoton production observables.
All the numerical results presented so far (with the exception of those
in Fig.~\ref{fig:dmnew})
use the values $p_H=25$~GeV and $p_S=22$~GeV. In this subsection we present 
some results and related comments on effects that are due to variations of
$p_H$ and $p_S$. In particular, we consider symmetric $p_T$ cut configurations
($p_H=p_S$) in addition to asymmetric configurations ($p_H > p_S$).

We consider the reference kinematical configuration (and theoretical setup)
used throughout this section, but we vary the photon $p_T$ cuts
$p_H$ and $p_S$. All the numerical results presented in this subsection refer
to the central value of the scales 
($\mu_R=\mu_F=\mu_0={\sqrt {M_{\gamma \gamma}^2 +  p_{T \gamma \gamma}^2}}\,\;$).

\begin{figure}[htb]
\begin{center}
\begin{tabular}{cc}
\includegraphics[width=0.48\textwidth]{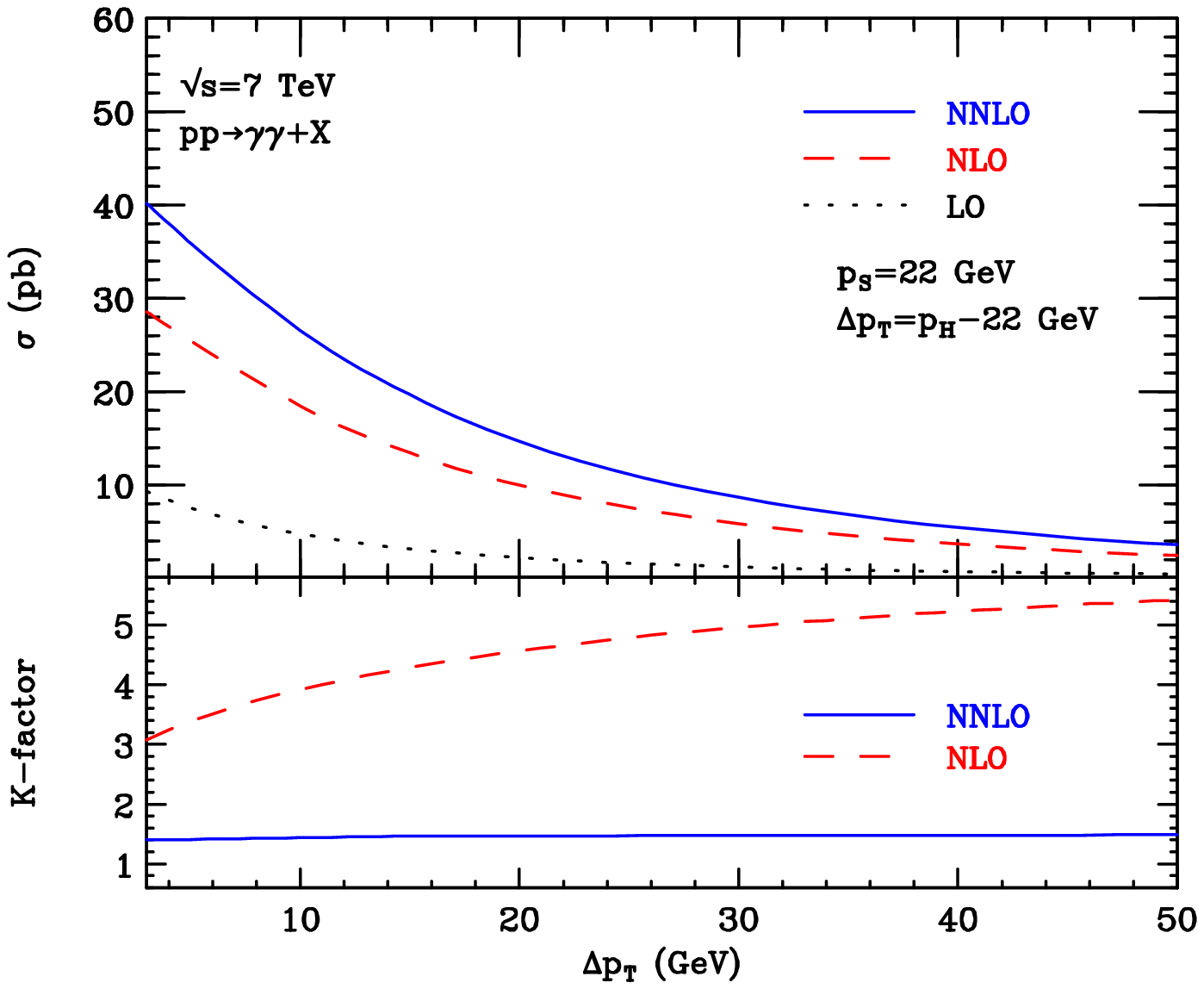}
& \includegraphics[width=0.48\textwidth]{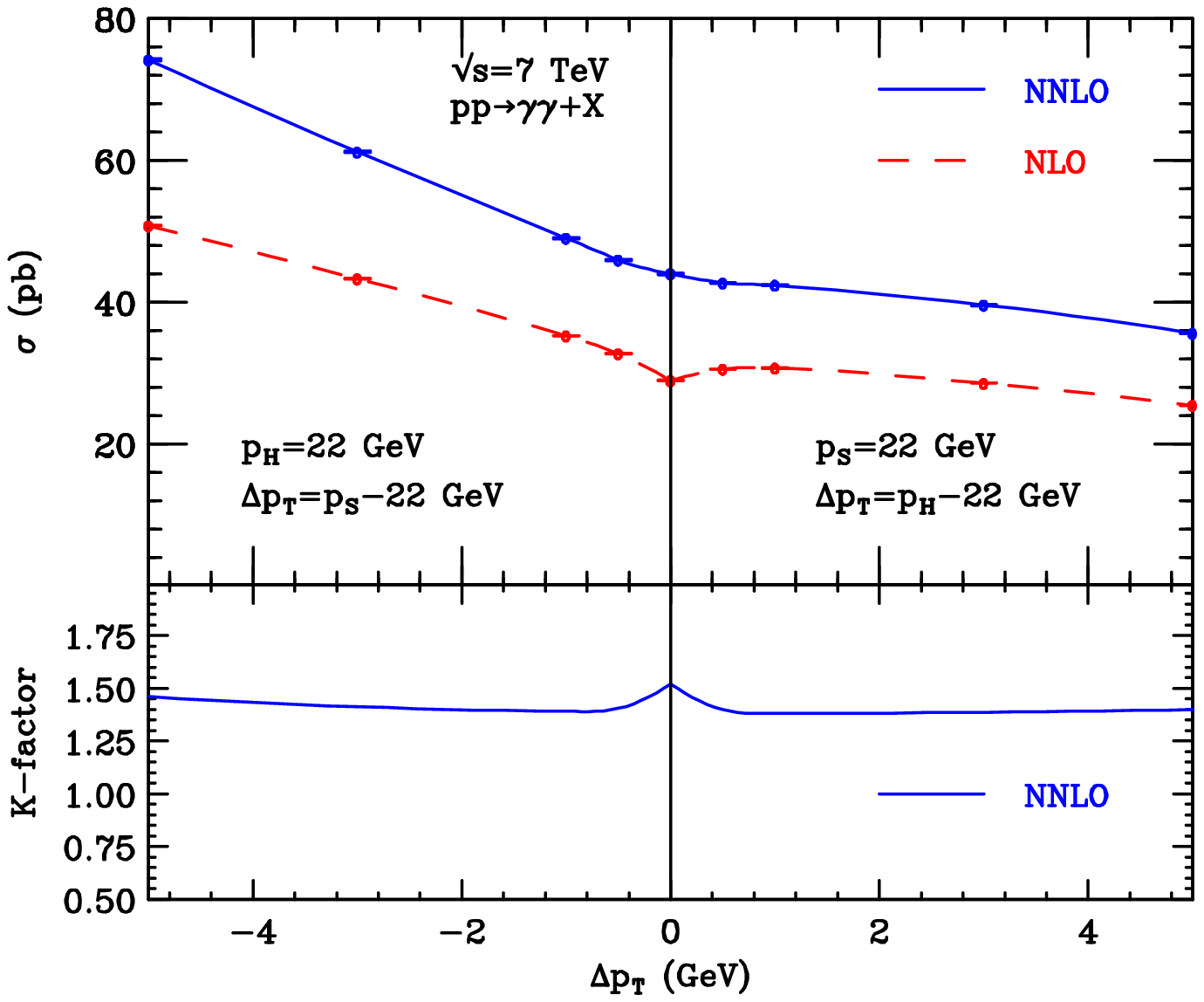}
\\
\end{tabular}
\end{center}
\caption{\label{fig:deltapt}
{\em Dependence on the photon $p_T$ cuts $p_H$ and $p_S$ 
($p_{T \gamma}^{hard} \geq p_H, 
\;p_{T \gamma}^{soft} \geq p_S$) of the total cross section for diphoton
production (the other kinematical cuts are the same as in 
Table~\ref{Table:total}).
The NLO (red dashed) and NNLO (blue solid) results are separately presented in the
regions of asymmetric (left panel) and nearly-symmetric (right panel) $p_T$ cuts.
The LO result (black dotted) is presented only in the case of asymmetric
$p_T$ cuts (left panel).
The lower subpanels present the corresponding NLO (red dashed) and 
NNLO (blue solid) $K$ factors.
}}
\end{figure}

We begin our quantitative study by fixing $p_S=22$~GeV and varying the value
of $p_H$. The dependence on $p_H$ is parametrized by 
$\Delta p_T = p_H - 22$~GeV.
The numerical results for the total cross section at LO, NLO and NNLO in the range
$3~{\rm GeV} < \Delta p_T < 50~{\rm GeV}$ are reported in 
Fig.~\ref{fig:deltapt}-left.
At each of these perturbative orders we note that the total cross section $\sigma(\Delta p_T)$
monotonically decreases by increasing $p_H$ (i.e, the value of $\Delta p_T$),
in agreement with physical expectations. We also note that the size of the NNLO
corrections, as given by the $K$ factor 
$K^{NNLO}(\Delta p_T) = \sigma^{NNLO}(\Delta p_T)/\sigma^{NLO}(\Delta p_T)$
(lower subpanel of Fig.~\ref{fig:deltapt}-left), is very weakly dependent on 
$\Delta p_T$: it varies in the range between 1.4 and 1.5\,.

This weak dependence on $\Delta p_T$ of the NNLO radiative corrections has to be contrasted with a corresponding strong dependence of the NLO radiative corrections.
In the results of Fig.~\ref{fig:deltapt}-left the NLO $K$ factor
$K^{NLO}(\Delta p_T) = \sigma^{NLO}(\Delta p_T)/\sigma^{LO}(\Delta p_T)$
is much larger than $K^{NNLO}(\Delta p_T)$ and it sizeably increases by 
increasing $\Delta p_T$. We have $K^{NLO} \sim 3$ at $\Delta p_T \sim 3$~GeV,
$K^{NLO} \sim 4$ at $\Delta p_T \sim 10$~GeV, and
$K^{NLO} \sim 5.4$ at $\Delta p_T \sim 50$~GeV. At the LO, values of 
$p_{T \gamma}^{soft}$ smaller than $p_H$ are kinematically forbidden, and they do not contribute to $\sigma^{LO}$. At the NLO the kinematical subregion with
$p_S < p_{T \gamma}^{soft} < p_H$ is allowed: by increasing its extension
(i.e. by increasing $\Delta p_T=p_H -p_S$) this subregion produces increasingly large
NLO corrections. The $p_{T \gamma}^{soft}$ region where 
$p_S < p_{T \gamma}^{soft} < p_H$ is kinematically allowed at both NLO and NNLO,
and the NNLO $K$ factor turns out to be weakly dependent on the asymmetry of the photon $p_T$ cuts.

We now move to consider the total cross section in configurations with symmetric
($p_H=p_S$) or nearly-symmetric ($p_H \sim p_S$) $p_T$ cuts. As a reference
symmetric configuration we consider the case with $p_H=p_S=22$~GeV.
Nearly-symmetric configurations can be obtained by either increasing 
$p_H$ at fixed $p_S=22$~GeV (we define $\Delta p_T = p_H - 22~{\rm GeV} > 0$) or
decreasing $p_S$ at fixed $p_H=22$~GeV 
(we define $\Delta p_T = p_S - 22~{\rm GeV} < 0$).
The NLO and NNLO results for the total cross section as a function of 
$\Delta p_T$ are reported in Fig.~\ref{fig:deltapt}-right. More precisely,
$\sigma(\Delta p_T)$ is computed at nine values of $\Delta p_T$,
namely $\Delta p_T({\rm GeV})=\{-5,-3,-1,-0.5,0,+0.5,+1,+3,+5\}$, and the results are
reported as data points in Fig.~\ref{fig:deltapt}-right 
(the continuous lines in  Fig.~\ref{fig:deltapt}-right are just a graphical
interpolation between the data points).

According to physical expectations, the total cross section 
$\sigma(\Delta p_T)$ should be a monotonically decreasing function of 
$\Delta p_T$. Indeed, the amount of physical diphoton events that contribute 
to $\sigma(\Delta p_T)$ decreases (increases) by increasing $p_H$ (decreasing
$p_S$). The results in Fig.~\ref{fig:deltapt}-right (especially the NLO results)
do not show such a physical behaviour. In particular, 
$\sigma^{NLO}(\Delta p_T)$ has a local maximum at $\Delta p_T \sim 1$~GeV
and a local minimum at $\Delta p_T = 0$. Considering the NLO results with
$\Delta p_T \geq 0$ in Fig.~\ref{fig:deltapt}-right, the unphysical behaviour
could be ascribed to either the local maximum (a cross section value that is
unphysically large at some finite, though small, value of $\Delta p_T$)
or the local minimum (a cross section value that is
unphysically small at $\Delta p_T=0$). However, the NLO cross section does not
show an evident unphysical (non-monotonic) behaviour at finite and negative
values of $\Delta p_T$. Therefore, we can conclude that the pathological
behaviour of the NLO results in Fig.~\ref{fig:deltapt}-right is located in the
region of symmetric (or nearly-symmetric) $p_T$ cuts ($|\Delta p_T| \to 0$).

\begin{figure}[htb]
\begin{center}
\begin{tabular}{cc}
\hspace*{-4.5mm}
\includegraphics[width=0.49\textwidth]{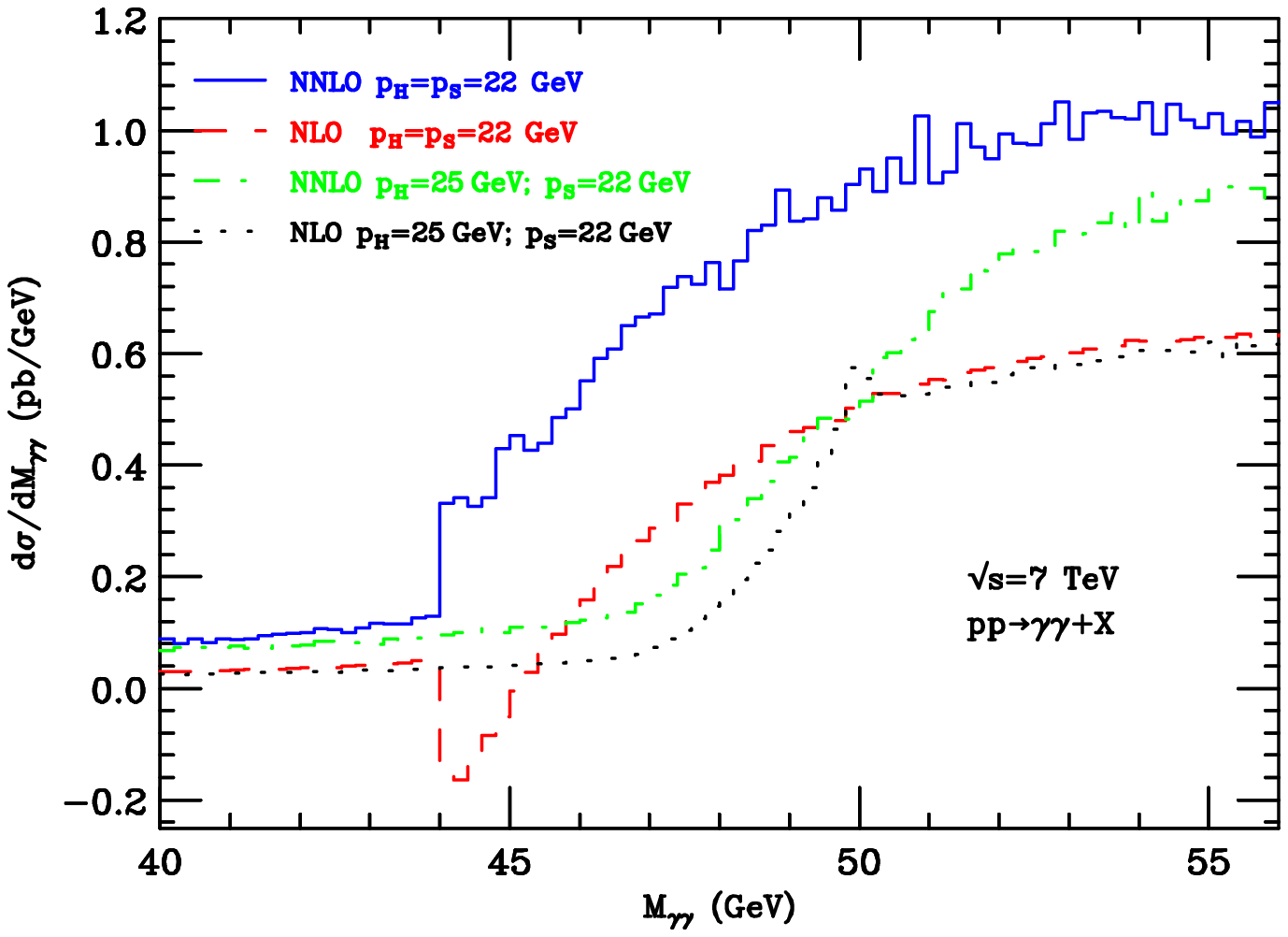}
& \includegraphics[width=0.46\textwidth]{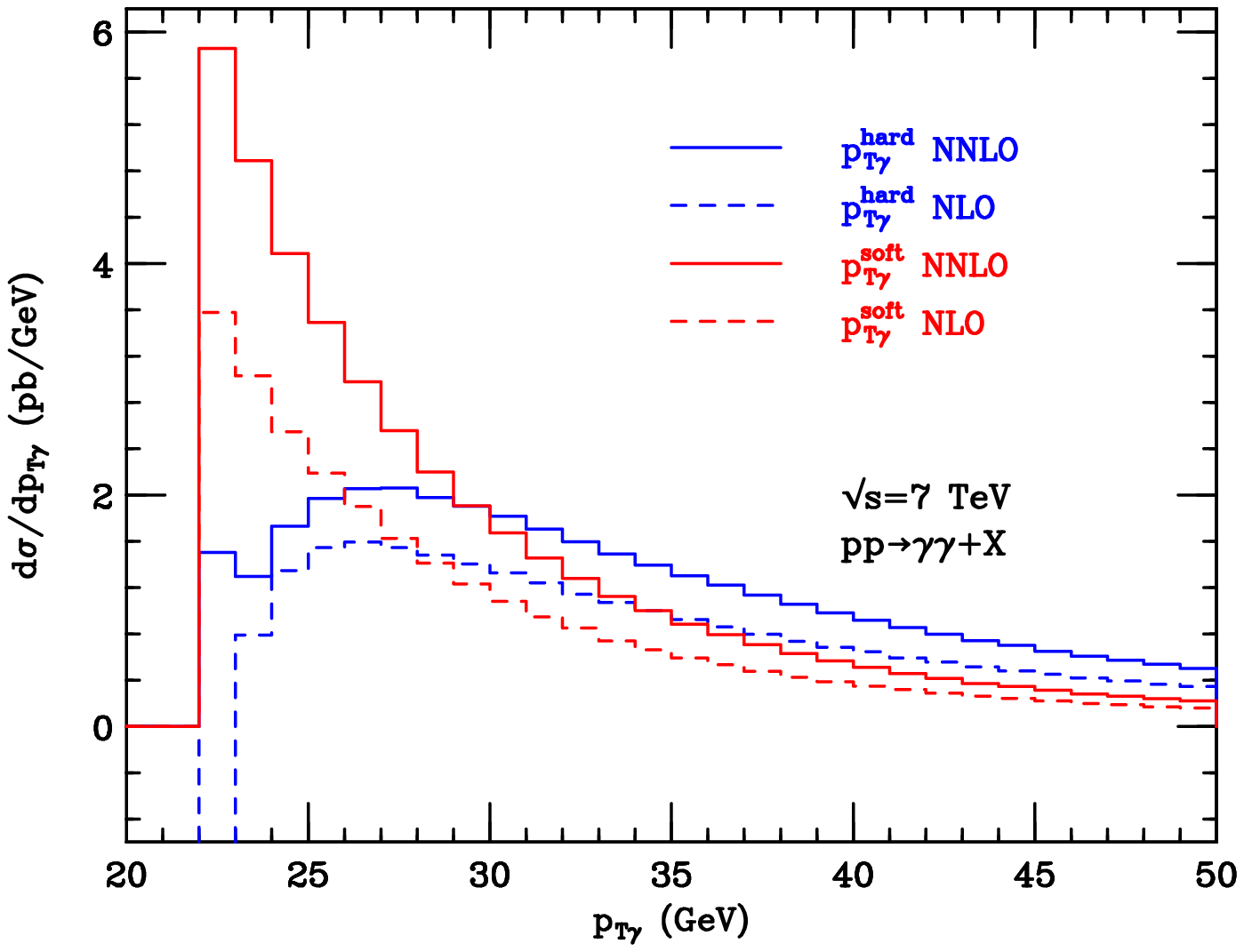}\\
\end{tabular}
\end{center}
\caption{\label{fig:symmMgg}
{\em 
Left panel: 
the differential cross section $d\sigma/dM_{\gamma \gamma}$ in 
two configurations with symmetric and asymmetric photon $p_T$ cuts.
NLO (red dashed) and NNLO (blue solid) results with symmetric cuts ($p_H=p_S=22$~GeV)
and NLO (black dotted) and NNLO (green dash-dotted) results with asymmetric cuts
($p_H=25$~GeV, $p_S=22$~GeV).
Right panel: NLO (dashed) and NNLO (solid) results for the differential cross
sections $d\sigma/dp_{T \gamma}^{hard}$ (blue) and 
$d\sigma/dp_{T \gamma}^{soft}$ (red)
of the harder and softer photon in the configuration with symmetric $p_T$ cuts,
$\,p_H=p_S=22$~GeV.
}}
\end{figure}

The unphysical behaviour of the fixed-order results for the  
diphoton total cross section in the presence of
symmetric $p_T$ cuts is expected, since a similar behaviour was first observed
and discussed in Ref.~\cite{Frixione:1997ks} in the context of dijet
photoproduction. Specifically, in the case of nearly-symmetric cuts
the NLO cross section behaves as
\beq
\label{totsym}
\sigma^{NLO}(p_H, p_S) - \sigma^{NLO}(p_H=p_S) 
\propto + \,(p_H - p_S) \;\ln^2(p_H - p_S) \;\;, 
\quad \quad \quad (p_H \simeq p_S) \;\;.
\eeq
This implies that the local minimum of $\sigma^{NLO}$ at $\Delta p_T = 0$
is unphysical. This also implies that $\sigma^{NLO}$ in 
Fig.~\ref{fig:deltapt}-right has a {\rm cusp} at $\Delta p_T = 0$, 
with a $\Delta p_T$-slope that diverges to $+\infty$ ($-\infty$) if 
$\Delta p_T$ tends to vanish from positive (negative) values of 
$\Delta p_T$. The NLO numerical results in Fig.~\ref{fig:deltapt}-right
are consistent with these features, although the cusp behaviour is not clearly
visible since it is located in a very narrow region of small values of 
$|\Delta p_T|$. The double-logarithmic behaviour in the right-hand side of
Eq.~(\ref{totsym}) exactly follows from the same reasoning as used in 
Ref.~\cite{Banfi:2003jj} in the context of dijet production. 
We note that the behaviour in the right-hand side of Eq.~(\ref{totsym})
is different from the single-logarithmic behaviour in the corresponding 
expression of Eq.~(2.9) of Ref.~\cite{Frixione:1997ks}. The NLO 
single-logarithmic contribution of Ref.~\cite{Frixione:1997ks} is produced
by {\em hard-collinear} radiation from the initial state (see Eq.~(2.8) therein).
The dominant double-logarithmic term in Eq.~(\ref{totsym}), which originates from
the NLO process in Eq.~(\ref{softtree}), is instead due to radiation that is both
{\em soft} and {\em collinear} to the direction of the initial-state colliding 
partons (related comments on this effect are postponed to our discussion 
of the results in Fig.~\ref{fig:symmMgg})

The unphysical behaviour of the total cross section at 
$p_H \simeq p_S$ persists at each subsequent perturbative orders and a physical
(smooth and monotonic) dependence of $\sigma(\Delta p_T)$ on $\Delta p_T$ 
can be recovered only by a proper all-order resummation of soft-gluon effects
\cite{Banfi:2003jj}.
Such a resummation is beyond the scope of the present paper and we limit
ourselves to comment on the quantitative reliability of the fixed-order diphoton
results that we have presented.

By direct inspection of the NLO and NNLO results in Fig.~\ref{fig:deltapt}-right,
we tend to conclude that the onset of the unphysical fixed-order
behaviour due to nearly-symmetric $p_T$ cuts occurs in a region of small values
of $\Delta p_T$, such as  $|\Delta p_T| \ltap 2$~GeV.
Therefore, in the case of asymmetric cuts with 
$p_H - p_S \gtap 3$~GeV, we argue that the unphysical behaviour has little 
effect on the total cross section. 

At smaller values of $p_H - p_S$ we can try to
quantify the effect of the unphysical behaviour. For instance, at each fixed
order we can assume that a `tentative' physical value of the cross section is in
the range between the values of $\sigma(\Delta p_T \sim -2~{\rm GeV})$
and $\sigma(\Delta p_T \sim +2~{\rm GeV})$ at the corresponding order. Then, we
can use the size of this range and the difference with respect to the
`unphysical' computed value of $\sigma(\Delta p_T)$ at small $|\Delta p_T|$
to assign a systematic theoretical uncertainty to this computed value. From the
results in Fig.~\ref{fig:deltapt}-right, this procedure leads to effects of about
10\%--15\% at NLO and of several percent at NNLO.
These quantitative effects can be regarded as a rough estimate of the uncertainty
due to the unphysical soft-gluon effects at fixed orders. We note that such
uncertainty quantitatively decreases by increasing the perturbative order.
This decrease is expected since the unphysical behaviour is located
in an increasingly smaller region of $\Delta p_T$ by increasing the perturbative
order (the effect is visible in the NLO and NNLO results of 
Fig.~\ref{fig:deltapt}-right). We also note that such uncertainty is
quantitatively similar to the scale dependence of the cross section at the
corresponding order (see Table~\ref{Table:total}).
More importantly for `practical' purposes, such uncertainty is definitely smaller
than the typical size (about 40\%) of the NNLO corrections
(the value of the NNLO $K$ factor in the lower subpanel of 
Fig.~\ref{fig:deltapt}-right increases very slightly, from 1.4 to 1.5, 
toward the region where $\Delta p_T \sim 0$). This implies that even in the case
of nearly-symmetric $p_T$ cuts  the bulk of the NNLO corrections to the total
cross section is due to hard-parton radiation rather than to unphysical
soft-gluon effects. This also implies (as we have discussed throughout this
section and, in particular, in Sect.~\ref{sec:datavsth}) 
that the perturbative uncertainty of the NNLO result for the total cross section
is dominated by the
effect of the large NNLO radiative corrections.

From our discussion of the results in Fig.~\ref{fig:deltapt}, we can make an 
overall comment about the total cross section at NNLO:
the size of the NNLO corrections and of the NNLO theoretical uncertainties
weakly depends on the amount $\Delta p_T$ of the asymmetry of the photon 
$p_T$ cuts. Some validation of this overall comment can be found in features 
of data/theory comparisons. Indeed, LHC measurements of diphoton total cross 
sections have been performed in configurations with different values of 
$\Delta p_T$ 
($\Delta p_T= 3$~GeV \cite{Aad:2012tba}, 
10~GeV \cite{Aaboud:2017vol} and 15~GeV \cite{Chatrchyan:2014fsa}),
and in all these configurations the comparison between data and NNLO results
(see Sect.~\ref{sec:datavsth} and 
Refs.~\cite{Aad:2012tba,Chatrchyan:2014fsa,Aaboud:2017vol,Campbell:2016yrh})
shows a similar degree of consistency.

We further study the effects produced by symmetric $p_T$ cuts by considering some
differential cross sections. Analogously to the case of the total cross section,
we consider the symmetric $p_T$ cut configuration with $p_H=p_S=22$~GeV and we
compute the $M_{\gamma \gamma}$-differential cross section 
$d\sigma/dM_{\gamma \gamma}$ and the inclusive transverse-momentum spectra 
$d\sigma/dp_{T \gamma}^{hard}$ and $d\sigma/dp_{T \gamma}^{soft}$
of the harder and softer photon.
The numerical results at NLO and NNLO are presented in Fig.~\ref{fig:symmMgg}.
Incidentally, we note that the main 
features of the results in Figs.~\ref{fig:deltapt} and \ref{fig:symmMgg}
do not depend on the specific value $p_H=p_S=22$~GeV of the symmetric $p_T$ cuts
(we have explicitly checked this by considering values of 
$p_H=p_S$ in the range between 20~GeV and 30~GeV).

In Fig.~\ref{fig:symmMgg}-left we report the results for 
$d\sigma/dM_{\gamma \gamma}$ in the $M_{\gamma \gamma}$ region that is close to the
LO threshold at $M_{\gamma \gamma} = M^{LO}_{\rm dir}$. For comparison, we 
present the results for two different configurations with asymmetric  
($p_H=25~{\rm GeV}, \,p_S=22~{\rm GeV}$) and symmetric 
($p_H=p_S=22~{\rm GeV}$) $p_T$ cuts.
We remark that the two results are obtained by only varying the value of $p_H$,
while all the other kinematical cuts and parameters of the calculation are
unchanged.

We first briefly comment on the case with asymmetric $p_T$ cuts.
The NLO and NNLO results in Fig.~\ref{fig:symmMgg}-left exactly correspond to
those in Figs.~\ref{fig:mlhc}-left and \ref{fig:mcdata}-left, the only difference
being the much smaller $M_{\gamma \gamma}$ bin size, which is equal to 0.2~GeV.
The behaviour of the LO and NLO results for $d\sigma/dM_{\gamma \gamma}$
has been discussed in detail at the end of Sect.~\ref{sec:diffNLO}
(see Fig.~\ref{fig:dmth}). In particular, the LO result has a threshold at 
$M_{\gamma \gamma} = M^{LO}_{\rm dir}=50$~GeV and the NLO result has an upward
double-side cusp at $M_{\gamma \gamma} = M^{LO}_{\rm dir}$. The NLO cusp behaviour
is due to soft-gluon radiation effects \cite{Catani:1997xc}. The `unphysical'
soft-gluon effects persist at the NNLO level and their dominant (at the formal
level) contribution leads to a negative double-logarithmic enhancement
($\propto - \ln^2(\epsilon_M)$) of the NLO cusp behaviour in Eq.~(\ref{nloth}).
This double-logarithmic contribution produces a downward double-side cusp
in the NNLO result at $M_{\gamma \gamma} = M^{LO}_{\rm dir}$. The NNLO cusp is very
narrow and its effect is (partly) smeared by the bin size in the results of
Fig.~\ref{fig:symmMgg}-left. Nonetheless, this effect is still visible in 
Fig.~\ref{fig:symmMgg}-left since the shapes of the NLO and NNLO results are quite
different in the vicinity of $M_{\gamma \gamma} \sim 50$~GeV.

The shape of $d\sigma/dM_{\gamma \gamma}$ is quite different in the two
configurations with asymmetric and symmetric $p_T$ cuts 
(Fig.~\ref{fig:symmMgg}-left). The LO result with symmetric $p_T$ cuts
is not shown in Fig.~\ref{fig:symmMgg}-left. Its shape is exactly
similar to that with asymmetric $p_T$ cuts (see Fig.~\ref{fig:dmth} and
Eq.~(\ref{loth}))
since the value of $p_S$ does not matter at the LO and, in particular,
$d\sigma^{LO}/dM_{\gamma \gamma}$ has its threshold at 
$M_{\gamma \gamma} = M^{LO}_{\rm dir}=44$~GeV (since $p_H=22$~GeV).
In the case of symmetric $p_T$ cuts, the NLO and NNLO values of 
$d\sigma/dM_{\gamma \gamma}$ are quite small below the LO threshold.
Just above the LO threshold ($M_{\gamma \gamma} \gtap M^{LO}_{\rm dir}$),
the NLO result is (relatively) large and {\em negative} and the NNLO result
is also (relatively) large but {\em positive}. 
This NLO behaviour is unphysical.
In particular, the 
relatively-large and negative differential cross section
$d\sigma^{NLO}/dM_{\gamma \gamma}$ at $M_{\gamma \gamma} \sim 44$~GeV
`explains' the unphysical behaviour of the NLO total cross section at
$\Delta p_T = 0$. In the case of symmetric (or nearly-symmetric) $p_T$ cuts
$d\sigma^{NLO}/dM_{\gamma \gamma}$ becomes negative at 
$M_{\gamma \gamma} \sim  M^{LO}_{\rm dir}$ and, after integration over 
$M_{\gamma \gamma}$, this negative contribution is responsible for the 
decreasing behaviour (see Fig.~\ref{fig:deltapt}-right)
of the NLO total cross section at small values of $\Delta p_T$ 
($0 \ltap \Delta p_T \ltap 1$~GeV).
We note that also $d\sigma/dM_{\gamma \gamma}$ (and not only the total cross
section) for symmetric $p_T$ cuts is physically expected to be larger than the
corresponding differential cross section for asymmetric $p_T$ cuts. This physical
expectation is not fulfilled by the NLO results of 
Fig.~\ref{fig:symmMgg}-left in the regions where 
$44~{\rm GeV} \ltap  M_{\gamma \gamma} \ltap 45.4~{\rm GeV}$ and
$M_{\gamma \gamma} \sim 50$~GeV (although both differential cross sections have 
a positive value
at $M_{\gamma \gamma} \sim 45.4~{\rm GeV}$ and $M_{\gamma \gamma} \sim 50$~GeV).

Despite their apparent shape difference, the fixed-order behaviour of
$d\sigma/dM_{\gamma \gamma}$ at $M_{\gamma \gamma} \sim M^{LO}_{\rm dir}$
for asymmetric and symmetric $p_T$ cuts is produced by the {\em same} underlying
mechanism \cite{Catani:1997xc}, as we are going to discuss below. 
In this sense,
the unphysical fixed-order behaviour of the total cross section at  
$p_H \simeq p_S$ can also be regarded as a consequence (after integration) of
unphysical soft-gluon effects for non-smooth differential distributions
\cite{Catani:1997xc} (e.g., $d\sigma/dM_{\gamma \gamma}$ and, also,
$d\sigma/dp_{T \gamma}^{hard}$ and $d\sigma/dp_{T \gamma}^{soft}$
as discussed below).

Analogously to the case of asymmetric $p_T$ cuts (see Eq.~(\ref{nloth})),
in the case of symmetric $p_T$ cuts we have examined the NLO shape of 
$d\sigma/dM_{\gamma \gamma}$ at $M_{\gamma \gamma} \sim M^{LO}_{\rm dir}$
in analytic form, and we find the dominant behaviour
\beq
\label{msym}
\frac{d\sigma^{NLO}}{dM_{\gamma \gamma}} = b_0 - 
{\sqrt {\epsilon_M}}   
\;b_{(+)} \,\ln^2\left( \frac{1}{\epsilon_M}\right)
\Theta(M_{\gamma \gamma} - M^{LO}_{\rm dir})  +\; \dots \;, 
\quad \quad
(p_H=p_S, M_{\gamma \gamma} \sim M^{LO}_{\rm dir}) \;,
\eeq
where $b_0$ and $b_{(+)}$ are positive constants (i.e., they do not depend on  
$M_{\gamma \gamma}$) and the dots in the right-hand side denote subdominant
contributions in the limit $M_{\gamma \gamma} \to M^{LO}_{\rm dir}$.
According to Eq.~(\ref{msym}), $d\sigma^{NLO}/dM_{\gamma \gamma}$ is finite
at $M_{\gamma \gamma} = M^{LO}_{\rm dir}$. Its behaviour just above the LO
threshold ($M_{\gamma \gamma} > M^{LO}_{\rm dir}$) is analogous to that in
Eq.~(\ref{nloth}) and, in particular, the first derivative of 
$d\sigma^{NLO}/dM_{\gamma \gamma}$ with respect to $M_{\gamma \gamma}$
(i.e., the slope of $d\sigma^{NLO}/dM_{\gamma \gamma}$) {\em diverges} to 
$- \infty$. Therefore, $d\sigma^{NLO}/dM_{\gamma \gamma}$ has an upward cusp at
$M_{\gamma \gamma} = M^{LO}_{\rm dir}$. At variance with Eq.~(\ref{nloth}),
the behaviour of the result in Eq.~(\ref{msym}) is smooth just below the LO
threshold ($M_{\gamma \gamma} < M^{LO}_{\rm dir}$) and, in particular, the slope
of $d\sigma^{NLO}/dM_{\gamma \gamma}$ is {\em finite} (subdominant terms of 
${\cal O}(\epsilon_M)$ are neglected in the right-hand side of Eq.~(\ref{msym}))
in this $M_{\gamma \gamma}$ region. Therefore, 
$d\sigma^{NLO}/dM_{\gamma \gamma}$ has a single-side cusp at 
$M_{\gamma \gamma} = M^{LO}_{\rm dir}$ (rather than a double-side cusp as in the
case of asymmetric $p_T$ cuts). 

The NLO quantitative results in Fig.~\ref{fig:symmMgg}-left 
for the symmetric $p_T$ cut configuration are consistent with
the analytic behaviour in Eq.~(\ref{msym}). We also note 
(Fig.~\ref{fig:symmMgg}-left) that the NLO value of 
$d\sigma/dM_{\gamma \gamma}$ at $M_{\gamma \gamma} = M^{LO}_{\rm dir}=44$~GeV
is very small and, in particular, this implies that $b_0$ in Eq.~(\ref{msym})
is much smaller than $a_0$ in Eq.~(\ref{nloth}). This is not unexpected since
in the case of symmetric $p_T$ cuts ($p_H=p_S$) the LO threshold  
$M^{LO}_{\rm dir}$ and the `approximate' threshold $M^{LO}$ for hard radiation
(see Eq.~(\ref{lothres})) 
actually coincides. As a consequence of the small value of
$b_0$, the upward single-side cusp drives $d\sigma^{NLO}/dM_{\gamma \gamma}$
to negative values in the region just above the LO threshold. By further
increasing $M_{\gamma \gamma}$, the physical (positive) behaviour of 
$d\sigma/dM_{\gamma \gamma}$ sets in and, consequently, 
$d\sigma^{NLO}/dM_{\gamma \gamma}$ has a local minimum (with a negative value)
in the vicinity of $M^{LO}_{\rm dir}$.

The NLO behaviours in Eqs.~(\ref{nloth}) and (\ref{msym}) (though they are
partly different) are directly related. Indeed, they are both due to the
non-smooth behaviour of $d\sigma^{LO}/dM_{\gamma \gamma}$ at 
$M_{\gamma \gamma} = M^{LO}_{\rm dir}$ that produces an unbalance between real
and virtual soft-gluon effects at higher perturbative orders 
\cite{Catani:1997xc}. The mechanism that leads to both 
Eqs.~(\ref{nloth}) and (\ref{msym}) is analogous (see the discussion that
accompanies Eq.~(\ref{nloth}) in Sect.~\ref{sec:diffNLO}) and the main important
difference regards the role of the {\em real} soft-emission contribution in the
process of Eq.~(\ref{softtree}) (owing to transverse-momentum conservation the
one-loop virtual contribution to the process $q{\bar q} \to \gamma \gamma$
is independent of the value of $p_S$ and, consequently, it is insensitive to
the difference between asymmetric and symmetric $p_T$ cuts).
The radiated soft gluon produces a transverse-momentum unbalance 
($p_{T \gamma}^{hard} \neq p_{T \gamma}^{soft}$) between the two photons
and it can preferably lead to either a decrease of $p_{T \gamma}^{soft}$ 
or an increase of $p_{T \gamma}^{hard}$ with respect to the LO configuration
($p_{T \gamma}^{hard} = p_{T \gamma}^{soft}= p_H$)
depending on the phase space that is available in the presence of the photon
$p_T$ cuts.
In the case of asymmetric $p_T$ cuts and 
$M_{\gamma \gamma} \sim M^{LO}_{\rm dir}$, the soft-gluon momentum recoil is
`absorbed' by the softer photon (i.e., the value of $p_{T \gamma}^{soft}$
decreases below its LO value $p_{T \gamma}^{soft}=p_H$, whereas 
$p_{T \gamma}^{hard} \simeq p_H$)
and this produces diphoton events with 
$M_{\gamma \gamma} < M^{LO}_{\rm dir}$ (see the accompanying comments to 
Eq.~(\ref{softtree})). In the case of symmetric $p_T$ cuts, the
momentum $p_{T \gamma}^{soft}$ of the softer photon cannot decrease since it
is always constrained to be
larger than $p_H=p_S$ (as in an LO configuration) and, consequently, the
soft-gluon momentum recoil is necessarily `absorbed' by the 
harder photon\footnote{This observation is also relevant 
for our subsequent discussion of the
results in Fig.~\ref{fig:symmMgg}-right.}.
The momentum $p_{T \gamma}^{hard}$ tends to increase above its threshold
value $p_H$ ($p_{T \gamma}^{hard} > p_H$), whereas 
$p_{T \gamma}^{soft} \simeq p_H$,
and this leads to diphoton events with
$M_{\gamma \gamma} > M^{LO}_{\rm dir}$. Therefore, below the LO threshold
($M_{\gamma \gamma} < M^{LO}_{\rm dir}$) there are no dominant real and virtual
soft-gluon effects (here $d\sigma^{NLO}/dM_{\gamma \gamma}$ is smooth since it
only receives contributions from hard radiation), whereas just above the LO
threshold ($M_{\gamma \gamma} > M^{LO}_{\rm dir}$) virtual soft-gluon effects
dominate since real soft-gluon radiation tends to produce diphoton events 
that are
sufficiently far from $M_{\gamma \gamma} = M^{LO}_{\rm dir}$. This real--virtual
kinematical mismatch produces the NLO double-logarithmic enhancement 
(see Eq.~(\ref{msym})) of the non-smooth behaviour of 
$d\sigma/dM_{\gamma \gamma}$ at $M_{\gamma \gamma} \gtap M^{LO}_{\rm dir}$.

In the case of symmetric $p_T$ cuts, the single-side cusp behaviour of 
$d\sigma/dM_{\gamma \gamma}$ at $M_{\gamma \gamma} \sim M^{LO}_{\rm dir}$
occurs at each perturbative order. Just above the LO threshold
the slope of $d\sigma/dM_{\gamma \gamma}$ alternatively diverges to $+ \infty$
or $- \infty$ at subsequent perturbative orders \cite{Catani:1997xc}. This
non-smooth order-by-order behaviour is removed by all-order-resummation of
soft-gluon effects \cite{Catani:1997xc}. After resummation, 
$d\sigma/dM_{\gamma \gamma}$ rapidly increases just above the LO threshold
but it has a finite slope at $M_{\gamma \gamma} \sim M^{LO}_{\rm dir}$. In the
context of fixed-order calculations, the unphysical shape of 
$d\sigma/dM_{\gamma \gamma}$ at $M_{\gamma \gamma} \sim M^{LO}_{\rm dir}$
is localized in a mass region whose size tends to decrease by increasing the
perturbative order. This is consistent with the results in 
Fig.~\ref{fig:symmMgg}-left. At the NNLO the behaviour of 
$d\sigma/dM_{\gamma \gamma}$ is qualitatively consistent with physical
expectations, with the sole exception of a narrow region very close to
$M_{\gamma \gamma}=44$~GeV, where the divergent (to $+ \infty$) slope of 
$d\sigma^{NNLO}/dM_{\gamma \gamma}$ produces an excess of diphoton events that
is eventually also responsible for the (quantitatively small) unphysical
(increasing) behaviour of the NNLO total cross section as $\Delta p_T \to 0$
(Fig.~\ref{fig:deltapt}-right).

We comment on the results in Fig.~\ref{fig:symmMgg}-right for the
transverse-momentum spectra $d\sigma/dp_{T \gamma}^{soft}$
and $d\sigma/dp_{T \gamma}^{hard}$ of the softer and harder photon.
At the LO we have $p_{T \gamma}^{soft}=p_{T \gamma}^{hard}$
(because of transverse-momentum conservation) and the two spectra are identical.
The LO spectrum (it is not shown in Fig.~\ref{fig:symmMgg}-right) monotonically
decreases in going from the `kinematical' (due to the $p_T$ cuts) lower limit
at  $p_{T \gamma}=p_H$ toward higher values of $p_{T \gamma}$.
In Fig.~\ref{fig:symmMgg}-right we show the NLO and NNLO spectra in the case of
the symmetric $p_T$ cut configuration with $p_H=p_S=22$~GeV. We note that the
shape of $d\sigma/dp_{T \gamma}^{soft}$ is qualitatively unchanged by increasing
the perturbative order, in agreement with the expected physical behaviour.
The NLO and NNLO results for $d\sigma/dp_{T \gamma}^{hard}$ are instead
definitely unphysical in the region that is close to the lower limit
$p_{T \gamma}^{hard}=p_H$. In particular, 
$d\sigma^{NLO}/dp_{T \gamma}^{hard}$ is negative in the first bin closest to 
$p_{T \gamma}^{hard}=p_H=22$~GeV. This unphysical behaviour is due to the same
soft-gluon effects that we have previously discussed in the context of the
near-threshold behaviour of $d\sigma/dM_{\gamma \gamma}$. In the case of
symmetric $p_T$ cuts, the real soft-gluon emission process in 
Eq.~(\ref{softtree}) mostly affects the transverse-momentum recoil of the harder
photon and the value of $p_{T \gamma}^{hard}$ tends to increase above its lower
limit at $p_H=22$~GeV. Therefore, in the region where 
$p_{T \gamma}^{hard} \simeq p_H$, virtual soft-gluon effects tend to be
unbalanced and this produces order-by-order perturbative instabilities of
$d\sigma/dp_{T \gamma}^{hard}$. 
Actually, in the limit 
$p_{T \gamma}^{hard} \to p_H$, the NLO result for 
$d\sigma/dp_{T \gamma}^{hard}$ logarithmically diverges 
to $- \infty$ \cite{Catani:1997xc},
\beq
\label{phard} 
\frac{d\sigma^{NLO}}{dp_{T \gamma}^{hard}} \,
\propto  - \;\ln^2(p_{T \gamma}^{hard} - p_S)
\;\;,
\quad \quad \quad \;\;\; \quad (p_{T \gamma}^{hard} \to p_H = p_S) \;\;,
\eeq 
and the NNLO result diverges to $+ \infty$.
This
divergent behaviour is consistent with the NLO and NNLO results in 
Fig.~\ref{fig:symmMgg}-right (the local minimum of 
$d\sigma^{NNLO}/dp_{T \gamma}^{hard}$ in the second bin closest to 22~GeV
is due to a numerical compensation between dominant and subdominant logarithmic
contributions).

This divergent behaviour of $d\sigma/dp_{T \gamma}^{hard}$ is also consistent
with the total cross section results in Fig.~\ref{fig:deltapt}-right. Indeed,
the total cross section can be obtained by integrating 
$d\sigma/dp_{T \gamma}^{hard}$ and, more specifically, the $\Delta p_T$-slope
of the total cross section is directly related to the $p_T$ spectra of the
photons. Independently of the perturbative order
and of the value of the $p_T$ cuts, we have
\beq
\label{sigmaH}
\frac{\partial \,\sigma(p_H, p_S)}{\partial \, p_H} = - \; 
\left. \frac{d\sigma}{dp_{T \gamma}^{hard}} \,\right|_{p_{T \gamma}^{hard}=p_H} 
\;\;,
\quad \quad \quad \;\;\; \quad (p_{T \gamma}^{soft} \geq p_S) \;\;.
\eeq 
This relation shows that the $\Delta p_T$-slope of the total cross section with
respect to variations of $p_H$ (e.g., provided $\Delta p_T > 0$ in 
Fig.~\ref{fig:deltapt}-right) is equal (but with the opposite sign) to the value
of $d\sigma/dp_{T \gamma}^{hard}$. 
In particular, in the case of symmetric $p_T$ cuts the behaviours in 
Eqs.~(\ref{totsym}) and (\ref{phard}) are fully consistent with each other, since
they are directly related throughout Eq.~(\ref{sigmaH}).

In the case of symmetric $p_T$ cuts, $p_{T \gamma}^{soft}$ is not directly
sensitive to soft-gluon radiation in the region where 
$p_{T \gamma}^{soft} \simeq p_H = p_S$, and there is no evident sign of
unphysical behaviour in the results of Fig.~\ref{fig:symmMgg}-right
for $d\sigma/dp_{T \gamma}^{soft}$.
In other words, the soft-gluon instabilities 
of $p_{T \gamma}^{hard}$ are smeared by the integration over 
$p_{T \gamma}^{hard}$ in the fixed-order computation of 
$d\sigma/dp_{T \gamma}^{soft}$.

As we have discussed in our comments on the behaviour of
$d\sigma/dM_{\gamma \gamma}$ at $M_{\gamma \gamma} \sim M^{LO}_{\rm dir}$,
the LO kinematical configuration with
$p_{T \gamma}^{soft}=p_{T \gamma}^{hard}=p_H$ is highly sensitive to soft-gluon
effects and, in going from symmetric to asymmetric $p_T$ cut configurations
the transverse momenta of the harder and softer photon exchange their role with
respect to soft-gluon sensitivity. Therefore, in the case of asymmetric   
$p_T$ cuts, we expect \cite{Catani:1997xc} an evident unphysical behaviour of 
$d\sigma/dp_{T \gamma}^{soft}$ (rather than $d\sigma/dp_{T \gamma}^{hard}$).

This expectation is confirmed by the results for 
$d\sigma/dp_{T \gamma}^{soft}$ and $d\sigma/dp_{T \gamma}^{hard}$
that are presented in Fig.~2 of Ref.~\cite{Catani:2011qz}
(they refer to the asymmetric $p_T$ cut configuration with $p_H=40$~GeV and 
$p_S=25$~GeV). The NLO and NNLO results for $d\sigma/dp_{T \gamma}^{soft}$
in Fig.~2-right of Ref.~\cite{Catani:2011qz} show perturbative instabilities in
the region where $p_{T \gamma}^{soft} \simeq p_H=40$~GeV. In particular,
the LO result for $d\sigma/dp_{T \gamma}^{soft}$ has an unphysical threshold at
$p_{T \gamma}^{soft} = p_H$ and, therefore \cite{Catani:2011qz}, the NLO result
has the double-logarithmic divergent behaviour
$d\sigma^{NLO}/dp_{T \gamma}^{soft} 
\propto  + \;\ln^2(p_H - p_{T \gamma}^{soft})$ if $p_{T \gamma}^{soft}$ tends
to $p_H$ in the region where $p_{T \gamma}^{soft} < p_H$.
Analogously to the case of Eq.~(\ref{phard}) for symmetric $p_T$ cuts,
this divergent behaviour of $d\sigma^{NLO}/dp_{T \gamma}^{soft}$ for 
asymmetric $p_T$ cuts is directly related to the unphysical behaviour of the 
total cross section with nearly-symmetric $p_T$ cuts. Indeed, we have
\beq
\label{sigmaS}
\frac{\partial \,\sigma(p_H, p_S)}{\partial \, p_S} = - \; 
\left. \frac{d\sigma}{dp_{T \gamma}^{soft}} \,\right|_{p_{T \gamma}^{soft}=p_S} 
\;\;,
\quad \quad \quad \;\;\; \quad (p_H > p_S) \;\;,
\eeq 
and this equation relates the value of $d\sigma/dp_{T \gamma}^{soft}$ to
the $\Delta p_T$-slope of the total cross section of
Fig.~\ref{fig:deltapt}-right in the region where $\Delta p_T < 0$.
In particular, by using Eqs.~(\ref{totsym}) and (\ref{sigmaS}), the divergent
behaviour of $d\sigma^{NLO}/dp_{T \gamma}^{soft}$ if 
$p_{T \gamma}^{soft} \to p_H$ with $p_S < p_H$
is fully consistent with the NLO behaviour of
the $\Delta p_T$-slope of the total cross section with nearly-symmetric $p_T$ 
cuts and $\Delta p_T < 0$. Analogously to the case of other soft-gluon sensitive
observables, the perturbative instabilities of 
$d\sigma^{NLO}/dp_{T \gamma}^{soft}$ for asymmetric $p_T$ cuts can be removed
by performing all-order soft-gluon resummation \cite{Cieri:2015rqa}.

The unphysical fixed-order behaviour of the total cross section for
nearly-symmetric $p_T$ cuts obviously affects also the behaviour of related
differential cross sections. Various differential cross sections are affected in
different ways. The results in Fig.~\ref{fig:symmMgg} show that 
$d\sigma/dM_{\gamma \gamma}$ and $d\sigma/dp_{T \gamma}^{hard}$
have evident unphysical behaviour in limited regions of $M_{\gamma \gamma}$
and $p_{T \gamma}^{hard}$, respectively. In the case of 
$d\sigma/d\cos \theta^{*}$ we expect `unphysical' effects of normalization (and,
possibly, shape) in the region of central values of $\cos \theta^{*}$, which is
mostly sensitive to the $M_{\gamma \gamma}$ region that is relatively close
to $M^{LO}_{\rm dir}$. The case of the differential cross section
$d\sigma/dp_{T \gamma}^{soft}$ in the region where $p_{T \gamma}^{soft} > p_H$
is somehow `special', since its fixed-order results do not show evident signs of
unphysical behaviour (Fig.~\ref{fig:symmMgg}-right). However, the integral of 
$d\sigma/dp_{T \gamma}^{soft}$ over the region with 
$p_{T \gamma}^{soft} \geq p_{T}^{{\rm min}}$ 
exactly corresponds to the total cross section in a configuration of symmetric
$p_T$ cuts with $p_H=p_S=p_{T}^{{\rm min}}$. Since the fixed-order result of
such total cross section is unphysical, it turns out that the overall
normalization (independently of the detailed shape) of  
$d\sigma/dp_{T \gamma}^{soft}$ is sensitive to soft-gluon perturbative
instabilities at {\em all} values of $p_{T \gamma}^{soft}$ with 
$p_{T \gamma}^{soft} > p_H$.

As discussed in Sect.~\ref{subsec:theory} the NNLO results presented in this paper are affected by a systematic uncertainty related to the finite value of the parameter ($q_{T \rm cut}$ or $r_{\rm cut}$) that is used in the numerical implementation of
the $q_T$ subtraction method. An estimate of these uncertainties for total cross sections is explicitly reported in
the NNLO results presented in Table~\ref{Table:total}.
At fixed value of $q_{T \rm cut}$ ($r_{\rm cut}$), the systematic
uncertainties on differential cross sections tend to be larger than the
corresponding uncertainties on fiducial (total) cross sections. Such uncertainties can be enhanced in the presence of soft-gluon instabilities (as in the case of observables
that we have discussed in this subsection),
since the instabilities are smeared by
the finite value of $q_{T \rm cut}$ ($r_{\rm cut}$). More detailed studies
are needed to assess the
numerical precision of the NNLO results in these situations.
Such studies, however, cannot improve the physical predictivity of
the NNLO result, since the latter is affected by sizeable
theoretical uncertainties produced by the same (unphysical) soft-gluon
effects.

\section{Summary}
\label{sec:summa}

In this paper we have considered diphoton production in hadronic collisions
at LHC energies and we have presented a study of QCD radiative corrections 
at NLO and NNLO.
At NLO we have performed a thorough analysis of photon isolation and its perturbative
QCD effects, comparing results obtained by using
smooth cone and standard cone isolation.
While the former facilitates theoretical calculations by removing the fragmentation component (and the ensuing effects of the poorly-known non-perturbative fragmentation functions of the photon), the latter is the isolation criterion that is typically used in experimental analyses. 
We have then extended our study to NNLO, where only smooth cone isolation results
can be presently obtained.

Our main results can be summarised as follows.
\begin{itemize}

\item We have shown that lowest-order results for diphoton production are affected by 
large radiative corrections at higher orders. The large radiative corrections are due to 
partonic subprocesses with high parton multiplicity in the final state. As a consequence,
the radiative corrections are typically positive and they enhance diphoton production 
rates and related kinematical distributions. Since the final-state partons can be produced
outside the photon isolation cones, it also follows that the large radiative corrections 
have a relatively-mild dependence on the photon isolation prescription. In particular,
the quantitative differences between smooth and standard isolation tend to decrease
upon the inclusion of radiative corrections.

\item We have presented a detailed comparison of fiducial cross sections and differential
distributions obtained within standard and smooth cone isolation at NLO.
In the case of tight isolation ($\ETmax\sim$~few~GeV) the comparison 
shows that the two isolation procedures lead to results that are consistent within the 
corresponding scale uncertainties.
In the case of moderate isolation ($\ETmax\sim$~10~GeV), the same features hold true for 
diphoton observables that are `effectively' computed at NLO accuracy. In `effective' 
LO regions (such as at small values of $M_{\gamma\gamma}$ or $\Delta\Phi_{\gamma\gamma}$)
the two isolation procedures lead to NLO differences that are much smaller than the size
of the NNLO corrections computed within smooth isolation.
We thus conclude that the smooth cone isolation provides a consistent theoretical 
framework to compute radiative corrections up to NNLO and that, if photon isolation is 
sufficiently tight, the ensuing predictions can be reliably compared with experimental 
measurements carried out by using standard cone isolation with the same 
values of the isolation parameters ($\ETmax$ and cone isolation radius $R$).

\item An alternative approximation scheme of standard cone isolation consists in 
complementing the NLO calculation of the direct component with the LO computation
of the fragmentation component. We have shown that such a scheme features an 
unphysical NLO dependence on $\ETmax$ and, therefore, it is not recommended.

\item The NNLO computation of fiducial and differential cross sections shows that 
the NNLO corrections are rather large in
the phase space regions where the calculation is an `effective' NNLO prediction.
The impact of the NNLO corrections can become huge in 
phase space regions where the calculation is `effectively' an NLO prediction, 
as for the cases of small values of $M_{\gamma\gamma}$, low values of 
$\Delta \Phi_{\gamma \gamma}$ and relatively-large values of $p_{T \gamma \gamma}$.
We have also observed that NLO and
NNLO uncertainties obtained through scale variations do not overlap. This is mainly 
due to the significant contribution of the $qg$ initial-state partonic channel, 
which, although parametrically 
suppressed by one power of $\as$ with respect to the LO $q{\bar q}$ 
contribution, is enhanced by the large ${\cal L}_{qg}$ luminosity. As a consequence, 
the true theoretical uncertainty is larger than the one obtained by performing 
customary scale variations. A more reliable
estimate of the perturbative 
uncertainties can be obtained by properly taking into account the differences between 
the NLO 
and NNLO results. For instance, the perturbative uncertainty at NNLO (NLO) can be defined
as the half difference between the central scale predictions at NNLO (NLO) and NLO (LO).

\item The comparison of the NNLO calculation to the experimental results shows a clear 
improvement with respect to NLO in the description of the LHC data. The data tend 
to overshoot the NNLO predictions but, if the perturbative uncertainty is properly taken 
into account, the LHC data are consistent with the NNLO results in both `effective' NNLO 
and `effective' NLO regions.

\item We have discussed and remarked how diphoton rates and kinematical distributions are
strongly affected by the selection cuts that are typically applied on the transverse 
momenta of the photons. These selection cuts also lead to unphysical thresholds and
ensuing perturbative instabilities in the fixed-order computations of the 
$M_{\gamma\gamma}$ distribution, of the transverse-momentum spectra of the photons and 
of related observables. We have discussed the behaviour of the NLO and NNLO results in 
these threshold regions, by presenting the logarithmic structure of the perturbative 
instabilities in analytic form. The effect of the perturbative instabilities is tamed by 
considering sufficiently smeared observables. Such fixed-order perturbative instabilities
can be eliminated only through a proper all-order 
resummation of the logarithmically-enhanced contributions.

\end{itemize}
  
\noindent {\bf Acknowledgements.} We are very grateful to Eric Pilon and 
Jean-Philippe Guillet for their help with the program \texttt{DIPHOX}.
LC would like to thank the INFN of Florence for kind hospitality, 
while parts of this project were carried out. This research was supported in part 
by Fondazione Cariplo under the grant number 2015-0761, by the Swiss National Science Foundation (SNF) under contract
200020-169041 and by the Research Executive Agency (REA) of the European Union under the Grant Agreement number PITN-GA-2012-316704 ({\itshape Higgstools}).


\end{document}